\documentclass[letterpaper,titlepage,11pt]{article}
\usepackage[colorlinks=true,urlcolor=blue,citecolor=magenta]{hyperref}
\usepackage{amssymb,amsmath,amsfonts}
\usepackage{epsfig}
\usepackage{color}
\usepackage{graphicx}
\usepackage{array}
\usepackage{epstopdf}
\usepackage{eso-pic}

\newcommand\BackgroundPic{%
\put(0,0){%
\parbox[b][\paperheight]{\paperwidth}{%
\vfill
\centering
\includegraphics[width=\paperwidth,height=\paperheight,%
keepaspectratio]{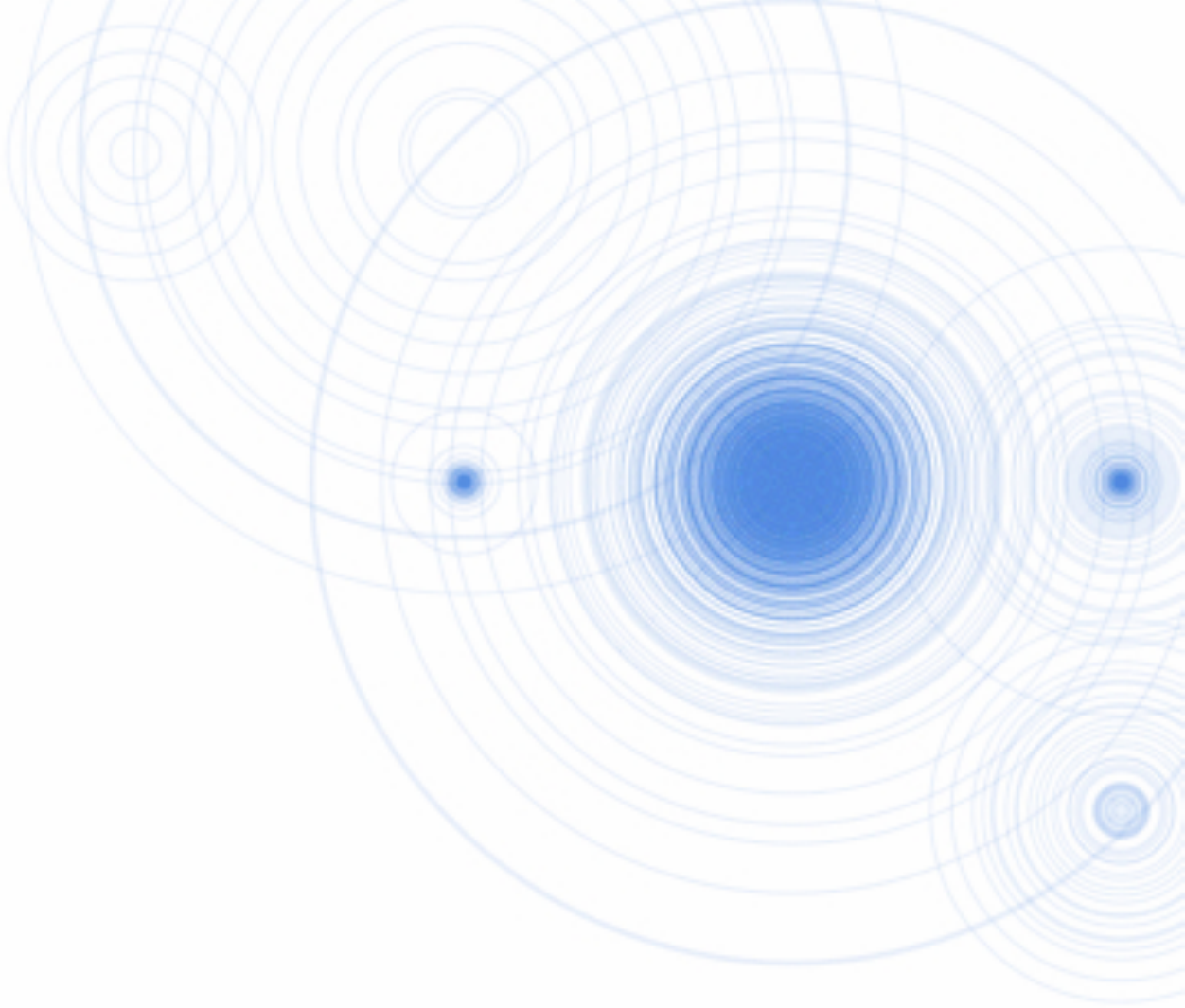}%
\vfill
}}}

\def\clock{{\count0=\time
           \divide\count0 60
           \ifnum\count0<10 0\fi\the\count0
           \multiply\count0 -60 \advance\count0 \time
           :\ifnum\count0<10 0\fi \the\count0
         }}
\newcommand{\timestamp}{{\small\vbox{\hbox{\tt\jobname.tex}
\hbox{\the\day/\the\month/\the\year, \clock}}}}

\newcommand{\be}{\begin{eqnarray}}
\newcommand{\ee}{\end{eqnarray}}
\newcommand{\beq}{\begin{eqnarray}}
\newcommand{\eeq}{\end{eqnarray}}

\newcommand{\beqa}{\begin{eqnarray}}
\newcommand{\eeqa}{\end{eqnarray}}

\let\oldsqrt\sqrt
\def\sqrt{\mathpalette\DHLhksqrt}
\def\DHLhksqrt#1#2{%
\setbox0=\hbox{$#1\oldsqrt{#2\,}$}\dimen0=\ht0
\advance\dimen0-0.2\ht0
\setbox2=\hbox{\vrule height\ht0 depth -\dimen0}%
{\box0\lower0.4pt\box2}}

\numberwithin{equation}{section}

\begin{document}
\AddToShipoutPicture*{\BackgroundPic}

\begin{titlepage}
 \vskip 1.8 cm

\centerline{\Huge \bf How Fluids Bend: the Elastic Expansion}
\vskip 0.6cm
\centerline{\Huge \bf for Higher-Dimensional Black Holes}
\vskip 1.5cm

\centerline{\large {\bf Jay Armas}}
\let\thefootnote\relax\footnote{\url{http://www.jacomearmas.com} }
\vskip 1.0cm

\begin{center}
\sl Albert Einstein Center for Fundamental Physics \\
\sl Institute for Theoretical Physics, University of Bern \\
\sl  Sidlerstrasse 5, 3012-Bern, Switzerland
\end{center}
\vskip 0.6cm

\centerline{\small\tt jay@itp.unibe.ch}

\vskip 1.3cm \centerline{\bf Abstract} \vskip 0.2cm \noindent
Hydrodynamics can be consistently formulated on surfaces of arbitrary co-dimension in a background space-time, providing the effective theory describing long-wavelength perturbations of black branes. When the co-dimension is non-zero, the system acquires fluid-elastic properties and constitutes what is called a fluid brane. Applying an effective action approach, the most general form of the free energy quadratic in the extrinsic curvature and extrinsic twist potential of stationary fluid brane configurations is constructed to second order in a derivative expansion. This construction generalizes the Helfrich-Canham bending energy for fluid membranes studied in theoretical biology to the case in which the fluid is rotating. It is found that stationary fluid brane configurations are characterized by a set of 3 elastic response coefficients, 3 hydrodynamic response coefficients and 1 spin response coefficient for co-dimension greater than one. Moreover, the elastic degrees of freedom present in the system are coupled to the hydrodynamic degrees of freedom. For co-dimension-1 surfaces we find a 8 independent parameter family of stationary fluid branes. It is further shown that elastic and spin corrections to (non)-extremal brane effective actions can be accounted for by a  multipole expansion of the stress-energy tensor, therefore establishing a relation between the different formalisms of Carter, Capovilla-Guven and Vasilic-Vojinovic and between gravity and the effective description of stationary fluid branes. Finally, it is shown that the Young modulus found in the literature for black branes falls into the class predicted by this approach - a relation which is then used to make a proposal for the second order effective action of stationary blackfolds and to find the corrected horizon angular velocity of thin black rings.

\end{titlepage}


\tableofcontents

\section{Introduction} \label{intro}
Higher-dimensional gravity has been shown to be fruitful as a testing ground for theories of hydrodynamics. In recent years many new properties, transport and response coefficients of (charged) fluids \cite{Bhattacharyya:2008jc, Bhattacharyya:2008mz, Erdmenger:2008rm, Banerjee:2008th, Armas:2011uf, Armas:2012ac} and superfluids \cite{Herzog:2011ec, Bhattacharya:2011eea, Bhattacharya:2011tra, Erdmenger:2011tj, Erdmenger:2012zu} have been uncovered through the study of long-wavelength fluctuations of black branes. These fluctuations can be along the worldvolume \cite{Camps:2010br, Caldarelli:2012hy, Gath:2013qya, Emparan:2013ila} or boundary directions \cite{Bhattacharyya:2008jc, Bhattacharyya:2008mz} where the fluid lives yielding the usual dynamics of viscous fluid flows, or transverse to it, originating instead genuine elastic behavior \cite{Armas:2011uf, Camps:2012hw, Armas:2012ac}.

The blackfold approach \cite{Emparan:2007wm, Emparan:2009cs, Emparan:2009at}, being the effective hydrodynamic theory that describes the long-wavelength perturbations of black branes, has taught us that the fluid system dual to the gravitational object needs not to live on the boundary of the space-time (fluid/gravity) but can also live on the horizon (membrane paradigm) or in an intermediate region \cite{Brattan:2011my, Emparan:2013ila}. In this regime, black holes and black branes should be seen as fluid branes \cite{Emparan:2009at, Armas:2012jg}: fluids living on dynamical surfaces of arbitrary co-dimension embedded in a background space-time.

The relation between the toy model of a fluid brane and gravitational physics can be thought of in the following way. Classically, black holes share many properties with other products of gravitational collapse, namely, they are characterized by a very dense distribution of matter but they differ from other stellar objects due to the formation of an event horizon. The crust of stars is composed of matter in a high pressure state, as the matter distribution that characterizes the black hole, and hence relativistic elasticity can be a good approximation for describing deformations of the star crust. It is then intuitively understandable why deformations of the matter source of a black hole can give rise to elastic behavior \cite{Binnington:2009bb, Poisson:2009qj, Kol:2011vg, Armas:2011uf}. On the other hand, it is also possible in some cases to perform inhomogeneous fluctuations of the event horizon without deforming the source leading to an effective viscous fluid behaviour \cite{Camps:2010br, Caldarelli:2012hy, Gath:2013qya, Emparan:2013ila}. These different possibilities are mapped onto the fluid brane toy model as transverse deformations to the surface where the fluid lives (elastic), or to fluctuations on the different thermodynamic quantities and fluid velocities that characterize the overall fluid configuration (fluid) \cite{Emparan:2009at, Armas:2012jg}. It is then possible to think of a black hole as having source degrees of freedom (elastic modes) and horizon degrees of freedom (hydrodynamic modes) which in general interact with each other when the entire system is subject to a perturbation. The natural question to ask is then: in which situations can the horizon be perturbed without perturbing the source and vice-versa? Focusing on stationary black brane configurations, in this work we partly answer this question when backreaction effects can be neglected.

Even though fluid branes have gained a role in gravitational physics, the physical problem of how a fluid living on a surface reacts to a deformation of the surface is a much older one dating back to the first observations of the strange biconcave shape of a red blood cell in the seventeenth century \cite{doi:10.1080/00018739700101488}. The models that described the shape of a cell, assuming a constant pressure along the surface, are the same models that describe the shape of a soap bubble, working under the principle of minimization of the surface area. However, in the 60's, Helfrich and Canham proposed what is now called the Helfrich-Canham bending energy for fluid membranes \cite{Canham197061, Helfrich1973}, consisting of introducing an additional contribution to the free energy of the cellular membrane of the form:
\beq \label{HCenergy}
\mathcal{F}[X^{\mu}]=\alpha\int_{A} dA\thinspace K^2~,
\eeq
where $K$ is the mean extrinsic curvature vector, $dA$ is the area element induced on the surface, $A$ is the area of the surface, and $\alpha$ is the modulus of rigidity. The inclusion of this term was enough to understand the biconcave shape of the red blood cell \cite{doi:10.1080/00018739700101488}, a fact that then led to the study of several properties of this fluid-elastic system \cite{Jay19731166, Jay1977169, Evans19893, 0305-4470-37-47-010, 0305-4470-35-30-302}. The rigidity term \eqref{HCenergy} has also found its role in particle physics when Polyakov \cite{Polyakov1986406}, and independently Kleinert \cite{Kleinert1986335}, added it to the Dirac-Nambu-Goto action in order to obtain an improved effective action for QCD. In this context, corrections of the type \eqref{HCenergy} are known as finite thickness corrections to brane actions.

In this paper, we generalize the Helfrich-Canham bending energy to account for possible pressure differences along the surface as well as for the possibility of the fluid being in stationary motion. Taking a brane-like stringy perspective, what we will be accomplishing is a systematic derivative correction to extremal and non-extremal brane actions, including all the possible correction terms quadratic in the extrinsic curvature, in the extrinsic twist potential and of second order in derivatives along the surface and obtaining the equations of motion for the resulting system. The first covariant formalism for obtaining the correct variations of surface tensors was developed by Carter in connection with his model of geodynamic-type (or stiff) strings and branes \cite{Carter:2000wv, Carter:1994yt, Carter:1997pb}. A more intuitive method that instead exploits the symmetries of the worldvolume surface was developed by Capovilla-Guven \cite{Capovilla:1994bs} and applied to different physical systems including lipid vesicles \cite{0305-4470-37-23-003, 0305-4470-38-41-002, 0305-4470-39-14-019}. On the other hand, a general framework for deriving the equations of motion of finite thickness probe branes based on a multipole expansion of the stress-energy tensor was developed by Vasilic-Vojinovic \cite{Vasilic:2006hs, Vasilic:2007wp}. Carter pointed out in \cite{Carter:1994yt} that the dynamics of geodynamic-type objects could be accounted for by adding dipole terms to the stress-energy tensor. Using mainly the formalism developed by Capovilla-Guven \cite{Capovilla:1994bs} and Vasilic-Vojinovic \cite{Vasilic:2007wp}, it is shown that this is indeed the case\footnote{Note that in the case of spinning corrections to the motion of point particles, it was already known that the equations of motion were captured by a multipole expansion of the stress-energy tensor \cite{Papapetrou:1951pa}. In the context of curvature corrections to cosmic strings, this was also derived in \cite{Letelier1992}. Here, we have generalized these results to arbitrary space-time dimension and for $p$-branes.}.

The upshot of the developments in hydrodynamics in connection with gravitational physics mentioned in the beginning of this section is that many systematic methods for constructing theories of fluid dynamics have been developed \cite{Banerjee:2012iz, Jensen:2012jh, Bhattacharyya:2012xi, Banerjee:2012cr, Dubovsky:2011sj, Bhattacharya:2012zx, Fukuma:2011pr}. Inspired by partition function methods \cite{Banerjee:2012iz, Jensen:2012jh, Bhattacharyya:2012xi, Banerjee:2012cr} and the effective action approach \cite{Dubovsky:2011sj, Bhattacharya:2012zx} we construct the free energy for generic stationary fluid flows living on dynamical surfaces to second order in a derivative expansion. We show that neutral fluids living in a surface of co-dimension greater than one are characterized by 3 elastic response coefficients, 3 hydrodynamic response coefficients and 1 spin response coefficient. Each of these response coefficients is associated with a particular term which is added to the action and from our analysis we conclude that the 3 elastic terms are coupled to 4 of the hydrodynamic terms due to geometric constraints. Contrary to the case of space-filling stationary fluid flows which are described by only 3 hydrodynamic response coefficients \cite{Banerjee:2012iz}, these fluids are described by a total of 7 independent response coefficients in a curved background. Using the generic form of the stress-energy tensor and dipole moment obtained from the action we match it with the Young modulus measured for black branes \cite{Armas:2011uf, Camps:2012hw} and hence propose a second order effective action for blackfolds which is then used to predict the corrected horizon angular velocity of thin black rings. We stress, however, that our construction is not the most general one to 2nd order in the derivative expansion. We will be ignoring terms that are constructed from the Levi-Civita symbol and which are dimension-dependent. The reason for this will be explained in Sec.~\ref{discussion}.

This paper is organized as follows. In Sec.~\ref{actions} we lay down the general framework for the variational calculus and the construction of the effective action in a derivative expansion. We then show how to iteratively account for higher order derivative corrections and obtain expressions that exhibit the coupling between the hydrodynamic and elastic modes. Sec.~\ref{poledipole} is quite technical but equally important. There we show how these corrections can be understood as a multipole expansion of the stress-energy tensor and proceed to construct conserved currents and charges. In Sec.~\ref{gravity} we match our results with the measurement of the Young modulus for black branes, firmly establishing what is meant by an elastic expansion for black holes. In Sec.~\ref{discussion} we comment on open issues and future work. We also include App.~\ref{geometry} where we describe in detail the notation used in this paper and collect different results of variational calculus and of geometry of embeddings. In App.~\ref{boundary} we analyze the boundary conditions for hydrodynamic modes.


\section{Effective action for (non)-extremal branes} \label{actions}
In this section we construct the most general effective action and corresponding free energy for a neutral stationary fluid brane quadratic in the extrinsic curvature and extrinsic twist potential to second order in a derivative expansion. We consider a $(p+1)$-dimensional surface with worldvolume topology $\mathcal{W}_{p+1}=\mathbb{R}\times \mathcal{B}_p$ embedded in a background space-time of $D=n+p+3$ dimensions with metric $g_{\mu\nu}(x^{\alpha})$ and space-time coordinates $x^\alpha$. The position of the surface is described by a set of mapping functions $X^{\mu}(\sigma^{a})$, where the set of coordinates $\sigma^a$ cover the embedded submanifold. The indices $\mu,\nu,\lambda...$ are space-time indices, the indices $a,b,c...$ denote directions along the worldvolume surface while the indices $i,j,k...$ denote transverse directions to the worldvolume. The worldvolume inherits an induced metric of the form $\gamma_{ab}=g_{\mu\nu}u^{\mu}_{a}u^{\nu}_{b}$, where $u^{\mu}_{a}=\partial_{a}X^{\mu}$, and if the worldvolume is bent one can assign to it an extrinsic curvature defined as ${K_{ab}}^{\mu}=\nabla_{a}u^{\mu}_{b}$~, where the covariant derivative $\nabla_{a}$ is defined in App.~\ref{geometry}. Any space-time vector $v^{\mu}$ can be decomposed into tangential and orthogonal components to the worldvolume surface such that $v^{\mu}=v^{a}u^{\mu}_{a}+v^{i}n^{\mu}_{i}$, where the set of tangential and orthogonal projectors satisfy $g_{\mu\nu}u^{\mu}_{a} n^{\nu}_{i}=0$. Given the set of orthogonal projectors one can define the extrinsic twist potential ${\omega_{a}}^{ij}=-{n^{j}}_{\mu}\nabla_{a}{n^{i\mu}}$, which is anti-symmetric in its two transverse indices, as well as the outer curvature ${\Omega_{ab}}^{ij}$ associated with it (see App.~\ref{geometry}). It is also useful to introduce projectors onto the worldvolume and onto its transverse space while keeping track of space-time indices. For that matter we introduce the first fundamental tensor $\gamma^{\mu\nu}$ defined as $\gamma^{\mu\nu}=\gamma^{ab}{u^{\mu}}_{a}{u^{\nu}}_{b}$~, which projects onto $\mathcal{W}_{p+1}$~, and the orthogonal projector ${\perp^{\mu\nu}}$ given by the expression ${\perp^{\mu\nu}}=g^{\mu\nu}-\gamma^{\mu\nu}$. Further, we assume the existence of a set of commuting worldvolume Killing vector fields that we take to be of the generic form:
\beq \label{KVF}
\textbf{k}^{a}\partial_{a}=\partial_{\tau}+\sum_{a=1}\Omega^{(a)}\partial\phi_{a}~,~\sigma^{a}=(\tau,\phi_{a},...)~.
\eeq
Here the set of angular velocities $\Omega^{(a)}$ is constant. The existence of such worldvolume Killing vector fields is a requirement for stationarity of the overall fluid configuration (see for example \cite{Caldarelli:2008mv}). We begin by reviewing the 0th order action and generalizing the analysis of \cite{Armas:2012jg} while simultaneously highlighting some of the elastic properties of fluid branes. The method for iteratively correcting the action then follows.


\subsection{0th order fluid-elastic system} \label{0thorder}
To 0th order in a derivative expansion the surface is described by an induced metric $\gamma_{ab}$ and the set of commuting Killing vector fields \eqref{KVF}. The only natural scalar invariant that can be constructed from these two is the modulus of the Killing vector defined as $\textbf{k}=|-\gamma_{ab}\textbf{k}^{a}\textbf{k}^{b}|^{1/2}$. Therefore, the action must be a functional of $\textbf{k}$\footnote{Note that $\lambda_0(\textbf{k})$ is also implicitly a function of the global temperature $T$. For explicit examples of non-extremal brane actions see \cite{Emparan:2009at, Grignani:2010xm, Caldarelli:2010xz, Emparan:2011hg, Armas:2012bk}.},
\beq \label{0action}
I\thinspace[X^{\mu}]=\int_{\mathcal{W}_{p+1}}\mathcal{L}\left(\sqrt{-\gamma},\textbf{k}\right)=\int_{\mathcal{W}_{p+1}}\!\!\!d^{p+1}\sigma\sqrt{-\gamma}~\lambda_0(\textbf{k})~~.
\eeq
In what follows, we will omit $d^{p+1}\sigma$ from our integrals. Our goal is to make a small deformation $\delta X^{\mu}$ of the worldvolume geometry along both tangential and orthogonal directions. To this aim we decompose the deformation as $\delta X^{\mu}=\Phi^{\mu}=\Phi^{a}u_{a}^{\mu}+\Phi^{i}n^{\mu}_{i}$. Under arbitrary small variations the metric changes by a Lie derivative such that
\beq \label{dgamma}
\delta\gamma_{ab}=2\nabla_{(a}\Phi_{b)}-2{K_{ab}}^{i}\Phi_i~~.
\eeq
Further, we assume the set of angular velocities in \eqref{KVF} to be held constant during the variation and hence the variations of the Killing vector fields $\delta k^{a}$ are zero. Therefore, the variation of the action \eqref{0action} can be written only in terms of the variation of the induced metric. This fact leads to a result of the form:
\beq \label{var0}
\delta I\thinspace[X^{\mu}]=\int_{\mathcal{W}_{p+1}}\sqrt{-\gamma}\left(\nabla_{a}\left(T^{ab}\Phi_{b}\right)-\Phi_{b}\nabla_{a}T^{ab}-T^{ab}{K_{ab}}^{i}\Phi_i\right)~~,
\eeq
where we have defined the monopole source of stress-energy tensor in the usual way,
\beq \label{st}
T^{ab}=\frac{2}{\sqrt{-\gamma}}\frac{\delta \mathcal{L}}{\delta \gamma_{ab}}~~.
\eeq
Before we analyze the explicit form of \eqref{st} note that in order to obtain the equations of motion from \eqref{var0} one must require $\delta I\thinspace[X^\mu]=0$. The first term in \eqref{var0} is a boundary term and can be written in the form
\beq
\int_{\partial \mathcal{W}_{p+1}}\sqrt{-h}T^{ab}\eta_a\Phi_b~~,
\eeq
where $\eta^a$ is a unit normal vector to the brane boundary and $h$ the determinant of the induced metric on the boundary. Hence, a well posed variational principle requires $T^{ab}\eta_a|_{\partial\mathcal{W}_{p+1}}=0$. The second term in \eqref{var0} results in an identity which is trivially satisfied as we will show below,
\beq \label{st01}
\nabla_{a}T^{ab}=0~~.
\eeq
In total, the variational principle \eqref{var0} yields the non-trivial equation of motion
\beq \label{st02}
T^{ab}{K_{ab}}^{i}=0~~.
\eeq
Eq.~\eqref{st01} expresses stress-energy tensor conservation and is responsible for the fluid behavior along worldvolume directions while Eq.~\eqref{st02} contains the elastic degrees of freedom which are manifested along transverse directions to the worldvolume. These equations will be analyzed in greater detail as we progress in this section.


\subsubsection*{The explicit form of the stress-energy tensor} 
To evaluate the stress-energy tensor \eqref{st} we need the variation of $\textbf{k}$ with respect to $\gamma_{ab}$, this is simply
\beq
\delta \textbf{k} = -\frac{\textbf{k}}{2} u^{a}u^{b}\delta \gamma_{ab}~,~u^{a}=\frac{\textbf{k}^a}{\textbf{k}}~.
\eeq
For an action of the type \eqref{0action} the stress-energy tensor takes the following form:
\beq \label{st0}
T^{ab}=T^{ab}_{(0)}=\lambda_0(\textbf{k})\gamma^{ab}-\lambda_0'(\textbf{k})\textbf{k}u^{a}u^{b}~.
\eeq
Here $\lambda_0'(\textbf{k})$ indicates a derivative with respect to $\textbf{k}$. This is easily recognized as the stress-energy tensor of a perfect fluid and indeed leads us to identify the set of normalized vectors $u^{a}u_{a}=-1$ as fluid velocities, while thermodynamics allows us to identify the pressure $P$ and energy density $\epsilon$ such that
\beq \label{id0}
P=\lambda_0(\textbf{k})~,~\epsilon+P=-\lambda_0'(\textbf{k})\textbf{k}~.
\eeq
This means that to 0th order, the action \eqref{0action} is the just the usual action for stationary perfect fluids \cite{Brown:1992kc} but now living on a submanifold in the ambient space-time. Assuming the Gibbs-Duhem relation $\epsilon+P=\mathcal{T}s$ and knowing that for such systems the constant global temperature $T$ is related to the local temperature via $T=\textbf{k}\mathcal{T}$ \cite{Caldarelli:2008mv}, using the identification \eqref{id0} one obtains an expression for the local entropy density
\beq \label{s0}
s=-\frac{1}{T}\lambda_0'(\textbf{k})\textbf{k}^2~.
\eeq
Given these identifications, the set of Eqs.~\eqref{st01} are interpreted as the fluid equations on the worldvolume $\mathcal{W}_{p+1}$. However, the stress-energy tensor \eqref{st0} satisfies the conservation equation \eqref{st01} regardless of any thermodynamic interpretation. Indeed, using \eqref{st0} in Eq.~\eqref{st01} results in 
\beq \label{0trick}
-\nabla^{b}\lambda_0(\textbf{k})+\frac{1}{2\thinspace\textbf{k}}\lambda_0'(\textbf{k})\nabla^{b}\textbf{k}^2=0~,
\eeq
which is trivially satisfied. In order to obtain the above, one needs to use the fact that the expansion $\theta\equiv \nabla_a u^a$ vanishes by virtue of the Killing equation $\nabla_{a}\textbf{k}_{b}=\nabla_{[a}\textbf{k}_{b]}$ and that for any worldvolume scalar or tensor $\mathbb{T}$, the Lie derivative along the Killing vector field vanishes $\pounds_{\textbf{k}}\mathbb{T}=0$.


\subsubsection*{The elasticity of fluid (mem)-branes}
Eq.~\eqref{st02} encodes the elastic degrees of freedom of the brane. In fact, note that it has a direct analog in terms of classical elasticity theory: it is the equation of motion obtained by varying the free energy of a thin stretched membrane when bending effects can be neglected \cite{Landau:1959te, Armas:2012jg}. For very thin membranes, the effect of bending is always subleading when compared to deformations caused by stretching or compression \cite{Landau:1959te}. To further see the connection with elastodynamics it is instructive to imagine the following scenario. Suppose that to a given fluid configuration satisfying Eq.~\eqref{st02} one applies a small deformation of the embedding surface in an arbitrary orthogonal direction $\Phi^i$. Prior to the deformation the metric $\bar{\gamma}_{ab}$ measured distances between fluid elements on the surface, while after the deformation the actual value of $\gamma_{ab}$ measures the new distances on $\mathcal{W}_{p+1}$. Therefore $\gamma_{ab}$ describes the state of strain of the brane and one can define the Lagrangian strain tensor $U_{ab}$\footnote{This is the usual definition of Lagrangian strain \cite{Quintana:1972} but now including the timelike direction as well.} as \cite{Armas:2012jg}
\beq \label{dU}
U_{ab}=-\frac{1}{2}\left(\gamma_{ab}-\bar\gamma_{ab}\right)~.
\eeq
For infinitesimal displacements along $\Phi^i$, the strain tensor changes by a Lie derivative and hence 
\beq
\pounds_{\Phi^{i}} U_{ab}\equiv dU_{ab}=-\frac{1}{2}d\gamma_{ab}={K_{ab}}^{i}\Phi_i~.
\eeq
Thus, we conclude that the extrinsic curvature tensor ${K_{ab}}^{i}$ measures the strain induced on the brane due to a deformation of the surface along orthogonal directions \cite{Armas:2011uf, Armas:2012jg}. Making the extra assumption of the existence of a background Killing vector field $\textbf{k}^{\mu}$ whose pullback onto the worldvolume coincides with the worldvolume Killing vector field $\textbf{k}^{a}$ allows us to write $\mathcal{T}s\dot{u}_{\mu}=-s\partial_{\mu}\mathcal{T}$. This, together with the form of the stress-energy tensor \eqref{st0} and the identification \eqref{id0}, allows us to rewrite Eq.~\eqref{st02} when contracted with the deformation vector $\Phi^i$ as
\beq \label{bulkeq}
dP=-Pd\mathcal{V}~,
\eeq
where we have defined the relative change in volume along an orthogonal direction as $d\mathcal V=(1/2)\gamma^{ab}d\gamma_{ab}$. Eq.~\eqref{bulkeq} allows for the definition of the modulus of hydrostatic compression $\mathcal{K}$ that measures the brane response to variations in volume such that
\beq
\frac{1}{\mathcal{K}}=\left(\frac{\partial \mathcal{V}}{\partial P}\right)_{T}=-\frac{1}{\lambda_0(\textbf{k})}~.
\eeq
Further elastic properties can be highlighted by using the first law of thermodynamics for the fluid $d\epsilon=\mathcal{T}ds$ and defining the solid density $\rho=\epsilon+P=-\lambda_0'(\textbf{k})\textbf{k}$ which along the directions $\Phi^{i}$ expresses the first law of thermodynamics for an elastic membrane that has been subject to hydrostatic compression $d\rho=\mathcal{T}ds-Pd\mathcal{V}$ \cite{Armas:2012jg}. To every fluid membrane one can also assign an elasticity tensor that measures the deformation of the stress-energy tensor \eqref{st0} due to surface deformations. In order to see this precisely one performs a small deformation of the equations of motion \eqref{st02} along the vector $\Phi^i$. To this aim, one requires the deformation of the extrinsic curvature tensor along orthogonal directions (see App.~\ref{geometry}),
\beq \label{perpK}
\delta_\perp {K_{ab}}^{i}={n^{i}}_{\mu}\nabla_{a}\left({\perp^{\mu}}_{\nu}\nabla_{b}(\Phi^j{n^{\nu}}_{j})\right)-{R^{i}}_{baj}\Phi^{j}-{K_{ac}}^{i}{K^{c}}_{bj}\Phi^{j}~~,
\eeq
where $R_{\mu\nu\lambda\rho}$ is the Riemann curvature tensor of the background geometry. Using this transformation rule, an infinitesimal deformation of Eq.~\eqref{st02} yields
\beq \label{eqlin}
E^{abcd}{K_{ab}}^{i}{K_{cd}}^{j}\Phi_j+T^{ab}{n^{i}}_{\mu}\nabla_{a}\nabla_{b}(\Phi^j{n^{\mu}}_{j})=T^{ab}{R^{i}}_{abj}\Phi^{j}~~,
\eeq
where we have defined the elasticity tensor for fluid branes $E^{abcd}$ through the Hookean relation $dT^{ab}=E^{abcd}dU_{cd}$, which for systems of the type \eqref{0action} takes the general form \cite{Armas:2012jg}
\beq \label{elasticity1}
E^{abcd}=2\left(\lambda_{0}(\textbf{k})\gamma^{a(c}\gamma^{d)b}-\left(\frac{\partial \lambda_0(\textbf{k})}{\partial \gamma_{ab}}\right)\gamma^{cd}-2\left(\frac{\partial^2 \lambda_0(\textbf{k})}{\partial\gamma_{ab}\partial \gamma_{cd}}\right)\right)~~.
\eeq
The structure of \eqref{elasticity1} is that of a material characterized by a varying modulus of compression. The case of Dirac-branes is obtained when $\lambda_0(\textbf{k})$ is constant and the two last terms vanish. The fact that the elasticity tensor is only probing compression and stretching is because we are working in the limit in which the fluid is confined to an infinitely thin surface as it will become clear in Sec.~\ref{poledipole}. We note that the linearized equation \eqref{eqlin} has been previously obtained by Carter in \cite{Carter:1997pb}, here we merely applied it to a fluid brane. It is interesting to note that Eq.~\eqref{eqlin} already encodes some of the structure of the equations of motion when bending effects are taken into account. The reason for this is explained in Sec.~\ref{poledipole}.


\subsubsection*{The free energy interpretation}
The 0th order action \eqref{0action} can be interpreted as the free energy of the fluid-elastic system \cite{Emparan:2009at, Caldarelli:2010xz}. After identifying the stress-energy tensor \eqref{st0} with that of a perfect fluid, then by defining the local Gibbs free energy density $\mathcal{G}$ and using \eqref{id0} one obtains
\beq
\mathcal{G}=\epsilon-\mathcal{T}s=-\lambda_0(\textbf{k})~.
\eeq
Wick rotating the integral over the worldvolume of the density $\mathcal{G}$ and integrating over the time circle of radius $\beta=1/T$ we obtain the total free energy $F$ of the system \eqref{0action}. This interpretation can also be realized at the level of the global charges. First note that the first term in the variational principle \eqref{var0} is a total divergence term. For the action \eqref{0action} to be invariant under an infinitesimal shift of worldvolume coordinates along a worldvolume Killing vector field, one must have that
\beq
\nabla_{a}\left(T^{ab}\textbf{k}_{b}\right)=0~.
\eeq
This is satisfied due to the Killing equation and the symmetry of $T^{ab}$. The set of surface currents $T^{ab}\textbf{k}_{b}$ is conserved and with those it is possible to construct a set of conserved surface charges (mass and angular momenta) of the form
\beq \label{MJ0}
M=\int_{\mathcal{B}_{p}}dV_{(p)}T^{ab}n_{a}\xi_b~~~,~~~J^{(a)}=-\int_{\mathcal{B}_{p}}dV_{(p)}T^{ab}n_{a}\chi_b^{(a)}~.
\eeq
In writing these expressions we have assumed that the worldvolume timelike Killing vector field $\xi^{a}\partial_a\equiv\partial_\tau$, whose norm is the redshift factor $R_0$, is hypersurface orthogonal with respect to $\mathcal{W}_{p+1}$. Further, we have introduced the spatial measure $dV_{(p)}$ on the worldvolume, a unit normal vector $n^{a}=\xi^a/R_0$ orthogonal to spacelike slices of $\mathcal{W}_{p+1}$ and defined the rotational Killing vector fields in \eqref{KVF} as $\chi^{(a)}\partial_a=\partial_{\phi^{a}}$. In Sec.~\ref{poledipole} we will show that this set of surface charges is indeed the set of global conserved charges associated with the fluid brane. Using the definition of the local entropy density \eqref{s0} it is straightforward to construct the global entropy of the system from the entropy current $J^{a}_{s}=su^{a}$:
\beq \label{0entropy}
S=-\int_{\mathcal{B}_{p}}dV_{(p)}J^{a}_{s}n_{a}~.
\eeq
Having defined the charges \eqref{MJ0} and the global entropy it is possible to verify that $F=M-\Omega^{(a)}J_{(a)}-TS$, where $\beta F=I_{\text{E}}$ and $I_{\text{E}}=-I$ is the Wick rotated (Euclidean) action. Hence the variational principle $d F=0$ with fixed $\Omega^{(a)}$ and global temperature $T$, requires the first law of thermodynamics to be satisfied \cite{Emparan:2009at, Caldarelli:2010xz}:
\beq
dM=\Omega^{(a)}dJ_{(a)}+TdS~.
\eeq


\subsection{General framework for higher order corrections} \label{framework}
We will now show how some of the ideas of the previous section can be pushed to second order in a derivative expansion. The general method, following \cite{Banerjee:2012iz, Bhattacharya:2012zx} , consists in adding all possible scalars constructed out of derivatives of worldvolume quantities to the action \eqref{0action}. We will split the type of corrections that can be supplemented to the action in three parts. Hydrodynamic corrections are those which only involve derivatives of fields that characterize the intrinsic geometry of the brane. These are, for example, terms proportional to $\nabla_{a}\textbf{k}$, $\nabla_{[a}\mathcal{\textbf{k}}_{b]}$, to the worldvolume Riemann tensor $\mathcal{R}_{abcd}$ or to the tangential projection of the background Riemann tensor $R_{abcd}$. Elastic corrections, on the other hand, are scalars proportional to the extrinsic curvature tensor ${K_{ab}}^{i}$. Note that by \eqref{dgamma} or \eqref{dU}, the extrinsic curvature tensor is a one-derivative term along transverse directions to the worldvolume. Finally, spin corrections are corrections proportional to the extrinsic twist potential ${\omega_{a}}^{ij}$~, which is also a one-derivative term. This means that we will consider a generic action of the form
\beq \label{action1}
I\thinspace[X^{\mu}]=\int_{\mathcal{W}_{p+1}}~\mathcal{L}(\sqrt{-\gamma},\gamma_{ab},\textbf{k}^{a},\nabla_{a},{K_{ab}}^{i},{\omega_{a}}^{ij})~.
\eeq
Since the Killing vectors $\textbf{k}^{a}$ are held constant during the variation, variations of \eqref{action1} can be analyzed solely through variations of the metric $\gamma_{ab}$, the extrinsic curvature tensor ${K_{ab}}^{i}$ and the extrinsic twist potential ${\omega_{a}}^{ij}$. For this purpose, we define the dipole moment ${\mathcal{D}^{ab}}_{i}$ and the spin current ${\mathcal{S}^{a}}_{ij}$ of the fluid-elastic system as\footnote{The reason for the interpretation of the quantities \eqref{dipolemoment} as the dipole moment and spin current will become clear in Sec.~\ref{poledipole}.}
\beq \label{dipolemoment}
{\mathcal{D}^{ab}}_{i}=\frac{1}{\sqrt{-\gamma}}\frac{\delta \mathcal{L}}{\delta {K_{ab}}^{i}}~~,~~{\mathcal{S}^{a}}_{ij}=\frac{1}{\sqrt{-\gamma}}\frac{\delta \mathcal{L}}{\delta {\omega_{a}}^{ij}}~~.
\eeq
To study small deformations of the geometry \eqref{action1} along both tangential and orthogonal directions to the worldvolume it is required the knowledge of the variation of ${K_{ab}}^{i}$ as well as of ${\omega_{a}}^{ij}$ along these directions. Using \eqref{perpK} for the transverse variations of ${K_{ab}}^{i}$, we have that for the tangential variations
\beq \label{deltaK}
\delta_{||} {K_{ab}}^{i}=\Phi^{c}{n^{i}}_{\rho}\nabla_c {K_{ab}}^{\rho}+2{K_{c(a}}^{i}\nabla_{b)}\Phi^{c}~,
\eeq
while for the extrinsic twist potential we have that (see App.~\ref{geometry})
\beq \label{deltaW}
\delta {\omega_{a}}^{ij}={n^{i}}_{\rho}{n^{j}}_{\lambda}\Phi^{b}\nabla_{b}{\omega_{a}}^{\rho\lambda}+{\omega_{b}}^{ij}\nabla_{a}\Phi^{b}-2{K_{ab}}^{[i}{n^{j]}}_{\rho}\nabla^{b}(\Phi^{k}{n_{k}}^{\rho})+{R^{ij}}_{ka}\Phi^{k}~~,
\eeq
where ${\omega_{a}}^{\lambda\rho}={n^{\lambda}}_{i}{n^{\rho}}_{j}{\omega_{a}}^{ij}$. Using the transformation rules \eqref{perpK},\eqref{deltaK} and \eqref{deltaW}, the total variation of \eqref{action1} takes the form
\beq
\begin{split} \label{var1}
\delta I\!=\!\!\int_{\mathcal{W}_{p+1}}\!\!\!\!\!\!\!\sqrt{-\gamma}\Big[ &\nabla_{a}\left(T^{ab}\Phi_b+{\mathcal{D}^{ac}}_{i}{K_{cb}}^{i}\Phi^{b}\!-\!\Phi_{i}{n^{i\mu}}\nabla_{b}{\mathcal{D}^{ab}}_{\mu}\!+\!{\mathcal{D}^{ab}}_{\mu}\nabla_b \Phi^{\mu}\!+\!{\mathcal{S}^{a}}_{ij}{\omega_{b}}^{ij}\Phi^{b}\!-\!2\thinspace{\mathcal{S}^{b}}_{ij}{K_{b}}^{ai}\Phi^{j}\right) \\ 
&+\Phi_{b}\left(-\nabla_{a}T^{ab}+{\mathcal{D}^{ac}}_{\rho}\nabla^{b}{K_{ac}}^{\rho}-2\thinspace\nabla_{a}\left({\mathcal{D}^{ac}}_{i}{{K^{b}}_c}^{i}\right)-{\mathcal{S}^{a}}_{ij}{{\Omega_{a}}^{bij}}-{\omega}^{b\rho\lambda}\nabla_{a}{\mathcal{S}^{a}}_{\rho\lambda}\right) \\ 
&+\Phi_i\left(-T^{ab}{K_{ab}}^{i}+{n^{i}}_{\rho}\nabla_{a}\nabla_{b}{\mathcal{D}^{ab\rho}}+{\mathcal{D}^{abj}}{R^{i}}_{ajb}-2\thinspace{n^{i}}_{\rho}\nabla_{b}\left({{\mathcal{S}^{a\rho}}_{j}} {K}^{abj}\right)+{\mathcal{S}}^{akj}{R^{i}}_{akj}\right)\Big]~,
\end{split}
\eeq
where we have used the definition of the outer curvature ${\Omega_{ab}}^{ij}$ (see App.~\ref{geometry}). As in the 0th order case \eqref{var0}, the variation yields a total divergence which can be integrated to a boundary term. However note that any component of the form $\nabla_{b}\Phi^{\mu}$ is not independent on the brane boundary, indeed one can decompose it such that $\nabla_{b}\Phi^{\mu}=\eta_b \eta^{a}\nabla_{a}\Phi^\mu+{v_{b}}^{\hat{a}}\nabla_{\hat a}\Phi^{\mu}$, where ${v_{a}}^{\hat{a}}$ are boundary coordinate vectors and the indices $\hat a,\hat b,...$ label boundary directions. The normal component to the brane boundary $\eta^a\nabla_{a}\Phi^\mu$ is independent but ${v_{a}}^{\hat{a}}\nabla_{\hat a}\Phi^{\mu}$ is not. Hence, assuming that the variations $\Phi^\mu$ vanish on the boundary of the brane boundary itself (if existing), a well defined variational principle requires
\begin{gather}
\nonumber
{\mathcal{D}}^{abi}\eta_a\eta_b|_{\partial\mathcal{W}_{p+1}}=0~~,~~{\mathcal{S}}^{aij}\eta_{a}|_{\partial\mathcal{W}_{p+1}}=0~~, \\ \label{bound1} \\ \nonumber
\left[\nabla_{\hat{a}}\left(\mathcal{D}^{abi}{n^{\mu}}_{i}\eta_a {v_{b}}^{\hat{a}}\right)-\eta_a\left(T^{ab}{u_{b}}^{\mu}+{\mathcal{D}^{ac}}_{i}{K_{c}}^{bi}{u_{b}}^{\mu}-{\perp^{\mu}}_{\rho}\nabla_{b}{\mathcal{D}^{ab\rho}}+2{n^{u}}_{j}{{\mathcal{S}^{bji}}}{{K_{b}}^{a}}_{i}\right)\right]|_{\partial\mathcal{W}_{p+1}}\!\!=0~~.
\end{gather}
Further, the second term in \eqref{var1} yields a set of non-trivial identities that must vanish. With the help of the Codazzi-Mainardi equation
\beq \label{CM}
{R^{i}}_{cba}={n^{i}}_{\rho}\left(\nabla_{b}{K_{ac}}^{\rho}-\nabla_{a}{K_{cb}}^{\rho}\right)~,
\eeq
and the Ricci integrability condition \cite{Carter:1992vb, Capovilla:1994bs}
\beq \label{RC}
{R_{ab}}^{ij}={\Omega_{ab}}^{ij}-{K_{ac}}^{i}{K_{b}}^{cj}+{K_{bc}}^{i}{K_{a}}^{cj}~~,
\eeq
it can be brought to the form
\beq\label{st11}
\nabla_{a}T^{ab}-{u_{\mu}}^{b}\nabla_a\nabla_c\mathcal{D}^{ac\mu}+2\thinspace{\mathcal{S}^{a}}_{ij}{K_{ac}}^{i}{K}^{bcj}=\mathcal{D}^{aci}{R^{b}}_{aic}+{\mathcal{S}^{aij}}{R^{b}}_{aij}-\omega^{bij}\nabla_{a}{\mathcal{S}^{a}}_{ij}~~.
\eeq
In Secs.~\ref{2nd} and \ref{2ndtwist} we will give a few examples of how this equation is satisfied for the actions we consider. Eq.~\eqref{st11} can be seen as the modified intrinsic dynamics of Eq.~\eqref{st01}. The reader may wonder if there is another definition of the stress-energy tensor for which Eq.~\eqref{st11} would just express its conservation. Indeed in Sec.~\ref{poledipole} we will show that it is possible to define the linear momentum which is always conserved in flat space. Finally, the variational principle $\delta I\thinspace[X^{\mu}]=0$ yields the non-trivial equation of motion:
\beq \label{st12}
T^{ab}{K_{ab}}^{i}={n^{i}}_{\rho}\nabla_{a}\nabla_{b}{\mathcal{D}^{ab\rho}}-2\thinspace{n^{i}}_{\rho}\nabla_{b}\left({\mathcal{S}_{a}}^{\rho j}{K^{ab}}_{j}\right)+{\mathcal{D}^{abj}}{R^{i}}_{ajb}+{\mathcal{S}}^{akj}{R^{i}}_{akj}~~.
\eeq
This is of course the modified version of the equation of motion \eqref{st02} due to the presence of dipole $\mathcal{D}^{abi}$ and spinning ${\mathcal{S}^{a}}_{ij}$ effects and, in fact, it is the generalization to arbitrary co-dimension and to curved backgrounds of the classical equation of motion of a deformed thin membrane when bending effects and rotation in transverse directions are taken into account. To see this precisely, let us focus on a flat background and consider the well studied case of the Helfrich-Canham bending energy \eqref{HCenergy} for co-dimension-1 surfaces in the absence of spinning effects, that is, ${\mathcal{S}^{a}}_{ij}=0$. In this case, as we will show below, we have $\mathcal{D}^{abi}=2\alpha \gamma^{ab}K^{i}$ where $K^i=\gamma^{ab}{K_{ab}}^{i}$ is the mean extrinsic curvature vector. The equation of motion \eqref{st12} can then be rewritten as
\beq \label{elasticity}
-2\thinspace\alpha\thinspace\nabla^{4}X^{i}+\mathcal{P}^{ab}{K_{ab}}^{i}=0~~,
\eeq
with $\mathcal{P}^{ab}=T^{ab}+2\alpha{K^{ab}}_{i}K^{i}$ and where $\nabla^{4}$ is the square of the Laplacian, or the biharmonic operator usually found in classical elasticity theory \cite{Landau:1959te}\footnote{Note that in classical elasticity theory $\mathcal{P}^{ab}$ is conserved while here, due to Eq.~\eqref{st11} it is not. However, the tensor $\mathcal{P}^{au}=\mathcal{P}^{ab}{u_{b}}^{\mu}-n^{\mu}_{i}\nabla_{a}\mathcal{D}^{abi}$ is conserved in flat space and $\mathcal{P}^{a\mu}{K_{a\mu}}^{i}=\mathcal{P}^{ab}{K_{ab}}^{i}$. Note also that for co-dimension-1 surfaces, the extrinsic twist ${\omega_{a}}^{ij}$ vanishes.}. Therefore, the theory described by Eqs.~\eqref{bound1}-\eqref{st12} is a general relativistic generalization of classical elastodynamics of thin membranes.

 It is also possible to construct surface currents and charges associated with the systems \eqref{action1}, however, since there are subtleties that can only be resolved when relating this approach to a multipole expansion of the stress-energy tensor, we will leave that for Sec.~\ref{poledipole}. On the other hand, the construction of the global entropy and entropy current as in \eqref{0entropy} to second order in the derivative expansion will be lacking in this work as it has not been developed for the systems we consider. Such endeavor is possible to accomplish and it will be presented in a future publication \cite{Armas:2014}.

For now, we will study the different contributions that can arise at each order in the derivative expansion. To each term involving one derivative we associate it with the expansion parameter $\varepsilon$. Two-derivative terms are of order $\mathcal{O}\left(\varepsilon^2\right)$. For clarity of explanation, we decompose the possible corrections to the intrinsic stress-energy tensor $T^{ab}$ into hydrodynamic $\Pi^{ab}$, elastic $\tau^{ab}$ and spin $\Theta^{ab}$ contributions such that
\beq
T^{ab}=T^{ab}_{(0)}+\Pi^{ab}+\tau^{ab}+\Theta^{ab}~.
\eeq
The higher order corrections $\Pi^{ab}$, $\tau^{ab}$, $\Theta^{ab}$ as well as the dipole moment $\mathcal{D}^{abi}$ and the spin current ${\mathcal{S}^{a}}_{ij}$ will be the sum of the contributions from the different scalars that can be added to the action \eqref{action1}. Generically,
\beq
\Pi^{ab}=\sum_\alpha \Pi^{ab}_{\alpha}~,~\tau^{ab}=\sum_\alpha \tau^{ab}_{\alpha}~,~\Theta^{ab}=\sum_\alpha \Theta^{ab}_{\alpha}~,~\mathcal{D}^{abi}=\sum_\alpha \mathcal{D}^{abi}_\alpha~,~{\mathcal{S}^{aij}}=\sum_\alpha {\mathcal{S}^{aij}_\alpha}~.
\eeq
Note however that the hydrodynamic corrections do not contribute to the dipole moment $\mathcal{D}^{abi}$ neither to the spin current $\mathcal{S}^{aij}$ by virtue of the definition \eqref{dipolemoment}.


\subsection{1st order action} \label{1st}
To 1st order in the derivative expansion the only possible terms that can be added are of hydrodynamic nature. This would be terms of the form:
\beq
\textbf{k}^{a}\nabla_{a}\textbf{k}~,~\nabla_{a}\textbf{k}^{a}~.
\eeq
However these terms vanish because $\textbf{k}^{a}$ is a Killing vector field. This is in agreement with the analysis of \cite{Banerjee:2012iz, Bhattacharya:2012zx} for both stationary and non-dissipative fluids. Indeed, for a generic dissipative fluid the entropy current to first order in derivatives is proportional to the square of the expansion $\theta\equiv \nabla_{a}u^{a}$ and to the square of the shear $\sigma^{ab}$\footnote{The shear tensor is defined as $\sigma^{ab}=P^{ac}P^{bd}\left(\nabla_{(c}u_{d)}-\frac{\theta}{p} \gamma_{cd}\right)$, where $P^{ab}$ is the projector along transverse directions to the fluid flows and defined as $P^{ab}=\gamma^{ab}+u^{a}u^{b}$.}. These two quantities vanish for stationary flows. One may also consider extrinsic curvature corrections to 1st order in derivatives but due to the transverse index in ${K_{ab}}^{i}$, any scalar built of the extrinsic curvature needs to be contracted with another copy of itself. Therefore, extrinsic curvature corrections only enter to second order in the derivative expansion. The same argument holds for terms proportional to the extrinsic twist potential\footnote{If we considered terms constructed from the Levi-Civita symbol then it would be possible to add a term of the form $\textbf{k}^{a}\epsilon_{ij}{\omega_{a}}^{ij}$. However, even though these are not considered in this paper, they are of physical interest, as they may be useful in the description of spinning black holes (see Sec.~\ref{discussion}).}. This is in agreement with the results obtained from gravity for the leading order corrections to thin black rings \cite{Emparan:2007wm} and blackfolds \cite{Camps:2012hw} for co-dimension surfaces higher than three for which a system of the type \eqref{action1} is supposed to be the correct description \cite{Armas:2011uf}. An exception to this argument of absence of extrinsic corrections to first order is the case of co-dimension-1 surfaces where we can simply omit the transverse index from ${K_{ab}}^{i}$. These cases will be dealt with in Sec.~\ref{cod1}. All in all, for surfaces of co-dimension greater than one we conclude that there are no corrections to first order in the derivative expansion.


\subsection{2nd order elastic corrections} \label{2nd}
To second order in the derivative expansion, focusing on the elastic corrections, we can add a total of 5 different terms to the action \eqref{action1} which are quadratic in the extrinsic curvature. These terms are of the following form:
\begin{gather}
\nonumber
\lambda_1(\textbf{k})K^{i}K_{i}~~,~~\lambda_2(\textbf{k})K^{abi}K_{abi}~~,~~\lambda_3(\textbf{k})\textbf{k}^{a}\textbf{k}^{b}{K_{ac}}^{i}{K^{c}}_{bi}~~, \\ \label{elastic2}
\\  \nonumber
\lambda_{4}(\textbf{k})\textbf{k}^{a}\textbf{k}^{b}{K_{ab}}^{i}K_{i}~~,~~\lambda_{5}(\textbf{k})\textbf{k}^{a}\textbf{k}^{b}\textbf{k}^{c}\textbf{k}^{d}{K_{ab}}^{i}K_{cdi}~~.
\end{gather}
The first term in \eqref{elastic2} is the generalization of the Helfrich-Canham bending energy \eqref{HCenergy} for surfaces of arbitrary co-dimension and for a non-trivial response coefficient  $\lambda_1(\textbf{k})$. The first two terms in \eqref{elastic2} for the case in which both $\lambda_1(\textbf{k})$ and $\lambda_2(\textbf{k})$ are constant build up Carter's model of stiff strings and branes \cite{Carter:1994yt} and have also been extensively studied by Guven \cite{Capovilla:1994bs, Arreaga:2000mr}. The other three terms have not been considered previously in the literature and they constitute the generalization of the Helfrich-Canham bending energy for stationary fluid (mem)-branes. However, the last two terms in \eqref{elastic2} are not independent and can be removed by a change of basis and a field redefinition as it will be explained in Sec.~\ref{relations}. Nevertheless, we consider their contribution as it will be convenient for later comparison with gravity results in Sec.~\ref{gravity}. We note that the reason why the second term in \eqref{elastic2} was not necessary to be added to the action of cellular membranes is that the first two terms are coupled to each other and in certain conditions, such as in flat space, they can be shown to be equivalent. We will explain this in detail in Sec.~\ref{hydro} when we study the coupling between modes. Decomposing each scalar in \eqref{elastic2} as $\lambda_\alpha(\textbf{k})\mathcal{L}_{\alpha}$, we summarize below the contributions of each term to $\tau^{ab}$ and $\mathcal{D}^{abi}$\footnote{Note that elastic corrections do not contribute to the spin current ${\mathcal{S}^{a}}_{ij}$.}:
\\ 
\renewcommand{\arraystretch}{1.5}
\begin{center} 
    \begin{tabular}{ | c | c | c |}
    \hline
    \color{red}{Scalar} & \color{red}{$\tau^{ab}_{\alpha}$} & \color{red}{$\mathcal{D}^{abi}_{\alpha}$} \\ \hline
    $\lambda_1(\textbf{k})\mathcal{L}_{1}$ & $\lambda_1(\textbf{k})\mathcal{L}_{1}\gamma^{ab}-\lambda_1'(\textbf{k})\textbf{k}\mathcal{L}_{1}u^{a}u^{b}-4\lambda_1(\textbf{k}){K^{ab}}_{i}{K}^{i}$ & $2\lambda_1(\textbf{k})\gamma^{ab}K^{i}$ \\   \hline
    $\lambda_2(\textbf{k})\mathcal{L}_{2}$ & $\lambda_2(\textbf{k})\mathcal{L}_{2}\gamma^{ab}-\lambda_2'(\textbf{k})\textbf{k}\mathcal{L}_{2}u^{a}u^{b}-4\lambda_2(\textbf{k}){K^{ac}}_{i}{{K^{b}}_{c}}^{i}$ & $2\lambda_2(\textbf{k})K^{abi}$ \\     \hline
    $\lambda_3(\textbf{k})\mathcal{L}_{3}$ & $\lambda_3(\textbf{k})\mathcal{L}_{3}\gamma^{ab}-\lambda_3'(\textbf{k})\textbf{k}\mathcal{L}_{3}u^{a}u^{b}-2\lambda_3(\textbf{k})\textbf{k}^{c}\textbf{k}^{d}{K^{a}}_{ci}{{K^{b}}_{d}}^{i}$ & $2\lambda_3(\textbf{k})\textbf{k}^d\textbf{k}^{(a}{{K^{b)}}_{d}}^{i}$ \\     \hline
    $\lambda_4(\textbf{k})\mathcal{L}_{4}$ & $\lambda_4(\textbf{k})\mathcal{L}_{4}\gamma^{ab}-\lambda_4'(\textbf{k})\textbf{k}\mathcal{L}_{4}u^{a}u^{b}-2\lambda_4(\textbf{k})\textbf{k}^{c}\textbf{k}^{d}{K^{ab}}_{i}{K_{cd}}^{i}$ & $\lambda_4(\textbf{k})\textbf{k}^a\textbf{k}^{b}K^{i}+\lambda_4(\textbf{k})\gamma^{ab}\textbf{k}^c\textbf{k}^{d}{K_{cd}}^{i}$ \\     \hline
    $\lambda_5(\textbf{k})\mathcal{L}_{5}$ & $\lambda_5(\textbf{k})\mathcal{L}_{5}\gamma^{ab}-\lambda_5'(\textbf{k})\textbf{k}\mathcal{L}_{5}u^{a}u^{b}$ & $2\lambda_5(\textbf{k})\textbf{k}^a\textbf{k}^{b}\textbf{k}^c\textbf{k}^d{K_{cd}}^{i}$ \\     \hline 
     \end{tabular}
         \label{tab:elastic} 
\end{center}
\vskip 0.3cm
We see that $\tau^{ab}$ is the sum of a total of 5 contributions $\tau^{ab}=\sum_{\alpha=1}^{5}\tau^{ab}_{\alpha}$ and contains a perfect fluid part (see \eqref{st0}) and an elastic part. When $\tau^{ab}$ is added to the 0th order contribution \eqref{st0} we decompose it into the total perfect fluid part $\mathcal{T}^{ab}$ and elastic part such that
\beq \label{Tdecelastic}
T^{ab}=T^{ab}_{(0)}+\tau^{ab}=\mathcal{T}^{ab}+ \mathcal{E}^{(acde}{K_{de}}^{i}{K^{b)}}_{ci}~~,
\eeq
where the elastic deformation to the worldvolume stress-energy tensor $\mathcal{E}^{abcd}$ has the form
\beq
 \mathcal{E}^{abcd}=-2\left(2\lambda_1(\textbf{k})\gamma^{ab}\gamma^{cd}+2\lambda_2(\textbf{k})\gamma^{ad}\gamma^{bc}+\lambda_3(\textbf{k})\gamma^{ad}\textbf{k}^b\textbf{k}^c+\lambda_4(\textbf{k})\gamma^{ab}\textbf{k}^c\textbf{k}^d\right)~~.
 \eeq
Rewriting the 0th order contribution \eqref{0action} as $\lambda_0(\textbf{k})\mathcal{L}_0$ with $\mathcal{L}_0=1$ and using the identification \eqref{id0}, the total fluid part $\mathcal{T}^{ab}$ can be written as
\beq
\mathcal{T}^{ab}=\mathcal{P}\gamma^{ab}+(\mathcal{E}+\mathcal{P})u^{a}u^{b}~~,
\eeq
where
\beq \label{id1}
\mathcal{P} =\sum_{\alpha=0}^{5}\lambda_{\alpha}(\textbf{k})\mathcal{L}_\alpha~~,~~\mathcal{E} + \mathcal{P}=-\sum_{\alpha=0}^{5}\lambda_{\alpha}'(\textbf{k})\textbf{k}\mathcal{L}_\alpha~~.
\eeq
Indeed, the quantity defined above as the modified pressure $\mathcal{P}$ is nothing more than the sum of all scalar contributions, that is, the 0th order scalar \eqref{0action} plus the 2nd order ones \eqref{elastic2}. The dipole moment $\mathcal{D}^{abi}$ is also the sum of 5 different contributions $\mathcal{D}^{abi}=\sum_{\alpha=1}^{5}\mathcal{D}^{abi}_{\alpha}$ and can be put into the elegant form
\beq \label{DI}
\mathcal{D}^{abi}=\mathcal{Y}^{abcd}{K_{cd}}^{i}~,
\eeq
where we have defined the Young modulus of the fluid brane as
\beq \label{YM}
\mathcal{Y}^{abcd}=2\left(\lambda_1\gamma^{ab}\gamma^{cd}+\lambda_2\gamma^{a(c}\gamma^{d)b}+\lambda_3\textbf{k}^{(a}\gamma^{b)(c}\textbf{k}^{d)}+\frac{\lambda_4}{2}\left(\gamma^{ab}\textbf{k}^{c}\textbf{k}^{d}+\gamma^{cd}\textbf{k}^a\textbf{k}^{b}\right) +\lambda_5\textbf{k}^{a}\textbf{k}^{b}\textbf{k}^{c}\textbf{k}^{d}\right)~.
\eeq
Here we have omitted the dependence of the response coefficients $\lambda_{\alpha}$ on $\textbf{k}$. The Young modulus \eqref{YM} is the physical quantity that encodes all the possible responses of stationary fluid branes to bending deformations. Furthermore, it exhibits all the symmetries of the usual elasticity tensor of a classical anisotropic crystal $\mathcal{Y}^{abcd}=\mathcal{Y}^{(ab)(cd)}=\mathcal{Y}^{cdab}$. As we will see in Sec.~\ref{gravity}, the Young modulus measured from gravity for bent black branes is a particular case of \eqref{YM}. Given the structures \eqref{DI}-\eqref{YM} and the identification \eqref{id0} we present two useful and equivalent rewritings of the action \eqref{action1} including only elastic corrections:
\beq \label{actionelastic}
I\thinspace[X^{\mu}]=\int_{\mathcal{W}_{p+1}}\sqrt{-\gamma}\left(P(\textbf{k})+\frac{1}{2}{\mathcal{D}^{ab}}_{i}{K_{ab}}^{i}\right)=\int_{\mathcal{W}_{p+1}}\sqrt{-\gamma}\left(P(\textbf{k})+\frac{1}{2}{\mathcal{Y}_{ab}}^{cd}{K^{ab}}_{i}{K_{cd}}^{i}\right)~.
\eeq
In Sec.~\ref{gravity}, we will study an example of this action. However we note that this action is exactly the type of action expected for the bending of thin membranes from classical elasticity theory \cite{Landau:1959te}, here we have presented a relativistic generalization of it.

\subsubsection*{Vanishing of the intrinsic equation of motion}
In Sec.~\ref{framework} we mentioned that the intrinsic equation of motion \eqref{st11} resulted in mere non-trivial identities for the actions we consider. One can verify that this is indeed the case for all corrections presented in the table above. Here we will just present two examples of how this is done. For the contribution $\lambda_1(\textbf{k})$, when introducing $\tau^{ab}_1$ and $\mathcal{D}^{abi}_1$ into Eq.~\eqref{st11} a part of the stress-energy tensor is conserved due to \eqref{0trick}, the remaining part is brought to the form
\beq
2\lambda_1(\textbf{k})K_{i}\left(\nabla^{b}K^{i}-\nabla_{a}K^{abi}\right)=2\lambda_1(\textbf{k})\gamma^{ac}K^{i}{R^{b}}_{aic}~.
\eeq
Using the Codazzi-Mainardi equation \eqref{CM}, the l.h.s. of the equation above can be seen to be equivalent to the r.h.s.. In fact, for the terms $\lambda_1(\textbf{k})$ and $\lambda_2(\textbf{k})$ one only needs to use \eqref{0trick} and \eqref{CM} to show the identity \eqref{st11}. We now take the case of the term $\lambda_5(\textbf{k})$. After using Eqs.\eqref{0trick} and \eqref{CM} we are left with
\beq \label{kill51}
 \lambda_5(\textbf{k})\textbf{k}^{a}\textbf{k}^{c}\textbf{k}^{e}{K_{cd}}^{i}{K_{aei}}\nabla^{b}\textbf{k}^{d}+\lambda_5(\textbf{k})\textbf{k}^{a}\nabla_{a}\left(\textbf{k}^{c}\textbf{k}^{d}\textbf{k}^{e}{K_{ed}}^{i}{{K_{c}}^{b}}_{i}\right)=0~.
 \eeq
Now we need to remember that the Lie derivative along any worldvolume Killing vector field of any worldvolume tensor must vanish, that is,
\beq \label{kill5}
\pounds_{\textbf{k}}\left(\textbf{k}^{c}\textbf{k}^{d}\textbf{k}^{e}{K_{ed}}^{i}{{K_{c}}^{b}}_{i}\right)=\textbf{k}^{a}\nabla_{a}\left(\textbf{k}^{c}\textbf{k}^{d}\textbf{k}^{e}{K_{ed}}^{i}{{K_{c}}^{b}}_{i}\right)-\textbf{k}^{c}\textbf{k}^{d}\textbf{k}^{e}{K_{ed}}^{i}{{K_{ac}}}_{i}\nabla^{a}\textbf{k}^{b}=0~.
\eeq
When Eq.~\eqref{kill5} is used, Eq.~\eqref{kill51} is trivially satisfied. This fact, together with the Killing equation, is enough to verify the identity \eqref{st11} for the terms $\lambda_3(\textbf{k})$ and $\lambda_4(\textbf{k})$.


\subsection{2nd order spin corrections} \label{2ndtwist}
Spin corrections to $\mathcal{O}\left(\varepsilon^2\right)$ in the derivative expansion are the terms that can be added to the action which are quadratic in the extrinsic twist potential. These do not contribute to the dipole moment $\mathcal{D}^{abi}$. Therefore, ignoring terms constructed from the Levi-Civita symbol, we can add two different scalars to the action
\beq \label{spinmodes}
\varpi_{1}(\textbf{k})\textbf{k}^{a}\textbf{k}^{b}{\omega_{aij}}{\omega_{b}}^{ij}~~,~~\varpi_{2}(\textbf{k}){\omega^{a}}_{ij}{\omega_{a}}^{ij}~~.
\eeq
Denoting each term by $\varpi_{\alpha}(\textbf{k})\mathcal{W}_{\alpha}$ we summarize below their contribution to the stress energy tensor $\Theta^{ab}$ and to the spin current $S^{aij}$:
\\ 
\renewcommand{\arraystretch}{1.5}
\begin{center} 
    \begin{tabular}{ | c | c | c |}
    \hline
    \color{red}{Scalar} & \color{red}{$ \Theta^{ab}_{\alpha}$} & \color{red}{$\mathcal{S}^{aij}_{\alpha}$} \\ \hline
    $\varpi_1(\textbf{k})\mathcal{W}_{1}$ &  $\varpi_1(\textbf{k})\mathcal{W}_{1}\gamma^{ab}-\varpi_1'(\textbf{k})\textbf{k}\mathcal{W}_{1}u^{a}u^{b}$ & $2\varpi_1(\textbf{k})\textbf{k}^{a}\textbf{k}^b{\omega_{b}}^{ij}$ \\   \hline
    $\varpi_2(\textbf{k})\mathcal{W}_{2}$ & $\varpi_2(\textbf{k})\mathcal{W}_{2}\gamma^{ab}-\varpi_2'(\textbf{k})\textbf{k}\mathcal{W}_{2}u^{a}u^{b}-2\varpi_2(\textbf{k})\omega^{aij}{\omega^{b}}_{ij}$ & $2\varpi_2(\textbf{k})\omega^{aij}$ \\   \hline
     \end{tabular}
\end{center} 
\vskip 0.3cm
The stress-energy tensor of these contributions contains a perfect fluid part that can be incorporated in the total pressure defined in \eqref{id1}. Defining a spin deformation tensor of the form
\beq
\mathcal{S}^{ab}=2\varpi_{1}(\textbf{k})\textbf{k}^{a}\textbf{k}^{b}+2\varpi_{2}(\textbf{k})\gamma^{ab}~~,
\eeq
the action \eqref{action1} including only the 0th order contribution \eqref{0action} and spin corrections can be written in two useful ways:
\beq \label{actionspin}
I\thinspace[X^{\mu}]=\int_{\mathcal{W}_{p+1}}\sqrt{-\gamma}\left(P(\textbf{k})+\frac{1}{2}{\mathcal{S}^{a}}_{ij}{\omega_{a}}^{ij}\right)=\int_{\mathcal{W}_{p+1}}\sqrt{-\gamma}\left(P(\textbf{k})+\frac{1}{2}\mathcal{S}^{ab}{\omega_{a}}^{ij}\omega_{bij}\right)~~.
\eeq
We will see however that the term $\varpi_{2}(\textbf{k})\mathcal{W}_{2}$ is not a desirable one in a physical theory.

\subsubsection*{Vanishing of the conservation equations}
As in Sec.~\ref{2nd} the vanishing of the conservation equation \eqref{st11} can be shown by following the same steps as in the previous section together with the definition of the outer curvature (see App.~\ref{geometry}) and the Ricci integrability condition \eqref{RC}. However, there is a subtlety in this demonstration. For the scalar $\varpi_1(\textbf{k})\mathcal{W}_{1}$, one can show that the last term in Eq.~\eqref{st11} vanishes independently, that is,
\beq \label{spinC}
{n^{i}}_{\rho}{n^{j}}_{\lambda}\nabla_{a}{\mathcal{S}}^{a\rho\lambda}_{1}=0~~,
\eeq
a result that follows from the fact that $\pounds_{\textbf{k}}\left(\textbf{k}^{b}{\omega_{b}}^{ij}\right)=\textbf{k}^{a}{n^{i}}_{\rho}{n^{j}}_{\lambda}\nabla_{a}\left(\textbf{k}^{b}{\omega_{b}}^{\rho\lambda}\right)=0$. However, even though the scalar $\varpi_{2}(\textbf{k})\mathcal{W}_{2}$ satisfies \eqref{st11}, it does not satisfy \eqref{spinC} individually. Eq.~\eqref{spinC} expresses conservation of angular momenta in transverse directions to the worldvolume, as we will see in Sec.~\ref{poledipole}. Therefore the term $\varpi_{2}(\textbf{k})\mathcal{W}_{2}$ is explicitly violating this conservation.


\subsection{2nd order hydrodynamic corrections and mode coupling} \label{hydro}
In this section we examine the possible 2nd order hydrodynamic corrections to the action \eqref{action1} and the stress-energy tensor $T^{ab}$ as well as the coupling between these modes and the ones found in the previous sections. For all these corrections the dipole moment $\mathcal{D}^{abi}$ vanishes. There are 7 types of scalars that can be constructed from the intrinsic geometry\footnote{Remember that since we are dealing with stationary flows both the expansion $\theta$ and shear $\sigma^{ab}$ are zero and hence any term proportional to these vanishes. These are, however, non-zero for generic non-dissipative flows \cite{Bhattacharya:2012zx}. Further, the reader may wonder why we have not considered couplings of the form ${R^{ij}}_{ij}$ or contractions with ${R^{i}}_{aib}$. The reason is that these terms are not well defined on the worldvolume and are related to geometric quantities with support on the transverse space. See \cite{Charmousis:2005ey} for a setting in which these terms play a role.}:
\begin{gather}
\nonumber
\upsilon_1(\textbf{k})\nabla_{a}\nabla^{a}\textbf{k}~~,~~\upsilon_2(\textbf{k})\mathcal{R}~~,~~\upsilon_3(\textbf{k})\textbf{k}^a\textbf{k}^b\mathcal{R}_{ab}~~, \\ \label{hydro2}
\\  \nonumber
\upsilon_{4}(\textbf{k})\nabla_{[a}\textbf{k}_{b]}\nabla^{[a}\textbf{k}^{b]}~~,~~\upsilon_{5}(\textbf{k})\nabla_{a}\textbf{k}\nabla^{a}\textbf{k}~~,~~\upsilon_{6}(\textbf{k}){{R^{a}}_{ba}}^{b}~~,~~\upsilon_{7}(\textbf{k})\textbf{k}^{a}\textbf{k}^{b}{{R^{c}}_{acb}}~~.
\end{gather}
Here we have introduced the Ricci tensor $\mathcal{R}_{ab}$ and Ricci scalar $\mathcal{R}$ of the worldvolume geometry. The terms $\upsilon_{6}(\textbf{k})$ and $\upsilon_{7}(\textbf{k})$ can be related to some of the other hydrodynamic terms and to the elastic modes in \eqref{elastic2} through the Gauss-Codazzi equation as we will explain below. This of course expresses the extra difficulty of the problem compared to \cite{Banerjee:2012iz, Bhattacharya:2012zx} since not only couplings to the worldvolume geometry need to be considered, but also couplings to the background geometry. Further, the terms $\upsilon_1(\textbf{k})$, $\upsilon_3(\textbf{k})$, $\upsilon_4(\textbf{k})$ and $\upsilon_5(\textbf{k})$ are not independent of each other and are related by on-shell equivalences. To see this note that
\beq \label{b1}
\begin{split}
\int_{\mathcal{W}_{p+1}}\sqrt{-\gamma}\upsilon_3(\textbf{k})\textbf{k}^a\textbf{k}^b\mathcal{R}_{ab}=&-\int_{\mathcal{W}_{p+1}}\sqrt{-\gamma}\upsilon_3(\textbf{k})\left[\nabla_{a},\nabla_{b}\right]\textbf{k}^a\textbf{k}^b \\
=&\int_{\mathcal{W}_{p+1}}\sqrt{-\gamma}\upsilon_3(\textbf{k})\left(\nabla_{a}\textbf{k}\nabla^{a}\textbf{k}+\textbf{k}\nabla_{a}\nabla^{a}\textbf{k}-\nabla_{[a}\textbf{k}_{b]}\nabla^{[a}\textbf{k}^{b]}\right)~.
\end{split}
\eeq
Moreover, the terms $\upsilon_1(\textbf{k})$ and $\upsilon_5(\textbf{k})$ are directly related to each other, modulo a boundary term, such that,
\beq\label{b2}
\int_{\mathcal{W}_{p+1}}\sqrt{-\gamma}\upsilon_1(\textbf{k})\nabla_{a}\nabla^{a}\textbf{k}=-\int_{\mathcal{W}_{p+1}}\sqrt{-\gamma}\upsilon_1'(\textbf{k})\nabla_{a}\textbf{k}\nabla^{a}\textbf{k} + \text{boundary term}~~.
\eeq
Therefore, using Eqs.~\eqref{b1}-\eqref{b2} we can eliminate the terms $\upsilon_4(\textbf{k})$ and $\upsilon_5(\textbf{k})$ and only consider the first three in \eqref{hydro2} \footnote{This is exactly the same type of analysis as in \cite{Bhattacharya:2012zx} but now applied to a submanifold embedded in a background space-time.}. The scalars \eqref{hydro2} yield boundary contributions that were not included in the general analysis of \eqref{var1}. For this reason we have prepared App.~\ref{boundary} where these issues are clarified. 

In order to perform orthogonal variations of \eqref{hydro2} it is only necessary to know how the worldvolume Christoffel symbols ${\gamma_{ab}}^{c}$ vary with respect to the induced metric (see App.~\ref{geometry}). Denoting each term by $\upsilon_\alpha(\textbf{k})\mathcal{V}_\alpha$, we summarize below the contribution of each of these 3 relevant hydrodynamic terms to the stress-energy tensor $\Pi^{ab}$:
\\
\newcolumntype{C}[1]{>{\centering\let\newline\\\arraybackslash\hspace{0pt}}m{#1}}
\renewcommand{\arraystretch}{1.7}
\begin{center} 
    \begin{tabular}{ | c | C{15cm} |}
    \hline
    \color{red}{Scalar} & \color{red}{$\Pi^{ab}_{\alpha}$} \\ \hline
    $\upsilon_1(\textbf{k})\mathcal{V}_1$ & $\upsilon_1(\textbf{k})\mathcal{V}_1\gamma^{ab}-\upsilon_1'(\textbf{k})\textbf{k}\mathcal{V}_{1}u^{a}u^{b}-\gamma^{ab}\nabla_{c}\left(\upsilon_1(\textbf{k})\nabla^{c}\textbf{k}\right)-\textbf{k}u^{a}u^{b}\nabla_{c}\nabla^{c}\upsilon_1(\textbf{k})-2\upsilon_1(\textbf{k})\nabla^{a}\nabla^{b}\textbf{k}+2\nabla^{(a}\left(\upsilon_1(\textbf{k})\nabla^{b)}\textbf{k}\right)$ \\   \hline
     $\upsilon_2(\textbf{k})\mathcal{V}_2$ & $\upsilon_2(\textbf{k})\mathcal{V}_2\gamma^{ab}-\upsilon_2'(\textbf{k})\textbf{k}\mathcal{V}_{2}u^{a}u^{b}-2\upsilon_2(\textbf{k})\mathcal{R}^{ab}+2\nabla^{a}\nabla^{b}\upsilon_2(\textbf{k})-2\gamma^{ab}\nabla_{c}\nabla^{c}\upsilon_2(\textbf{k})$ \\   \hline
      $\upsilon_3(\textbf{k})\mathcal{V}_3$ & $\upsilon_3(\textbf{k})\mathcal{V}_3\gamma^{ab}-\upsilon_3'(\textbf{k})\textbf{k}\mathcal{V}_{3}u^{a}u^{b}-\nabla_{c}\nabla^{c}\left(\upsilon_3(\textbf{k})\textbf{k}^{a}\textbf{k}^{b}\right)+\gamma^{ab}\nabla_{c}\nabla_{d}\left(\upsilon_3(\textbf{k})\textbf{k}^{c}\textbf{k}^{d}\right)-2\nabla_{c}\nabla^{(a}\left(\upsilon_3(\textbf{k})\textbf{k}^{b)}\textbf{k}^{c}\right)$ \\   \hline
     \end{tabular}
      \label{tab:hydro}
\end{center} 
\vskip 0.3cm
Looking at the table above, we see that again there are contributions of the perfect fluid form which can be included in the definition of $\mathcal{P}$ and $\mathcal{E}$ in \eqref{id1}, and other new contributions which are second order in derivatives of the intrinsic variables. It is also possible to show that all such contributions satisfy the identity \eqref{st11}. The only necessary ingredients to show this are Eq.~\eqref{0trick}, the Codazzi-Mainardi equation \eqref{CM}, the fact that the worldvolume Einstein tensor is divergenceless $\nabla_{a}\mathcal G^{ab}=\nabla_{a}\left(\mathcal{R}^{ab}-(1/2)\gamma^{ab}\mathcal{R}\right)=0$ and stationarity of the overall configuration.

\subsubsection*{Coupling between elastic and hydrodynamic modes}
Here we study the coupling between the elastic modes \eqref{elastic2} and the hydrodynamic modes \eqref{hydro2}. The most important piece of knowledge is the Gauss-Codazzi equation \cite{Carter:1992vb},
\beq \label{GC}
R_{abcd}=\mathcal{R}_{abcd}-{K_{ac}}^{i}{K_{bdi}}+{K_{ad}}^{i}{K_{bci}}~.
\eeq
relating the intrinsic geometry to the extrinsic geometry. We begin by contracting \eqref{GC} with the induced metric $\gamma^{bd}$, this leads to
\beq \label{GCg}
\gamma^{bd}R_{abcd}=\mathcal{R}_{ac}-{K_{ac}}^{i}K_{i}+{K_{ab}}^{i}{K^{b}}_{ci}~.
\eeq
If further contracted with $\textbf{k}^{a}\textbf{k}^c$, this equation expresses the coupling between the hydrodynamic terms $\upsilon_3(\textbf{k}),\upsilon_7(\textbf{k})$ and the two elastic contributions $\lambda_3(\textbf{k}),\lambda_4(\textbf{k})$. Indeed we can see why we can eliminate $\upsilon_7(\textbf{k})$ in favor of the other three terms. Hence, if $\gamma^{bd}R_{abcd}$ is non-zero, only 2 of these 4 terms can be regarded as independent\footnote{Note that the elastic contribution proportional to $\lambda_4(\textbf{k})$ can be gauged away as it will be explained in Sec.~\ref{relations}.}, while if $\mathcal{R}_{ac}$ or $\gamma^{bd}R_{abcd}$ is vanishing, only 1 of the 3 remaining terms can be regarded as independent. However, if both the background and the worldvolume are flat, the two elastic terms left are exactly equivalent to each other and hence both can be gauged away as explained in Sec.~\ref{relations}. If we now contract \eqref{GCg} with $\gamma^{ac}$ we obtain
\beq
\gamma^{ac}\gamma^{bd}R_{abcd}=\mathcal{R}-{K}^{i}K_{i}+{K^{ab}}_{i}{K_{ab}}^{i}~.
\eeq
This equation, in turn, expresses the coupling between the hydrodynamic terms $\upsilon_1(\textbf{k}),\upsilon_{6}(\textbf{k})$ and the two elastic constributions $\lambda_1(\textbf{k}),\lambda_2(\textbf{k})$. Therefore, if the background is curved, only 3 of these 4 terms can be regarded as independent and again, if ${{R^{a}}_{ba}}^{b}$ or $\mathcal{R}$ vanish, only 2 of the 3 remaining terms are independent. If both the background and the worldvolume are flat only 1 of the 2 elastic contributions is independent. For the fluid membranes studied in theoretical biology for which the Helfrich-Canham bending energy \eqref{HCenergy} was crucial to understand, the term $\upsilon_2(\textbf{k})=constant$, even though entering at the same order in derivatives, is not added to the action as it is a purely topological invariant. Moreover, in flat space, and since the two elastic contributions $\lambda_1(\textbf{k})$ and $\lambda_2(\textbf{k})$ are coupled through the relation \eqref{GC}, there was no need to consider $\lambda_2(\textbf{k})$ in describing red blood cells. On the other hand, the hydrodynamic contribution $\upsilon_1(\textbf{k})$ is independent and can be excited without exciting the remaining ones. Furthermore, the spin modes \eqref{spinmodes} are decoupled from the hydrodynamic and elastic modes to order $\mathcal{O}\left(\varepsilon^2\right)$.

Summarizing our results, for surfaces of co-dimension greater than one, if the background and the worldvolume are curved, the fluid-elastic system \eqref{action1} admits a 7 parameter family of stationary fluid branes\footnote{We are taking into account the contribution $\varpi_{1}(\textbf{k})$ but not $\varpi_{2}(\textbf{k})$ since the latter is pathological (see Sec.~\ref{2ndtwist}). We are also not taking into account the contributions $\lambda_{4}(\textbf{k})$ and $\lambda_{5}(\textbf{k})$ since it can be removed by a change of basis, see Sec.~\ref{relations}.}. On the other hand, if either the background or the worldvolume are flat, there exists a 5 parameter family of stationary fluid branes. Finally, if both the background and the worldvolume are flat, there exists a 3 parameter family described by the response coefficients $\lambda_{1}(\textbf{k}),~\varpi_{1}(\textbf{k}),~\upsilon_1(\textbf{k})$. The presence of elastic degrees freedom introduces further non-trivial response coefficients than those analyzed in \cite{Banerjee:2012iz, Jensen:2012jh, Bhattacharya:2012zx} for both stationary and non-dissipative space-filling fluid flows.


\subsection{Fluid membranes and co-dimension-1 surfaces} \label{cod1}
As mentioned in Sec.~\ref{1st}, in the case of co-dimension-1 surfaces there are non-trivial elastic corrections to 1st order in the derivative expansion. Since the transverse space is only 1-dimensional, the extrinsic curvature tensor can be written as $K_{ab}^{\perp}=K_{ab}$. We omit the transverse index and only use it if necessary. For this reason, there are two non-trivial and independent elastic contributions to 1st order in derivatives
\beq \label{1modes}
\vartheta_1(\textbf{k})K~~,~~\vartheta_2(\textbf{k})\textbf{k}^{a}\textbf{k}^{b}K_{ab}~~.
\eeq
These terms contribute to $T^{ab}$ and $\mathcal{D}^{ab\perp}$ while the equations of motion \eqref{bound1}-\eqref{st12} remain the same but have only one component along the transverse space. The first term in \eqref{1modes} had been previously considered in the literature \cite{Capovilla:1994bs} but it does not play a role in real fluid membranes since the mean extrinsic curvature is required to be kept constant due to experimental constraints, that is, the variation of \eqref{action1} must be supplemented with local constraints \cite{0305-4470-37-23-003}. The second term in \eqref{1modes} has not been previously considered in the literature and is accounting for the fact that the fluid may be rotating. However, it can be removed by a change of basis as it will be explained in Sec.~\ref{relations}. Nevertheless, we consider its contribution. Denoting each scalar in \eqref{1modes} by $\vartheta_\alpha(\textbf{k})\mathcal{C}_{\alpha}$ and their contributions to $T^{ab}$ as $\tilde \tau^{ab}$, we summarize the results below:
\\ 
\renewcommand{\arraystretch}{1.5}
\begin{center} 
    \begin{tabular}{ | c | c | c |}
    \hline
    \color{red}{Scalar} & \color{red}{$\tilde \tau^{ab}_{\alpha}$} & \color{red}{$\mathcal{D}^{ab \perp}_{\alpha}$} \\ \hline
    $\vartheta_1(\textbf{k})\mathcal{C}_{1}$ & $\vartheta_1(\textbf{k})\mathcal{C}_{1}\gamma^{ab}-\vartheta_1'(\textbf{k})\textbf{k}\mathcal{C}_{1}u^{a}u^{b}-2\vartheta_1(\textbf{k}){K^{ab}}$ & $\vartheta_1(\textbf{k})\gamma^{ab}$ \\   \hline
    $\vartheta_2(\textbf{k})\mathcal{C}_{2}$ & $\vartheta_2(\textbf{k})\mathcal{C}_{2}\gamma^{ab}-\vartheta_2'(\textbf{k})\textbf{k}\mathcal{C}_{2}u^{a}u^{b}$ & $\vartheta_2(\textbf{k})\textbf{k}^{a}\textbf{k}^b$ \\   \hline
     \end{tabular}
\end{center} 
\vskip 0.3cm
As one can see, there is a total contribution to the pressure and energy densities defined in \eqref{id1}. Summing up the two different contributions to the dipole moment, one concludes that for co-dimension-1 surfaces, it is necessary to add to the action composed of the terms \eqref{elastic2} and \eqref{hydro2} a 1st order contribution of the form
\beq
I_{(1)}\thinspace[X^{\mu}]=\int_{\mathcal{W}_{p+1}}\sqrt{-\gamma}\left(\vartheta_1(\textbf{k})\gamma^{ab}+\vartheta_2(\textbf{k})\textbf{k}^{a}\textbf{k}^{b}\right)K_{ab}=\int_{\mathcal{W}_{p+1}}\sqrt{-\gamma}~\mathcal{D}^{ab}_\perp K_{ab}~~.
\eeq
Here, $\mathcal{D}^{ab}_{\perp}$ itself measures the most general response of a stationary fluid brane due to a 1st order bending. To second order there is also another set of contributions that can be added to the action which mix the fluid and elastic behavior\footnote{The extrinsic twist potential does not play a role in co-dimension-1 surfaces since it vanishes.}. These are
\begin{gather}
\nonumber
\vartheta_3(\textbf{k})\textbf{k}^{a}\textbf{k}^{b}\textbf{k}^{c}\nabla_{a}K_{bc}~~,~~\vartheta_4(\textbf{k})\textbf{k}^{b}\nabla_{a}{K^{a}}_{b}~~,~~\vartheta_5(\textbf{k})\textbf{k}^a\nabla_{a}K~~, \\ \label{cod12}
\\  \nonumber
\vartheta_{6}(\textbf{k})\textbf{k}^{a}{K^{b}}_{a}\nabla_{b}\textbf{k}~~,~~\vartheta_{7}(\textbf{k})\gamma^{ac}\textbf{k}^{b}{R^{\perp}}_{cba}~~.
\end{gather}
However, the term $\vartheta_{5}(\textbf{k})$ vanishes since $\textbf{k}^{a}$ is a Killing vector field while the contributions $\vartheta_{4}(\textbf{k})$ and $\vartheta_{6}(\textbf{k})$ can be related to the term $\vartheta_3(\textbf{k})$ by on-shell equivalences as in Eq.~\eqref{b2}. Furthermore, the last contribution $\vartheta_{7}(\textbf{k})$ can be related to two other terms as we will explain below. It is then only sufficient to consider the first contribution in \eqref{cod12}. For this we need new machinery. For simplicity, we examine an action of the form:
\beq \label{actionCOD}
I\thinspace[X^{\mu}]=\int_{\mathcal{W}_{p+1}}~\mathcal{L}(\sqrt{-\gamma},\gamma_{ab},\textbf{k}^{a},\nabla_{a}{K_{bc}})~.
\eeq
To study terms proportional to $\nabla_{a}{K_{bc}}$ it is necessary to define a new moment $\mathcal{D}^{abc}_{\perp}$ of quadrupole nature:
\beq
{\mathcal{D}^{abc}_{\perp}}=\frac{1}{\sqrt{-\gamma}}\frac{\delta\mathcal{L}}{\delta \nabla_{a}{K_{bc}}^{\perp}}~~.
\eeq
Using App.~\ref{geometry} for the correct variations we find the modified intrinsic and extrinsic dynamics of the fluid-elastic system:
\beq \label{stcod1}
\nabla_{a}T^{ab}+3\nabla_{a}\mathcal{D}^{adc}\nabla_{d}{K^{b}}_{c}+2\mathcal{D}^{adc}\nabla_{d}\nabla_{a}{K^{b}}_{c}=2\mathcal{D}^{adc}{{{R^{e}}_{da}}^{b}}{K_{ec}}+\nabla_{a}\mathcal{D}^{adc}{{{R^{\perp}}_{cd}}^{b}}~~,
\eeq
\beq\label{stcod2}
\begin{split}
&T^{ab}{K_{ab}}-\left(3\nabla_{a}\mathcal{D}^{abc}+2\nabla_{a}\mathcal{D}^{bac}\right)K_{bd}{K^{d}}_{c}+2\nabla_{d}\mathcal{D}^{abc}{K_{ab}}{K^{d}}_c \\
&-2\mathcal{D}^{abc}\left(K_{ad}\nabla_{b}{K^{d}}_{c}+{K_{bd}}\nabla_{a}{K^{d}}_{c}-K_{ab}\nabla_{d}{K^{d}}_{c}\right)=\nabla_{a}\mathcal{D}^{abc}{R^{\perp}}_{bc\perp} ~~.
\end{split}
\eeq
The boundary conditions are summarized in App.~\ref{boundary}. Inspecting Eqs.~\eqref{stcod1}-\eqref{stcod2} it is easy realized that it does not fall into the class described by the equations of motion \eqref{st11}-\eqref{st12}. Indeed, for this type of corrections to be properly accounted for, it would be necessary to understand quadrupole corrections to brane actions in the spirit of Sec.~\ref{poledipole}. Below, we summarize the contributions to the stress-energy tensor and corresponding quadrupole moment of the term $\vartheta_3(\textbf{k})$: 
\\ 
\renewcommand{\arraystretch}{1.5}
\begin{center} 
    \begin{tabular}{ | c | c | c |}
    \hline
    \color{red}{Scalar} & \color{red}{$\tilde \tau^{ab}_{\alpha}$} & \color{red}{$\mathcal{D}^{abc\perp}_{\alpha}$} \\ \hline
    $\vartheta_3(\textbf{k})\mathcal{C}_{3}$ & $\vartheta_3(\textbf{k})\mathcal{C}_{3}\gamma^{ab}-\vartheta_3'(\textbf{k})\textbf{k}\mathcal{C}_{3}u^{a}u^{b}$ & $\vartheta_3(\textbf{k})\textbf{k}^{a}\textbf{k}^{b}\textbf{k}^{c}$ \\   \hline
     \end{tabular}
\end{center} 
\vskip 0.3cm
Therefore the only contribution to the quadrupole moment can be written as
\beq
\mathcal{D}^{abc}_{\perp}=\vartheta_3(\textbf{k})\textbf{k}^{a}\textbf{k}^{b}\textbf{k}^{c}~,
\eeq
and encodes the most general response to corrections proportional to $\nabla_{a}K_{bc}$ for co-dimension-1 surfaces. Therefore, for co-dimension-1 surfaces, one needs to add to the action \eqref{action1}, a 2nd order contribution of the form
\beq
I_{(2)}\thinspace[X^{\mu}]=\int_{\mathcal{W}_{p+1}}\sqrt{-\gamma}\mathcal{D}^{abc}_{\perp}\nabla_{a}K_{bc}~.
\eeq
We end this section by noting that the terms $\vartheta_4(\textbf{k})$, $\vartheta_5(\textbf{k})$ and $\vartheta_7(\textbf{k})$ are coupled to each other. To see this we contract \eqref{CM} with $\gamma^{ac}\textbf{k}^{b}$ in order to obtain:
\beq
\gamma^{ac}\textbf{k}^{b}{R^{\perp}}_{cba}=\textbf{k}^{b}\left(\nabla_{b}{K}-\nabla_{a}{K_{b}}^{a}\right)~.
\eeq
This is why we did not need to study the term $\vartheta_7(\textbf{k})$ separately. Further note that the term $\textbf{k}^{b}\nabla_{b}{K}$ vanishes since $\textbf{k}^{b}$ is a Killing vector field and hence if the background Riemann tensor vanishes, so does the contribution of the term $\vartheta_3(\textbf{k})$ since it is related by on-shell equivalences to the term $\vartheta_4(\textbf{k})$.  To summarize, the family of curved fluid branes of co-dimension-1 embedded in a curved background is parametrized by a total of 8 independent response coefficients.

\subsubsection*{Two-dimensional fluid membranes}
Here we perform a counting of the independent parameters of two-dimensional stationary fluid membranes of co-dimension-1 in a flat background. This particular case is of general interest as it accounts for the response coefficients of two-dimensional spatial surfaces of cellular membranes embedded in three-dimensional Euclidean space. For two-dimensional surfaces, the induced Riemann curvature tensor is characterized entirely by the worldvolume Ricci scalar such that
\beq
\mathcal{R}_{abcd}=\frac{\mathcal{R}}{2}\left(\gamma_{ac}\gamma_{bd}-\gamma_{ad}\gamma_{bc}\right)~~,
\eeq
which implies in particular that $\mathcal{R}_{ab}=(1/2)\mathcal{R}\gamma_{ab}$. This in turn implies that the hydrodynamic term $\upsilon_3(\textbf{k})\textbf{k}^a\textbf{k}^b\mathcal{R}_{ab}$ is redundant and can be expressed in terms of the contribution $\upsilon_2(\textbf{k})\mathcal{R}$ in \eqref{hydro2}.
Furthermore, we have that for two-dimensional surfaces without boundaries the Gaussian energy for constant $\upsilon_2(\textbf{k})$ yields
\beq \label{GE}
\int_{A}dA\thinspace\upsilon_2(\textbf{k})\thinspace\mathcal{R}=8\pi\alpha_{G}(1-g)~~,
\eeq
where $g$ is the genus of the surface and $\upsilon_2(\textbf{k})=\alpha_{G}$ for constant $\alpha_G$ \cite{0305-4470-37-23-003}. In such cases the term $\upsilon_2(\textbf{k})\mathcal{R}$ is purely topological and need not be considered in the effective action. This is the reason why the contribution $\upsilon_2(\textbf{k})\mathcal{R}$ is not considered in the effective action of fluid membranes \cite{0305-4470-37-23-003} neither, for example, when considering finite thickness corrections to the string action \cite{Polyakov1986406, Kleinert1986335}. In such situations, the effective action, due to the Gauss-Codazzi equation \eqref{GC}, is only described by the response coefficients $\lambda_1(\textbf{k}),~\upsilon_1(\textbf{k}),~\vartheta_1(\textbf{k})$. For non-trivial response coefficient $\upsilon_2(\textbf{k})$, the Gaussian energy \eqref{GE} is not necessarily a topological invariant and needs to be considered in the effective action. In this case, the family of stationary fluid membranes is characterized by the response coefficients $\lambda_1(\textbf{k}),~\lambda_2(\textbf{k}),~\lambda_3(\textbf{k}),~\upsilon_1(\textbf{k}),~\vartheta_1(\textbf{k})$.


\section{Multipole expansion as derivative corrections} \label{poledipole}
In this section we establish a relation between the action formalism of Sec.~\ref{framework} and a multipole expansion of the stress-energy tensor for curved branes. This expansion, to pole-dipole order, is sufficient to capture all the corrections studied in Secs.~\ref{2nd}-\ref{hydro} but not enough to capture the quadrupole corrections studied in Sec.~\ref{cod1}. Here, we focus on the pole-dipole order and leave the extension to pole-quadrupole order for future work. The relation between this expansion and the corrections of the previous sections is particularly useful as it gives physical meaning to the corrections themselves and it allows to establish a precise link between gravity and the effective description of fluid branes. 


\subsection{Equations of motion} \label{motion}
The multipole expansion consists in expanding the stress-energy tensor of a curved brane in a Dirac-delta series, in the same spirit as in electromagnetism the electric current is expanded in order to describe finite thickness dielectric effects \cite{Vasilic:2007wp}:
\beq \label{pdstress}
\hat T^{\mu\nu}(x^{\alpha})\!=\!\int_{\mathcal{W}_{p+1}}\!\!\!\!\!\!\!d^{p+1}\sigma\sqrt{-\gamma}\left(B^{\mu\nu}(\sigma^{a})\frac{\delta^{D}(x^{\alpha}-X^{\alpha}(\sigma^a))}{\sqrt{-g}}-\nabla_{\rho}\left(B^{\mu\nu\rho}(\sigma^{a})\frac{\delta^{D}(x^{\alpha}-X^{\alpha}(\sigma^a))}{\sqrt{-g}}\right)+...\right)~.
\eeq
The stress-energy tensor \eqref{pdstress} is symmetric, since it should be coupled to gravity, which implies $B^{\mu\nu}=B^{(\mu\nu)}$ as well as $B^{\mu\nu\rho}=B^{(\mu\nu)\rho}$. The structure $B^{\mu\nu}$ is a monopole source of stress-energy while the structure $B^{\mu\nu\rho}$ is a dipole source and encodes the finite thickness effects. Introducing higher order structures in the expansion \eqref{pdstress} results in higher order moments such as the quadrupole moment. Further, to each structure in the expansion we associate an order parameter $\tilde\varepsilon$ such that $B^{\mu\nu}=\mathcal{O}\left(1\right)$ and $B^{\mu\nu\rho}=\mathcal{O}\left(\tilde\varepsilon\right)$. More importantly, note that the expansion parameter $\tilde\varepsilon$ is not the same as the expansion parameter $\varepsilon$ introduced in Sec.~\ref{actions} with the purpose of keeping track of the derivative expansion. For example, in the case of the hydrodynamic expansion for unbent black branes \cite{Camps:2010br, Caldarelli:2012hy, Gath:2013qya, Emparan:2013ila}, one can iteratively correct the metric to arbitrary order in $\varepsilon$ while still being at order $\mathcal{O}\left(1\right)$ in $\tilde\varepsilon$. However, if the deformations of black branes involve strains along transverse directions to the worldvolume the corrections will be encoded in $B^{\mu\nu\rho}$ \cite{Armas:2011uf, Camps:2012hw, Armas:2012ac}, as well as in higher order structures, inducing a multipole expansion in $\tilde\varepsilon$. In particular, corrections proportional to one copy of the extrinsic curvature are encoded in $B^{\mu\nu\rho}$. Note, however, that if $B^{\mu\nu\rho}$ contains a one-derivative term, then the worldvolume effective theory that it gives rise to is of order $\mathcal{O}\left(\varepsilon^2\right)$. This means, for example, that deformations of black branes can acquire dipole moments $B^{\mu\nu\rho}$ to order $\mathcal{O}\left(\varepsilon\right)$ without affecting the effective theory to that same order. This is due to the fact that the covariant derivative in \eqref{pdstress} is acting on $B^{\mu\nu\rho}$. This digression will become clearer as we progress in this section.

The equations of motion for an object characterized by a stress-energy tensor of the form \eqref{pdstress} can be obtained at the probe level, and in the absence of other external background fields, by solving the conservation equation
\beq \label{stC}
\nabla_{\nu}\hat{T}^{\nu\mu}=0~~,
\eeq
using the methods described in \cite{Vasilic:2007wp}. To express these equations in a nice form it is useful to decompose the structure $B^{\mu\nu\rho}$ into tangential and orthogonal constituents to the worldvolume $B^{\mu\nu\rho}=2{u^{(\mu}}_{b}B^{\nu)\rho b}_{\perp}+{u^{\mu}}_{a}{u^{\nu}}_{b}B^{\rho ab}_{\perp}$ with $B^{(\mu\nu)a}_{\perp}=B^{\mu[ab]}_{\perp}=0$\footnote{The symbol $\perp$ here means that the space-time indices of the tensor are transverse to the worldvolume. For example, ${\gamma^{\lambda}}_{\rho}B^{\rho ab}_{\perp}=0$.} and to introduce the variables\footnote{Here we have gauged away the parallel components ${u^{\rho}}_{a}B^{\mu\nu a}$ since by the 'extra symmetry 1' \cite{Vasilic:2007wp} they can be gauged away everywhere on the worldvolume. On the boundary, these components may be non-zero but if we assume the absence of extra sources on the boundary they vanish. In any case, this extra structure can be trivially incorporated into our description by redefining $B^{ab}\to\tilde B^{ab}=B^{ab}-{u^{a}}_{\rho}{u^{b}}_{\lambda}\nabla_{c}B^{\rho\lambda c}$ and adding an extra term to the boundary conditions \cite{Vasilic:2007wp}.}
\beq \label{sn}
S^{\mu\nu a}=B^{\mu\nu a}_{\perp}+{u^{[\mu}}_{b}B^{\nu]ab}_{\perp}~~,~~N^{\mu\nu a}={u^{(\mu}}_{b}B^{\nu)ba}_{\perp}~~,~~m^{ab}=B^{ab}-{u^{a}}_{\rho}{u^{b}}_{\lambda}\nabla_{c}N^{\rho\lambda c}~~.
\eeq
In terms of these, the equations of motion can be written as \cite{Vasilic:2007wp}
\beq\label{pd1}
{\perp^{\mu}}_{\lambda}{\perp^{\nu}}_{\rho}\nabla_{a}S^{\lambda\rho a}=0~~,
\eeq
\beq \label{pd2}
\nabla_{b}\left(m^{ab}{u_{a}}^{\mu}-2{u^{b}}_{\lambda}\nabla_{a}S^{\mu\lambda a}+{u^{\mu}}_{c}{u^{c}}_{\rho}{u^{b}}_{\lambda}\nabla_{a}S^{\rho\lambda a}\right)=S^{\lambda\rho c}{R^{\mu}}_{c\lambda\rho}~~.
\eeq
Eq.~\eqref{pd1} expresses the conservation of transverse angular momenta. In the point particle case, these equations reduce to the ones derived by Papapetrou for spinning point particles \cite{Papapetrou:1951pa}. Eqs.~\eqref{pd1}-\eqref{pd2} must be supplemented by a similar decomposition of $B^{\mu\nu}$ as $B^{\mu\nu}=B^{\mu\nu}_{\perp}+2{u^{(\mu}}_{b}B^{\nu)b}_{\perp}+{u^{\mu}}_{a}{u^{\nu}}_{b}B^{ab}$~, where the first two sets of coefficients are related to the dipole structure via the relations \cite{Vasilic:2007wp}
\beq
B^{\mu\nu}_{\perp}={\perp^{\mu}}_{\lambda}{\perp^{\nu}}_{\rho}\nabla_{a}N^{\lambda\rho a}~~,~~B^{\mu a}_{\perp}={u^{a}}_{\lambda}{\perp^{\mu}}_{\rho}\nabla_{b}\left(S^{\lambda\rho b}+N^{\lambda\rho b}\right)~~.
\eeq
These relations imply that the only free coefficients in the theory are $B^{ab}$ and $B^{\mu\nu\rho}$. Eq.~\eqref{pd2} can be rewritten in a simpler form. In order to do so, we decompose the term inside the parenthesis on the l.h.s. into tangential and orthogonal parts to the worldvolume and define a linear momentum tensor of the form\footnote{The reason for attributing the name of 'linear momentum' to \eqref{linM} is explained in App.~\ref{geometry}.}
\beq \label{linM}
\mathcal{P}^{\nu\mu}=\left(m^{ab}-{u^{a}}_{\rho}{u^{b}}_{\lambda}\nabla_{c}S^{\rho\lambda c}\right){u^{\mu}}_{a}{u^{\nu}}_{b}+2{\gamma^{\nu}}_{\rho}{\perp^{\mu}}_{\lambda}\nabla_{c}S^{\rho\lambda c}~~.
\eeq
The linear momentum \eqref{linM} is not necessarily symmetric. Further, it is tangential in its first index but not in its second, that is, ${\perp^{\rho}}_{\nu}\mathcal{P}^{\nu\mu}=0$. Using this definition, the equation of motion \eqref{pd2} can be recast \emph{\`{a} la Carter}\footnote{This sentence should be pronounced with French accent.} \cite{Carter:1997pb}:
\beq \label{pdc}
{\gamma^{\lambda}}_{\nu}\nabla_{\lambda}\mathcal{P}^{\nu\mu}=S^{\lambda\rho c}{R^{\mu}}_{c\lambda\rho}~~,
\eeq
where the term on the r.h.s. can be seen as a force $\mathcal{F}^{\mu}=S^{\lambda\rho c}{R^{\mu}}_{c\lambda\rho}$ acting on the worldvolume due to the coupling to the background Riemann tensor. Moreover, note that when the Riemann curvature tensor vanishes, the linear momentum $\mathcal{P}^{\mu\nu}$ is conserved along the surface. The set of Eqs.~\eqref{pdc} can be split into two sets by projecting along the worldvolume directions with ${u^{b}}_{\mu}$ and orthogonally to the worldvolume with ${n^{i}}_{\mu}$, leading to,
\beq \label{pdeq1}
\nabla_{a}\mathcal{P}^{ab}-\mathcal{P}^{ai}{K^{b}}_{ai}=S^{\lambda\rho c}{R^{b}}_{c\lambda\rho}~~,
\eeq
\beq \label{pdeq2}
\mathcal{P}^{ab}{K_{ab}}^{i}+\nabla_{a}\mathcal{P}^{ai}+{{\omega_{a}}^{i}}_{j}\mathcal{P}^{aj}=S^{\lambda\rho c}{R^{i}}_{c\lambda\rho}~~.
\eeq
Here we have used the definitions $\mathcal{P}^{ab}={u^{a}}_{\nu}{u^{b}}_{\mu}\mathcal{P}^{\nu\mu}$ and $\mathcal{P}^{bi}={u^{b}}_{\nu}{n^{i}}_{\mu}\mathcal{P}^{\nu\mu}$. These equations are written in the same fashion as in the work of Guven et al. \cite{Arreaga:2000mr}, however, we have generalized it for pole-dipole branes and for curved backgrounds. Now note that in the case of absence of dipole effects $S^{\lambda\rho c}=\mathcal{P}^{ai}=0$~, Eqs.~\eqref{pdeq1}-\eqref{pdeq2} reduce to those obtained to 0th order in the expansion \eqref{st01}-\eqref{st02} upon the identification $B^{ab}=T^{ab}_{(0)}$~. The space-time stress-energy tensor that gives rise to the worldvolume theory \eqref{st01}-\eqref{st02} is the one given in \eqref{pdstress} with $B^{\mu\nu\rho}=0$. This is what is meant by the fluid being confined to an infinitely thin surface since the stress-energy tensor $\hat{T}^{\mu\nu}$ is localized there to order $\mathcal{O}\left(1\right)$.

Before we understand the relation between the multipole expansion and the action variations of the previous section, it is important to provide the boundary conditions that arise from solving Eq.~\eqref{stC}. In terms of the linear momentum \eqref{linM} these can be written as
\begin{gather}\nonumber 
S^{\mu\nu a}\eta_a \eta_{\nu}|_{\partial\mathcal{W}_{p+1}}=0~~,~~{\perp^{\mu}}_{\lambda}{\perp^{\nu}}_{\rho}S^{\lambda\rho a}\eta_a|_{\partial\mathcal{W}_{p+1}}=0~~, \\ \label{pdb}\\ \nonumber
\left[2\nabla_{\hat a}\left(S^{\mu\nu a}\eta_a v^{\hat a}_{\nu}\right)-\eta_b\mathcal{P}^{b\mu}\right]|_{\partial\mathcal{W}_{p+1}}=0~~,
\end{gather}
where we have used the definition of the boundary vectors introduced in Sec.~\ref{framework}.


\subsection{Relation to the action principle} \label{relations}
To connect the results of the previous section with those of Sec.~\ref{actions} it is necessary to clarify the physical meaning of the different components of the structure $B^{\mu\nu\rho}$. This has been done in \cite{Armas:2011uf} and we review it here. If we focus on flat space-time and on uniform $p$-branes extended along all $x^{0},...,x^{p}$ directions we can evaluate the total angular momentum on the transverse plane labelled by the indices $\mu,\nu$ as:
\beq \label{transverse}
{J^{\mu\nu}_{\perp}}=\int_{\Sigma}d^{D-1}x\left(\hat{T}^{0\mu}x^{\nu}-\hat{T}^{0\nu}x^{\mu}\right)=2 \int_{\mathcal{B}_p}dV_{(p)}B^{\mu\nu0}_{\perp}~~,
\eeq
where $\Sigma$ is a spatial slice of the background space-time. This leads to the introduction of a spin current $j^{a\mu\nu}$ such that $j^{a\mu\nu}=2B^{\mu\nu a}_{\perp}$. Further, we can evaluate the worldvolume dipole moment of the brane as
\beq \label{dipole}
D^{ab\rho}=\int_{\Sigma}d^{D-1}x\hat{T}^{\mu\nu}{u_{\mu}}^{a}{u_{\nu}}^{b}x^{\rho}=\int_{\mathcal{B}_p}dV_{(p)}B^{\rho ab}_{\perp}~~,
\eeq
leading to the introduction of a worldvolume dipole density $d^{ab\rho}=B^{\rho ab}_{\perp}$. This last one, as we will see below, can be interpreted as the bending moment of the brane. Given these definitions we can rewrite the tensors $S^{\mu\nu a}$ and $N^{\mu\nu a}$ introduced in \eqref{sn} as
\beq \label{newv}
S^{\mu\nu a}=\frac{1}{2}j^{a\mu\nu}-d^{ab[\mu}{u^{\nu]}}_{b}~~,~~N^{\mu\nu a}=d^{ab(\mu}{u^{\nu)}}_{b}~~.
\eeq
We will now focus on different aspects that relate the multipole expansion \eqref{pdstress} with the action \eqref{action1}.

\subsubsection*{The dipole moment}
Here we consider the case where the transverse angular momenta \eqref{transverse} vanishes, that is, $j^{a\mu\nu}=0$. In this case, the linear momentum \eqref{linM} can be written as
\beq \label{pdipole}
\mathcal{P}^{\nu\mu}=\left(\hat{T}^{ab}-d^{cbi}{K^{a}}_{ci}\right){u^{\mu}}_{a}{u^{\nu}}_{b}+{u^{\nu}}_{b}{\perp^{\mu}}_{\lambda}\nabla_{a}d^{ab\lambda}~~,
\eeq
where we have defined the symmetric tensor $\hat{T}^{ab}=B^{ab}+2d^{(aci}{K^{b)}}_{ci}$. With this, the equations of motion \eqref{pdeq1}-\eqref{pdeq2} can be written as\footnote{These equations have been previously obtained in \cite{Armas:2011uf} but in a different form.}
\beq
\nabla_{a}\hat{T}^{ab}=-{u^{b}}_{\mu}\nabla_{a}\nabla_{c}d^{ac\mu}+d^{aci}{R^{b}}_{aci}~~,
\eeq
\beq \label{eqdipole}
\hat{T}^{ab}{K_{ab}}^{i}=-{n^{i}}_{\mu}\nabla_a\nabla_b d^{ab\mu}-d^{abj}{R^{i}}_{ajb}~~.
\eeq
Comparing this with Eqs.~\eqref{st11},\eqref{st12} when the spin current vanishes, that is ${\mathcal{S}^{a}}_{ij}=0$, we find that both are equivalent upon identifying
\beq \label{idelastic}
\hat{T}^{ab}=T^{ab}~~,~~d^{abi}=-\mathcal{D}^{abi}~~,
\eeq
which in turn implies $B^{ab}=T^{ab}-2d^{(aci}{K^{b)}}_{ci}$. The boundary conditions \eqref{pdb} and \eqref{bound1} can also be seen to be equivalent upon the same identification. With this we have shown that all corrections to brane actions quadratic in the extrinsic curvature can be accounted for by a multipole expansion to pole-dipole order of the stress-energy tensor. Moreover, the conservation equation \eqref{pd1} reduces to the integrability condition
\beq \label{int}
d^{ab[\nu}{K_{ab}}^{\mu]}=0~~.
\eeq
This condition is automatically satisfied for the actions we consider due to the form of $\mathcal{D}^{abi}$ given in \eqref{DI}. Indeed, the form of \eqref{DI} is what is expected for the bending moment of thin membranes \cite{Landau:1959te, Armas:2011uf}, an interpretation which is now justified due to Eq.~\eqref{dipole}. In fact, when a rod is bent, a varying concentration of matter across the transverse directions induces a bending moment which is proportional to the Lagrangian strain \eqref{dU} \cite{Landau:1959te, Armas:2011uf}. Further, note that this analogy with classical elasticity is direct when one deals with co-dimension-1 surfaces for which the transverse index in the extrinsic curvature can be omitted \cite{Landau:1959te}.

\subsubsection*{The spin current}
Now we consider the case for which the worldvolume dipole moment vanishes $d^{abi}=0$ but the transverse momenta is non-zero. In this case, the linear momentum \eqref{linM} takes the form
\beq \label{pspin}
\mathcal{P}^{\nu\mu}=B^{ab}{u^{\mu}}_{a}{u^{\nu}}_{b}+{u^{\nu}}_{b}{\perp^\mu}_{\rho}j^{c\rho i}{K^{b}}_{ci}~~.
\eeq
In turn, the intrinsic equation of motion \eqref{pdeq1} reads
\beq\label{intspin}
\nabla_{a}B^{ab}=-\frac{1}{2}{j^a}_{ij}{\Omega_{a}}^{bij}~~,
\eeq
while the extrinsic equation \eqref{pdeq2} takes the form \cite{Armas:2011uf}
\beq \label{eqspin}
B^{ab}{K_{ab}}^{i}=-{n^{i}}_{\mu}\nabla_{b}\left({j_{a}}^{\mu j}{K^{ab}}_{j}\right)+\frac{1}{2}j^{akj}{R^{i}}_{akj}~~.
\eeq
Comparison of this last equation with Eq.~\eqref{st12} when dipole effects are absent $\mathcal{D}^{abi}=0$ leads us to identify
\beq \label{idspin}
B^{ab}=T^{ab}~~,~~j^{aij}=2\mathcal{S}^{aij}~~.
\eeq
This identification is sufficient for the intrinsic dynamics \eqref{st11} and boundary conditions \eqref{bound1} to match those given by Eq.~\eqref{intspin} and Eq.~\eqref{pdb} when Eq.~\eqref{pd1} is imposed. The conservation equation \eqref{pd1} reduces instead to the conservation equation of the spin current
\beq \label{scons}
\frac{1}{2}{\perp^{\mu}}_{\lambda}{\perp^{\nu}}_{\rho}\nabla_{a}j^{a\lambda\rho}=0~~.
\eeq
This is the reason why we have discarded the term $\varpi_{2}(\textbf{k})$ in Sec.~\ref{2ndtwist} as it does not satisfy this equation. Given this identification we have shown that corrections quadratic in the extrinsic twist potential\footnote{This also holds when the corrections are proportional to only one copy of the extrinsic twist potential.} can be accounted for by a multipole expansion of the stress-energy tensor and that these are interpreted as the fluid-elastic system acquiring motion in transverse directions to the worldvolume by means of Eq.~\eqref{transverse}. Both the integrability condition \eqref{int} and the spin conservation equation \eqref{scons} may be derived as constraint equations from the action \eqref{action1} in the spirit of \cite{Arreaga:2000mr}. We leave this exercise for when we analyze in detail spinning corrections to black holes \cite{Armas:2014}.

\subsubsection*{The `extra symmetry 2'}
The stress-energy tensor \eqref{pdstress}, truncated to order $\tilde \varepsilon$, enjoys of a perturbative symmetry coined the `extra symmetry 2' by the authors of \cite{Vasilic:2007wp} which results from the freedom of shifting the worldvolume surface by a small amount, that is,
\beq \label{def_extra2}
X^{\mu}(\sigma^{a})\to \tilde X^{\mu}(\sigma^{a})=X^{\mu}(\sigma^{a})+\tilde\varepsilon^{\mu}(\sigma^a)~~,
\eeq
where $\tilde\varepsilon^{\mu}$ is an infinitesimal shift vector of order $\tilde \varepsilon$. The interpretation of this symmetry can be understood if one remembers that the expansion \eqref{pdstress} delocalizes the stress-energy tensor by giving the surface a finite thickness. Since the thickness is finite, the worldvolume can be placed anywhere inside the surface \cite{Vasilic:2007wp}. This symmetry acts on the structures $B^{\mu\nu}$ and $B^{\mu\nu\rho}$ as\footnote{In order to obtain this transformation rule from \eqref{pdstress} one should use \eqref{dgamma} and the fact that $\partial_\rho B^{\mu\nu\rho}=0$ since $B^{\mu\nu\rho}$ is a function of $\sigma^a$ and not of the space-time coordinates. Further, note that any scalar or tensor which is a function of the space-time coordinates does not transform under a shift of the worldvolume surface. See Ref.~\cite{Vasilic:2007wp}.}
\beq \label{extra2}
\begin{split}
&\delta_2 B^{\mu\nu}=-B^{\mu\nu}{u^{a}}_{\rho}\nabla_{a}\tilde\varepsilon^{\rho}-2B^{\lambda(\mu}{\Gamma^{\nu)}}_{\lambda\rho}\tilde\varepsilon^{\rho}~~,\\
&\delta_2 B^{\mu\nu\rho}=-B^{\mu\nu}\tilde\varepsilon^{\rho}~~,
\end{split}
\eeq
and leaves the stress-energy tensor \eqref{pdstress} invariant to order $\tilde\varepsilon$. Along worldvolume directions, this transformation coincides with worlvolume reparametrizations\footnote{This is true except at the boundary when extra boundary sources are present \cite{Vasilic:2007wp}.} but it is non-trivial along transverse directions. To see this directly in the equations of motion we separate the transformation into parallel and orthogonal parts such that $\tilde\varepsilon^{\mu}=\tilde\varepsilon^{i}{n^{\mu}}_{i}+{u^{\mu}}_{a}\tilde\varepsilon^{a}$. Then, according to \eqref{extra2} we have that along orthogonal directions
\beq \label{tr2}
\delta^{\perp}_2 m^{ab}=-\left({u^{c}}_{\mu}m^{ab}+{u^{(a}}_{\mu}m^{b)c}\right)\nabla_{c}\left(\tilde\varepsilon^{i}{n^{\mu}}_{i}\right)~~,~~\delta^{\perp}_2d^{abi}=-m^{ab}\tilde\varepsilon^{i}~~,~~\delta_2^{\perp}j^{aij}=\mathcal{O}\left(\tilde\varepsilon^{2}\right)~~.
\eeq
Further note that due to the identifications \eqref{idelastic} and \eqref{idspin} the stress-energy tensor $T^{ab}$ transforms as $\delta_2 T^{ab}=\delta_2m^{ab}-m^{c(a}{K^{b)}}_{ci}\tilde\varepsilon^{i}=m^{ab}\tilde\varepsilon^{i}K_{i}$. Since the transformation of $j^{aij}$ is of higher order it suffices to look at the equation of motion \eqref{eqdipole}. Using the transformation rule \eqref{tr2} together with \eqref{perpK}\footnote{When using Eq.~\eqref{perpK}, $\Phi^{i}$ should be replaced by $\tilde\varepsilon^{i}$.} it is straightforward to check that the equation of motion \eqref{eqdipole} is invariant under the `extra symmetry 2'. This can also be seen at the level of the action \eqref{actionelastic}. In this case, note that $m^{ab}$ can be replaced by $T^{ab}_{(0)}$ due to the identification \eqref{idelastic} and \eqref{Tdecelastic}.  Therefore the variation of the dipole moment can be written as $\delta^{\perp}_2d^{abi}=-T^{ab}_{(0)}\tilde\varepsilon^{i}$. This means that, by virtue of the definition of the dipole moment \eqref{dipolemoment} and the identification \eqref{idelastic}, a term of the form $T^{ab}_{(0)}\tilde\varepsilon_{i}{K_{ab}}^{i}$ should appear in the action \eqref{actionelastic} now evaluated on the surface $\tilde{X}^{\mu}(\sigma^{a})$. If we now want to write the action \eqref{actionelastic} in terms of quantities evaluated on the surface $X^{\mu}(\sigma^{a})$ we decompose $\tilde X^{\mu}(\sigma^{a})=X^{\mu}(\sigma^{a})+\tilde\varepsilon^{\mu}(\sigma^a)$ inducing a transformation in $\sqrt{-\gamma}P(\textbf{k})$. Using Eq.~\eqref{dgamma} this is simply
\beq \label{trP}
\delta_2^{\perp}\left(\sqrt{-\gamma}P(\textbf{k})\right)=-T^{ab}_{(0)}\tilde\varepsilon_{i}{K_{ab}}^{i}~~.
\eeq
Clearly, the action \eqref{action1} is transforming with the opposite sign as in \eqref{tr2} and hence is invariant under the `extra symmetry 2'\footnote{Note that the inclusion of the hydrodynamic contributions \eqref{hydro2} does not spoil the invariance of the action \eqref{action1} under the `extra symmetry 2'. This is because since these scalars only contribute to the monopole stress-energy tensor, their variation is of order $\mathcal{O}\left(\tilde\varepsilon^2\right)$.}. Now consider the case in which the expansion parameter $\tilde\varepsilon^{\mu}$ is controlled by the radius of curvature of the worldvolume. In this case we can write $\tilde\varepsilon^{i}(\sigma^a)=\tilde{k}(\sigma^a)K^{i}$ for some arbitrary function $\tilde k(\sigma^a)$, leading to $\delta^{\perp}_2d^{abi}=-T^{ab}_{(0)}\tilde k(\sigma^{a})\gamma^{cd}{K_{cd}}^{i}$. This means that the Young modulus \eqref{YM} picks up a gauge dependent part of the form
\beq \label{Ytilde}
\begin{split}
{\tilde{\mathcal{Y}}}^{abcd}&=\tilde k(\sigma^{a})\left(T^{ab}_{(0)}\gamma^{cd}+T^{cd}_{(0)}\gamma^{ab}\right) \\
&=\tilde k(\sigma^{a})\left(2\lambda_{0}(\textbf{k})\gamma^{ab}\gamma^{cd}-\lambda'_{0}(\textbf{k})\textbf{k}(u^{a}u^{b}\gamma^{cd}+u^{c}u^{d}\gamma^{ab})\right)~~,
\end{split}
\eeq
where we have used the explicit form of $T^{ab}_{(0)}$ given in Eq.~\eqref{st0}. This indicates that the elastic contributions $\lambda_{1}(\textbf{k})$ and $\lambda_{4}(\textbf{k})$ given in \eqref{elastic2} pick up gauge dependent terms. From another perspective, for surfaces of non-vanishing mean extrinsic curvature, this can also be understood as the pressure $P(\textbf{k})$ picking up a dependence quadratic in the extrinsic curvature such that,
\beq \label{Ptransform}
P(\textbf{k})\to\tilde{P}(\textbf{k})=P(\textbf{k})-T^{ab}_{(0)}\gamma^{cd}\tilde k(\sigma^a){K_{ab}}^{i}{K_{cd}}_{i}~~.
\eeq
It is worth noting that since the variation of $j^{aij}$ is of higher order, the equation of motion \eqref{eqspin} is not invariant under the `extra symmetry 2'. This in fact means that the truncation $d^{abi}=0$ of the pole-dipole equations \eqref{pd1}-\eqref{pd2} is not gauge invariant. To make it gauge invariant one should add an extra $\tilde{d}^{abi}$ term of the form $\tilde{d}^{abi}=-m^{ab}\tilde\varepsilon^{i}$. Relating this with the action \eqref{actionspin} which accounts for spin corrections, it means that one should in fact add a term of the form $(1/2)\tilde{\mathcal{Y}}^{abcd}{K_{ab}}^{i}{K_{cdi}}$ in case of non-vanishing mean extrinsic curvature or a term $-\tilde d^{abi}{K_{abi}}$ generically. However, we can always choose a gauge (surface) for which $P$ is only dependent on $\textbf{k}$, which is the gauge choice leading to \eqref{eqspin}, or equivalently, choose a gauge for which $d^{abi}=0$ in the case of purely spin corrections.


\subsubsection*{Alternative basis for elastic modes and field redefinition}
In Sec.~\ref{2nd} we mentioned that the two last terms in \eqref{elastic2} were not independent of the remaining terms which could be seen by choosing a different basis and a field redefinition. To see this precisely let us define the 0th order elastic equation of motion \eqref{st02} as $\mathcal{E}^{i}_{(0)}=T^{ab}_{(0)}{K_{ab}}^{i}$~, where $T^{ab}_{(0)}$ is given in \eqref{st0}. Now note that due to the form of $T^{ab}_{(0)}$ we can rewrite the two last contributions in \eqref{elastic2} as
\beq \label{extraextra}
\begin{split}
\lambda_{4}(\textbf{k})\textbf{k}^{a}\textbf{k}^{b}{K_{ab}}^{i}K_{i}=&\lambda_{4}(\textbf{k})\tilde f(\textbf{k})\left(K^{i}K_{i}-\lambda_{0}^{-1}(\textbf{k})\mathcal{E}^{i}_{(0)}K_{i}\right)~~,\\
\lambda_{5}(\textbf{k})\textbf{k}^{a}\textbf{k}^{b}\textbf{k}^{c}\textbf{k}^{d}{K_{ab}}^{i}{K_{cdi}}=&\lambda_{5}(\textbf{k})\tilde f^2(\textbf{k})\left(K^{i}K_{i}-2\lambda_{0}^{-1}(\textbf{k})\mathcal{E}^{i}_{(0)}K_{i}+\lambda_{0}^{-2}(\textbf{k})\mathcal{E}^{i}_{(0)}\mathcal{E}_{(0)i}\right)~~,
\end{split}
\eeq
where we have defined $\tilde f(\textbf{k})=(\lambda_{0}(\textbf{k})\textbf{k})/\lambda_{0}'(\textbf{k})$. We see that these two terms can be expressed in the basis $K^{i}K_{i},~\mathcal{E}^{i}_{(0)}K_{i},~\mathcal{E}^{i}_{(0)}\mathcal{E}_{(0)i}$. However, note that adding any term to the action proportional to $\mathcal{E}^{i}_{(0)}\mathcal{E}_{(0)i}$ will result in a set of equations of motion proportional to $\mathcal{E}^{i}_{(0)}$ and hence vanish on-shell for any configuration. This means in fact that the term proportional to $\lambda_{5}(\textbf{k})$ in \eqref{extraextra} is redundant. Additional terms added to the action of the form $\mathcal{E}^{i}_{(0)}K_{i}$ do not in general lead to trivial equations of motion as one can see by considering the action
\beq\label{extraact}
I\thinspace[X^{\mu}]=\int_{\mathcal{W}_{p+1}}\sqrt{-\gamma}\lambda_{6}(\textbf{k})\mathcal{E}^{i}_{(0)}K_{i}~~.
\eeq
Evaluating the monopole stress-energy tensor and dipole moment we obtain
\beq \label{monoextra}
T^{ab}=\left(\lambda_{6}(\textbf{k})\gamma^{ab}-\lambda_{6}'(\textbf{k})\textbf{k}u^{a}u^{b}\right)\mathcal{E}^{i}_{(0)}K_{i}-2\lambda_{6}(\textbf{k})\mathcal{E}^{i}_{(0)}{K^{ab}}_{i}+\lambda_{6}(\textbf{k})E^{abcd}{K_{cd}}^{i}K_{i}~~,
\eeq
as well as ${\mathcal{D}}^{abi}=\lambda_{6}(\textbf{k})\gamma^{ab}\mathcal{E}^{i}_{(0)}+\lambda_{6}(\textbf{k})T^{ab}_{(0)}K^{i}$. In the expression for the monopole stress-energy tensor \eqref{monoextra} we have used the definition of the elasticity tensor introduced in \eqref{elasticity1}. Now note that when these quantities are introduced in the equation of motion \eqref{st12} with ${\mathcal{S}^{a}}_{ij}=0$ all the terms proportional to $\mathcal{E}^{i}_{(0)}$ vanish on-shell and we are left with the linearized equation \eqref{eqlin} with $\Phi^{i}=\lambda_{6}(\textbf{k})K^{i}$. Eq.~\eqref{eqlin} is non-trivial and not proportional to $\mathcal{E}^{i}_{(0)}$ even in flat space and neither when $\lambda_{6}(\textbf{k})$ or  $\lambda_{0}(\textbf{k})$ are constant. This means that the first term in \eqref{extraextra} cannot be removed by a change of basis. Furthermore, the action \eqref{extraact} explains why the structure of Eq.~\eqref{eqlin} already takes into account effects due to bending. However, any contribution of the form \eqref{extraact} can be removed by a redefinition of the pressure through the transformation \eqref{Ptransform} where $\tilde\varepsilon^{i}=\tilde k(\sigma^{a}) K^{i}=\lambda_{6}(\textbf{k})K^{i}$. Indeed, this gives an interpretation of the `extra symmetry 2' as a field redefinition of order $\mathcal{O}\left(\tilde\varepsilon\right)$ since in fact this transformation is defined by the redefinition of $X^{\mu}(\sigma^{a})$ via \eqref{def_extra2}. Concluding, both terms written in \eqref{extraextra} are not physical, neither is the contribution $\vartheta_2(\textbf{k})$ given in \eqref{1modes}. Finally, note that the elasticity tensor is the only non-zero contribution to the stress-energy tensor \eqref{monoextra}. This seems to be in contradiction with the transformation rule for the `extra symmetry 2' \eqref{tr2} where no terms proportional to the elasticity tensor appear. The reason for this is due to the fact that the pole-dipole formalism \eqref{pdstress} needs to be generalized by promoting the tensors $B^{\mu\nu},B^{\mu\nu\rho}$ to functions of the mapping functions $X^{\mu}(\sigma)$ instead of just the worldvolume coordinates $\sigma^{a}$. This is done in a later paper \cite{Armas:2013goa} eliminating such apparent contradiction.


\subsection{Construction of conserved charges} \label{currents}
In this section we show how to construct conserved currents and charges for systems obeying the pole-dipole equations of motion \eqref{pd1}-\eqref{pd2}. We begin by using the technique developed by Carter for geodynamic-type branes \cite{Carter:1994yt}. This consists in finding a conserved surface current $\mathcal{P}^{\nu}_{\textbf{k}}$~, by definition purely tangential ${\perp^{\mu}}_{\nu}\mathcal{P}^{\nu}_{\textbf{k}}=0$, such that\footnote{Note that here we have allowed the Killing vector field $\textbf{k}^\mu$ to also have non-zero components along the transverse directions to the worldvolume.}
\beq \label{conP}
{\gamma^{\lambda}}_{\nu}\nabla_{\lambda}\mathcal{P}^{\nu}_{\textbf{k}}=0~~.
\eeq
The surface charges constructed from such current would then be conserved charges of the system. We will now show that the same type of ansatz as the one used by Carter can be extended to a large class of pole-dipole branes. We take the surface current to be of the form
\beq \label{ansatzP}
\mathcal{P}^{\nu}_{\textbf{k}}=\mathcal{P}^{\nu\mu}\textbf{k}_{\mu}+\Sigma^{\nu\mu\rho}\nabla_{\mu}\textbf{k}_{\rho}~,
\eeq
for an arbitrary space-time Killing vector field $\textbf{k}^{\mu}$ and with ${\perp^{\lambda}}_{\nu}\Sigma^{\nu\mu\rho}=0$. Here we have used the definition of the linear momentum \eqref{linM}. We now introduce the ansatz \eqref{ansatzP} into Eq.~\eqref{conP} and use Eq.~\eqref{pdc}, leading to
\beq
\textbf{k}_{\mu}\left(S^{\lambda\rho c}{u^{\nu}}_{c}+\Sigma^{\nu\lambda\rho}\right){R^{\mu}}_{\nu\lambda\rho}+\left(\mathcal{P}^{[\nu\mu]}+\nabla_c\Sigma^{c[\nu\mu]}\right)\nabla_{\nu}\textbf{k}_{\mu}=0~~,
\eeq
where we have used the fact that for any Killing vector field $\nabla_{\nu}\nabla_{\mu}\textbf{k}_{\rho}=R_{\rho\mu\nu\lambda}\textbf{k}^{\lambda}$. Looking at the term proportional to the Riemann tensor, it seems that the correct choice of $\Sigma^{\nu\mu\rho}$ is
\beq \label{sigma}
\Sigma^{\nu\mu\rho}=-S^{\mu\rho c}{u^{\nu}}_{c}~~,
\eeq
as long as $\mathcal{P}^{[\nu\mu]}=-\nabla_c\Sigma^{c[\nu\mu]}$ is satisfied. We now focus on the class of pole-dipole branes for which the integrability condition \eqref{int} is satisfied. In this case, a simple exercise using \eqref{linM} allows one to show that
\beq
\mathcal{P}^{[\nu\mu]}=\nabla_cS^{[\nu\mu]c}-{\perp^{\mu}}_{\lambda}{\perp^{\nu}}_{\rho}\nabla_{a}S^{\lambda\rho a}=\nabla_cS^{[\nu\mu]c}~~,
\eeq
where we have used Eq.~\eqref{pd1} to eliminate the second term on the r.h.s. above\footnote{It may be useful to write the l.h.s. of  Eq.~\eqref{pd1} explicitly in terms of the variables introduced in \eqref{newv}. This is simply the sum of Eq.~\eqref{scons} with Eq.~\eqref{int}.}.  Hence, the choice \eqref{sigma} leads to a conserved surface current of the form \eqref{ansatzP}. For the case derived by Carter in \cite{Carter:1994yt}, where $j^{a\mu\nu}=0$, we have that $\Sigma^{\nu\mu\rho}=-d^{\nu\mu\rho}$. Furthermore, note that the integrability condition \eqref{int} is satisfied for all the actions we considered in Sec.~\ref{actions}.

Carter suggested in \cite{Carter:1994yt} that charges obtained from \emph{the relevant surface integral} of the current \eqref{ansatzP} are the conserved charges of the pole-dipole brane. The naive way to implement this would be similar to the 0th order case \eqref{MJ0}, that is, a generic charge $\mathcal{Q}_{\textbf{k}}$ associated with a background Killing vector field is simply given by the integral of the surface current \eqref{ansatzP} over spatial slices of the worldvolume
\beq \label{Qs}
|\mathcal{Q}_{\textbf{k}}|=\int_{\mathcal{B}_{p}}dV_{(p)}\mathcal{P}^{\nu}_{\textbf{k}}n_{\nu}~~.
\eeq
However, we can test if this is indeed the case, since we have the full space-time stress-energy tensor \eqref{pdstress}. Since \eqref{pdstress} is symmetric, any current of the form $P^{\nu}_{\textbf{k}}=\hat{T}^{\nu\mu}\textbf{k}_{\mu}(x^{\alpha})$ will be conserved in the full space-time, that is,
\beq \label{consP}
\nabla_{\nu}P^{\nu}_{\textbf{k}}=0~~.
\eeq
The conservation of the space-time currents \eqref{consP} must have a counterpart in terms of conserved currents of the worldvolume theory since the effective description of the dynamics of curved branes to pole-dipole order is given in terms of a worldvolume theory. In order to obtain the corresponding worldvolume conserved currents one solves \eqref{consP} by contracting it with an arbitrary scalar function $f(x^{\mu})$ of compact support and integrating it over space-time according to the method developed in \cite{Vasilic:2007wp}. This results in the worldvolume currents \cite{Armas:2014}
\beq \label{connew}
\hat{P}_{\textbf{k}}^{a}=\left(B^{ab}\textbf{k}_{b}+u^{a}_{\mu}{\perp_{\nu}}^{\lambda}~\textbf{k}_{\lambda}\nabla_{c}\left(B^{\mu\rho\nu}u_{\rho}^{c}\right)+B^{a\mu\rho}\nabla_{\rho}\textbf{k}_{\mu}+B^{\mu\nu\rho}\textbf{k}_{\mu}u_{\nu}^{b}{K^{a}}_{b\rho}\right)~~,
\eeq
which satisfy the conservation equation $\nabla_{a}\hat{P}^{a}_{\textbf{k}}=0$.

Under the assumptions that the timelike Killing vector field $\xi^{\mu}$ is hypersurface orthogonal with respect to the space-time metric $g_{\mu\nu}$ and that $\xi^{\mu}$ is parallel to the worldvolume timelike Killing vector field $\xi^{a}$, i.e. $\xi^{\mu}={u^{\mu}}_{a}\xi^{a}$ , which is also assumed to be hypersurface orthogonal with respect to the worldvolume metric, we can write the total conserved charge as\footnote{Note that the expression for the conserved charges Eq.~\eqref{Qp} differs from the one obtained in \cite{Armas:2011uf}. The expression obtained in \cite{Armas:2011uf} is not invariant under the `extra symmetry 2'.}
\beq \label{Qp}
|\hat{\mathcal{Q}}_{\textbf{k}}|=\int_{\mathcal{B}_p}dV_{(p)}\hat{P}^{a}_{\textbf{k}}n_{a}~~,
\eeq
where all quantities involved in \eqref{Qp} should be evaluated on the worldvolume surface $x^{\alpha}=X^{\alpha}(\sigma^{a})$. Therefore, the difference between the charges \eqref{Qs} and \eqref{Qp} is given by
\beq \label{dQ}
|\mathcal{Q}_{\textbf{k}}|-|\hat{\mathcal{Q}}_{\textbf{k}}|=0~~.
\eeq
A few comments are now in place. From the above we see that the space-time currents introduced in \eqref{consP} correspond to the surface currents introduced in \eqref{conP} and hence the space-time charges agree with the surface charges computed using surface currents. Furthermore, this allows us to identify $\mathcal{P}_{\textbf{k}}^{a}=\hat{P}_{\textbf{k}}^{a}$. Moreover, the currents \eqref{ansatzP} are not invariant under the `extra symmetry 2' transformation and in fact, due to the transformation rule \eqref{extra2}, transform as
\beq 
\mathcal{P}_{\textbf{k}}^{a}\to\tilde{\mathcal{P}}_{\textbf{k}}^{a}=\mathcal{P}_{\textbf{k}}^{a}-B^{ab}\textbf{k}_{b}u^{c}_{\rho}\nabla_{c}\tilde\varepsilon^{\rho}_{\perp}~~.
\eeq
However, the charges computed using \eqref{Qs} are invariant under the same transformation rule for any choice of $\textbf{k}$. In order to see this explicitly one should use \eqref{extra2} in \eqref{Qp} together with \eqref{dgamma}. It is worth noting that as in \eqref{MJ0}, the choice of Killing vector field $\textbf{k}$ in Eq.~\eqref{Qp} results in either the total mass, angular momentum along worldvolume directions or angular momentum along transverse directions to the worldvolume of the fluid-elastic system. Moreover, in the limit where both dipole and spin effects are turned off, the charges computed from Eq.~\eqref{Qp} agree with those computed from \eqref{MJ0}. As a final comment, we note that the conserved surface currents \eqref{ansatzP} can be obtained directly from \eqref{action1} by requiring the action to be invariant under space-time translations along Killing directions. We show how this is done for the case $j^{a\mu\nu}=0$ in App.~\ref{geometry}.


\section{Matching with gravity} \label{gravity}
In this section we apply some of the results of the previous sections to the bending of neutral black branes in pure Einstein gravity. It is well known that the stress-energy tensor of black $p$-branes takes the form of a perfect fluid \cite{Emparan:2009at} and if the brane is subject to a long-wavelength perturbation then the stress-energy tensor \eqref{st0} receives corrections at each order in the derivative expansion. If the type of perturbation is along the worldvolume directions then the corrections are generically of dissipative nature and do not fall into the class studied in Sec.~\ref{actions}. However, if the perturbations are along transverse directions and lead to stationary configurations then the system \eqref{action1} could potentially describe the effective behavior of the perturbed brane. Indeed, this is the case to 0th order \eqref{0action} as demonstrated in \cite{Emparan:2007wm, Camps:2012hw}. The interest in this type of stationary perturbations comes from the several applications of the blackfold method \cite{Emparan:2007wm, Emparan:2009cs, Emparan:2009at} to the construction of higher-dimensional black hole solutions. The method consists in taking the metric of a boosted black $p$-brane and wrapping it over a submanifold $\mathcal{W}_{p+1}$ of characteristic curvature $R$. The end product of such methodology is the perturbative construction of higher-dimensional black holes for which their near horizon geometry is that of a bent boosted black $p$-brane to a certain order $\varepsilon=r_0/R$ in the perturbative expansion. Here $r_0$ is the thickness of the brane, which for Schwarzschild branes coincides with the horizon radius. To be more precise, it is instructive to write the metric of a boosted Schwarzschild black $p$-brane in $D=n+p+3$ space-time dimensions
\beq \label{ds0}
ds^{2}_{(0)}=\left(\gamma_{ab}(X^{\mu}(\sigma^{a}))+\frac{r_0^{n}(\sigma^{a})}{r^n}u_{a}(\sigma^{a})u_b(\sigma^{a})\right)d\sigma^{a}d\sigma^{b}+\frac{dr^2}{1-\frac{r_0^n(\sigma^{a})}{r^n}}+r^2d\Omega^{2}_{(n+1)}~~+~~....~~.
\eeq
Here we have promoted the horizon radius $r_0$ as well as the boost velocities $u^{a}$ and brane worldvolume metric $\gamma_{ab}$ to slowly varying functions of $\sigma^{a}$ over the submanifold $\mathcal{W}_{p+1}$. When promoting the various fields to functions of the worldvolume coordinates, the metric \eqref{ds0} is in general no longer a solution of Einstein equations and should be corrected by including terms proportional to the derivatives of $r_0$, $u^{a}$ and $\gamma_{ab}$. To 0th order in $\varepsilon$ the metric of the black hole solution constructed from wrapping Schwarzschild $p$-branes is given by \eqref{ds0} and since, as it will be explained below, its stress-energy tensor is of the perfect fluid form, the effective dynamics are described by a system of the form \eqref{0action} with equations of motion \eqref{st01}-\eqref{st02}. In fact, this has been shown to be the case directly from Einstein equations \cite{Emparan:2007wm, Camps:2012hw}. Here we are interested in perturbations which are first order in derivatives of the metric $\gamma_{ab}$ along transverse directions. In this case, to first order in $\varepsilon$, a small perturbation $h_{\mu\nu}$ should be added to the metric \eqref{ds0}. Generically, it can be put into the following form \cite{Camps:2012hw}
\beq \label{ds1}
\begin{split}
ds^2_{(1)}=&\left(\eta_{ab}-2{K_{ab}}^{\hat{i}}r\cos\theta+\frac{r_0^n}{r^n}u_{a}u_{b}\right)d\sigma^{a}d\sigma^{b}+\frac{dr^2}{1-\frac{r_0^n}{r^n}}+r^2d\theta^2+r^2\sin^2\theta d\Omega^{2}_{(n)} \\
&+h_{\mu\nu}(r,\theta)dx^{\mu}dx^{\nu}+\mathcal{O}\left(r^2/R^2\right)~~.
\end{split}
\eeq
Here we have used the label $\hat i$ to indicate that the perturbation is taken along a single transverse direction to the worldvolume and also introduced $\eta_{ab}$, which is the flat metric on the worldvolume. As explained in \cite{Gorbonos:2004uc, Camps:2012hw}, perturbations in each individual transverse direction decouple from each other at this order. The important piece of knowledge about the perturbation $h_{\mu\nu}$ is that it is a dipole type perturbation $h_{\mu\nu}(r,\theta)=\cos\theta \hat{h}_{\mu\nu}(r)$ and that all of its components are proportional to the extrinsic curvature ${K_{ab}}^{\hat i}$ \cite{Camps:2012hw}. For the metric \eqref{ds1} to be regular it must satisfy the equations of motion \eqref{st01}-\eqref{st02}, which can be shown through the usage of the method of matched asymptotic expansion \cite{Emparan:2007wm, Camps:2012hw}. The metric \eqref{ds1} is valid to order $\mathcal{O}(\varepsilon)$ and promoting all fields to functions of $\sigma^{a}$ as in \eqref{ds0} allows for iteratively correcting the metric of black holes constructed from wrapping Schwarzschild $p$-branes. However, as argued in Sec.~\ref{1st} and shown in \cite{Emparan:2007wm, Camps:2012hw}, there are no corrections to the asymptotic charges, angular velocities, entropy and temperature to order $\mathcal{O}(\varepsilon)$ from \eqref{ds1}. However, as explained in the beginning of Sec.~\ref{motion}, introducing corrections proportional to the extrinsic curvature induces a multipole expansion of the stress-energy tensor \eqref{pdstress}. Therefore, even though there are no corrections to the charges, the metric acquires a bending moment \cite{Armas:2011uf, Camps:2012hw, Armas:2012ac} to this order, which we will analyze below.

There is an exception to the absence of corrections to the charges to this order, which is the case $n=1$, related to the presence of backreaction effects \cite{Emparan:2007wm, Camps:2012hw}. As argued in \cite{Armas:2011uf}, there are two types of corrections that $\eqref{ds0}$ can be subject to: backreaction corrections and curvature corrections. Newtonian estimates \cite{Armas:2011uf} indicate that curvature corrections become more important when $n>2$. Curvature corrections (or elastic) are included in the formalism of Sec.~\ref{actions} since the resulting equations of motion satisfy stress-energy conservation \eqref{stC}, while backreaction effects are not. In the cases where backreaction is subleading, the perturbation \eqref{ds1} is said to be a pure bending. Therefore, it is only for the cases $n>2$ that one should expect an effective description of the form \eqref{action1}. The procedure of iteratively correcting the metric by introducing strains along transverse directions to the worldvolume has only been completed to order $\mathcal{O}\left(\varepsilon\right)$ leading to the metric \eqref{ds1} and, due to the current state of affairs, there is no data available to $\mathcal{O}\left(\varepsilon^2\right)$ in this long-wavelength perturbation. Nevertheless, it is possible to obtain information about \eqref{action1} from the metrics \eqref{ds0} and \eqref{ds1} as we will describe below.


\subsection{0th order metric: thermodynamic fluid variables}
To 0th order the metric describing the fluid-elastic system is the one presented in \eqref{ds0}. The stress-energy tensor obtained from the ADM formalism or the Brown-York prescription is of the perfect fluid form \eqref{st0} where the pressure and energy density take the following form \cite{Emparan:2009cs}
\beq \label{pbf}
P=-\frac{\Omega_{(n+1)}}{16\pi G}r_0^{n}~~,~~\epsilon=\frac{\Omega_{(n+1)}}{16\pi G}(n+1)r_0^{n}~~.
\eeq
The local fluid temperature $\mathcal{T}$ and local entropy density $s$ can be obtained from \eqref{ds0} by reading off the surface gravity and the horizon area respectively,
\beq \label{tbf}
\mathcal{T}=\frac{n}{4\pi r_0}~~,~~s=\frac{\Omega_{(n+1)}}{4 G} r_0^{n+1}~~.
\eeq
The Gibbs-Duhem relation $\epsilon +P=\mathcal{T}s$ and the first law of thermodynamics $d\epsilon=\mathcal{T}ds$ are obeyed due to the form of \eqref{pbf}-\eqref{tbf}. Due to the relation between global temperature and local temperature $T=\textbf{k}\thinspace\mathcal{T}$ mentioned above Eq.~\eqref{s0}, one can find the relation between the thickness $r_0$ and $\textbf{k}$ such that $(4\pi T)r_0=n\textbf{k}$ and hence the dependence of the pressure P given in \eqref{pbf} on $\textbf{k}$~,
\beq \label{pkbf}
P(\textbf{k})=-\frac{\Omega_{(n+1)}}{16\pi G}\left(\frac{n}{4\pi T}\right)^{n}\textbf{k}^{n}~~.
\eeq
The dependence of $P$ on $\textbf{k}$ is in fact enough to predict the energy density and entropy through Eq.~\eqref{st0} and Eq.~\eqref{s0}~. Using the identification \eqref{id0} one has that $\lambda_{0}(\textbf{k})=P(\textbf{k})$, and hence that $\lambda_0'(\textbf{k})=nP(\textbf{k})/\textbf{k}$. Therefore, using \eqref{st0} we find
\beq \label{stbf}
T^{ab}_{(0)}=P(\textbf{k})\gamma^{ab}-nP(\textbf{k})u^{a}u^{b}~~,
\eeq
which is the stress-energy tensor for the fluid \eqref{pbf}-\eqref{tbf} with equation of state $\epsilon=-(n+1)P$. To 0th order, we have full predictability of black hole masses, angular momenta and entropy using \eqref{MJ0} and \eqref{0entropy} since we know all the local microscopic properties of the fluid \eqref{pbf}-\eqref{tbf}~. This has been used to show the existence of several new black hole solutions \cite{Emparan:2009vd, Armas:2010hz, Caldarelli:2010xz, Emparan:2011hg}.

\subsection{1st order metric: Young modulus}
As mentioned above, there are no corrections to the asymptotic charges and local thermodynamic potentials to order $\mathcal{O}\left(\varepsilon\right)$ of the metric \eqref{ds1} when $n>1$, however, the bending moment $\mathcal{D}^{abi}$ given in \eqref{DI} is a $\mathcal{O}(\varepsilon)$ correction since it is only proportional to one copy of ${K_{ab}}^{i}$. The bending moment can be measured from the metric \eqref{ds1} using the methods of \cite{Armas:2011uf} and briefly, consists in finding the stress-energy tensor of the form \eqref{pdstress} that sources the metric \eqref{ds1}. This was first done in \cite{Armas:2011uf} for neutral black strings bent into a circle and later generalized for Schwarzschild black $p$-branes bent into an arbitrary shape in \cite{Camps:2012hw}. The bending moment found in \cite{Camps:2012hw} is of the form \eqref{DI} as expected from classical elasticity theory with a Young modulus given by\footnote{Note that in \cite{Camps:2012hw} the Young modulus $\tilde Y^{abcd}$ is related to $\mathcal{Y}^{abcd}$ via the relation $\mathcal{Y}^{abcd}=\tilde Y^{abcd}$.}
\beq \label{YMbf}
\begin{split}
\mathcal{Y}^{abcd}\thinspace =\thinspace &-P(\textbf{k})\thinspace r_0^2(\textbf{k})\thinspace\xi(n)\left(\frac{1}{n+2}\gamma^{a(c}\gamma^{d)b}+2u^{(a}\gamma^{b)(c}u^{d)}+\frac{3n+4}{n+2}u^{a}u^{b}u^{c}u^{d}\right) \\
&+ k\thinspace r_0^2(\textbf{k})\thinspace P(\textbf{k})\thinspace\xi(n)\left(2\gamma^{ab}\gamma^{cd}-n\left(u^{a}u^{b}\gamma^{cd}+u^{c}u^{d}\gamma^{ab}\right)\right)~~,
\end{split}
\eeq
where $k$ is a constant and the function $\xi(n)$ takes the form
\beq
\xi(n)=\frac{n \tan(\pi/n)}{\pi}\frac{\Gamma\left(\frac{n+1}{n}\right)^4}{\Gamma\left(\frac{n+2}{n}\right)^2}~~.
\eeq
Note that the function $\xi(n)$ evaluates to zero when $n=1$ and diverges when $n=2$, in agreement with the expectation that elastic corrections are subleading when compared to bakreaction corrections for the cases $n=1,2$. One of the key results in this work is the prediction of the structure of \eqref{YMbf} from \eqref{YM}. Indeed comparison of \eqref{YMbf} with \eqref{YM} using the results of Sec.~\ref{relations} leads to the identification
\begin{gather} \nonumber
\lambda_{1}(\textbf{k})= k\thinspace r_0^2(\textbf{k})\thinspace P(\textbf{k})\thinspace\xi(n)~~,~~\lambda_{2}(\textbf{k})=-\frac{P(\textbf{k})\thinspace r_0^2(\textbf{k})\thinspace\xi(n)}{2(n+2)}~~,~~\lambda_{3}(\textbf{k})=-\frac{P(\textbf{k})\thinspace r_0^2(\textbf{k})\thinspace\xi(n)}{\textbf{k}^2}\\  \label{idbf} \\ \nonumber
\lambda_{4}(\textbf{k})= \frac{k\thinspace r_0^2(\textbf{k})\thinspace n\thinspace P(\textbf{k})\thinspace\xi(n)}{\textbf{k}^2}~~,~~\lambda_{5}(\textbf{k})=-\frac{3n+4}{2(n+2)}\frac{P(\textbf{k})\thinspace r_0^2(\textbf{k})\thinspace\xi(n)}{\textbf{k}^4}~~.
\end{gather}
A few remarks are now in place. Note that the elastic contributions $\lambda_{1}(\textbf{k})$ and $\lambda_{4}(\textbf{k})$ are gauge dependent. Inspecting the second line of \eqref{YMbf}, we see that it has the same structure as the gauge dependent part of the Young modulus defined in \eqref{Ytilde}. This comparison allows us to identify $\tilde k(\sigma^{a})= -k\thinspace r_0^2(\textbf{k}) \thinspace \xi(n)$. Furthermore, it is not surprising that the scalars associated with $\lambda_{1}(\textbf{k})$ and $\lambda_{4}(\textbf{k})$ are only gauge dependent. This is because since the metric \eqref{ds1} is valid only to order $\mathcal{O}\left(\varepsilon\right)$, it cannot probe the hydrodynamic contributions \eqref{hydro2} since these are second order corrections to $T^{ab}$. Therefore, since $\mathcal{R}_{abcd}$ can be neglected to this order and since the measurement of \eqref{YMbf} was done in flat space, according to Eq.~\eqref{GC} we have that, $\mathcal{L}_{1}=\mathcal{L}_{2}$ and  $\mathcal{L}_{3}=\mathcal{L}_4$. It is worth noting that measuring \eqref{pdstress} from the metric \eqref{ds1} does not uncover any of the corrections to $\tau^{ab}$ or $\Pi^{ab}$ given in Sec.~\ref{2nd} and Sec.~\ref{hydro} since these are of order $\mathcal{O}\left(\varepsilon^2\right)$. In fact, besides the measurement of $\mathcal{D}^{abi}$, from \eqref{ds1} one also obtains $B^{ab}$ as defined in \eqref{sn} which takes the same form as in \eqref{stbf}. In other words, to order $\mathcal{O}\left(\varepsilon\right)$, the perturbation \eqref{ds1} does not affect the monopole contribution to the stress-energy tensor \eqref{pdstress}. This means, for example, that one cannot see the full invariance of \eqref{pdstress} under the `extra symmetry 2' to order $\mathcal{O}\left(\varepsilon\right)$.


\subsection{Elastic corrections to black rings}
As mentioned towards the end of Sec.~\ref{gravity}, the procedure of iteratively correcting the metric \eqref{ds0} in a derivative expansion has not been completed to order $\mathcal{O}\left(\varepsilon^2\right)$. However, the measurement of $\mathcal{D}^{abi}$ is all that is necessary to predict the structure of $\tau^{ab}$ using the table given in Sec.~\ref{2nd}. One may assume that if the deformation to the metric \eqref{ds0} is a pure bending then the only excited modes are those presented in \eqref{elastic2} and the hydrodynamic contributions $\upsilon_2(\textbf{k})$ and $\upsilon_3(\textbf{k})$\footnote{There is no Gauss-Codazzi-type equation that relates any of the elastic contributions \eqref{elastic2} to the hydrodynamic contribution $\upsilon_1(\textbf{k})$ so if the deformation is pure bending this term is not, in principle, excited.}. However, we can turn the hydrodynamic contributions off by considering the simple embedding of a string bent into a ring, for which $\mathcal{R}_{abcd}=0$, placed in a flat background. This means that the elastic contributions $\lambda_{1}(\textbf{k})$ and $\lambda_{4}(\textbf{k})$ continue to only be gauge dependent. Further, due to the homogeneity and isotropy of the ring solution we take $\textbf{k}$ to be constant along the worldvolume, which means that all the contributions from the hydrodynamic term $\upsilon_1(\textbf{k})$ will vanish (see table in Sec.~\ref{hydro}). Given these facts, what we would like to test is if 1st order data is enough to predict black hole charges to 2nd order in the derivative expansion, at least for specific cases. For this purpose one would require the knowledge of the thermodynamic quantities \eqref{pbf} and, in particular, of the dependence of the pressure on $\textbf{k}$ to order $\mathcal{O}\left(\varepsilon^2\right)$. From here on we will assume that this functional dependence to order $\mathcal{O}\left(\varepsilon^2\right)$ is the same as that given in \eqref{pkbf}. The effective action for this configuration would then just be that given by \eqref{actionelastic} with the identification \eqref{idbf}. Using the definition of the modified pressure \eqref{id1}, this is simply
\beq \label{actionbf}
I\thinspace[X^{\mu}]=\int_{\mathcal{W}_{p+1}}\sqrt{-\gamma}\thinspace\mathcal{P}\left(\textbf{k},{K_{ab}}^{i}\right)~~.
\eeq
We parametrize the flat background as
\beq
ds^2=-dt^2+dr^2+r^2d\phi^{2}+\sum_{i=1}^{n+1}dx^2_{i}~~,
\eeq
and place the ring at $r=R$ such that $\sigma^{0}=\tau=t~,~,\sigma^{1}=\phi$ and $x^{i}=0$. The induced metric is just $\gamma_{ab}d\sigma^{a}d\sigma^{b}=-d\tau^2+R^2d\phi^2$. The only non-vanishing component of the extrinsic curvature is given by ${K_{\phi\phi}}^{r}=-R$. This implies, due to the form of the linear momentum \eqref{pdipole} that, $\mathcal{P}^{ai}=0$ and the equations of motion \eqref{pdeq2} reduce to
\beq \label{mr}
\mathcal{P}^{ab}{K_{ab}}^{i}=0~~.
\eeq
Since that there is only one non-vanishing component of the extrinsic curvature, \eqref{mr} is actually just $\mathcal{P}^{\phi\phi}=0$. Instead of computing all the components of the stress-energy tensor involved in $\mathcal{P}^{\phi\phi}$ using the table given in Sec.~\ref{2nd}, it is more practical to simply compute the scalar $\mathcal{P}\left(\textbf{k},{K_{ab}}^{i}\right)$ and vary \eqref{actionbf}. In order to do so, we choose a gauge (surface) for which $k=0$ in \eqref{YMbf} (see Sec.~\ref{relations} for the invariance of the result under gauge choices). Using \eqref{YM} together with \eqref{idbf} we find
\beq
\mathcal{P}\left(\textbf{k},{K_{ab}}^{i}\right)=\left(P(\textbf{k})+\frac{\lambda_{2}(\textbf{k})}{R^2}+\lambda_{3}(\textbf{k})\Omega^2+\lambda_{5}(\textbf{k})\Omega^4R^2\right)~~.
\eeq
Varying now the action \eqref{actionbf} leads to the equation of motion \eqref{mr}. This can be solved perturbatively in the manner $\Omega=\Omega_{(0)}+\Omega_{(2)}\varepsilon^2$, where $\varepsilon^2=r_0^2/R^2$. Here $\Omega_{(0)}$ for this configuration can be obtained from \eqref{0action} together with \eqref{tbf} and is given by the relation $\Omega_{(0)}R=1/\sqrt{n+1}$ \cite{Emparan:2009vd}. Introducing this decomposition in the equations of motion \eqref{mr} leads to a solution to $\Omega_{(2)}$ of the form
\beq \label{omega2}
\Omega_{(2)}=\frac{(n-4)\sqrt{n+1}}{2\thinspace n^2(n+2)\thinspace R}\xi(n)\thinspace\varepsilon^2~~.
\eeq
Given the corrected rotation velocity $\Omega$ we can proceed and compute the total mass and angular momentum using \eqref{Qs}. For this we need the conserved surface current \eqref{ansatzP}, which for this particular case takes the form
\beq
\mathcal{P}^{\nu}_{\textbf{k}}=\mathcal{P}^{\nu\mu}\textbf{k}_{\mu}-d^{\nu[\mu\rho]}\nabla_{\mu}\textbf{k}_{\rho}~~.
\eeq
Thus, the mass of the fluid-elastic system is given by
\beq
M=\int_{\mathcal{B}_{p}}dV_{(p)}\thinspace\mathcal{P}^{\tau\tau}=2\pi\thinspace R\thinspace \mathcal{T}^{\tau\tau}=2\pi \thinspace R\thinspace \textbf{P}_{(0)}\left((n+2)-\frac{2(n+1)}{n}\thinspace\xi(n)\thinspace\varepsilon^2\right)~~,
\eeq
Where $\textbf{P}_{(0)}$ is the modulus of $P(\textbf{k})$ given in \eqref{pkbf}, when $\Omega=\Omega_{(0)}$. The angular momentum along the direction $\phi$ reads
\beq
\begin{split}
J\!=\!-\!\int_{\mathcal{B}_{p}}\!\!\!\!dV_{(p)}\!\!\left(\thinspace\mathcal{P}^{ab}\xi_{a}\chi_{b}^{(\phi)}+\frac{1}{2}d^{a\phi r}\xi_{a}\partial_{r}\chi_{\phi}^{(\phi)}\right)\!=\!2\pi R^2\textbf{P}_{(0)}\thinspace\sqrt{n+1}\left(1+\frac{(n+1)(n+12)}{2\thinspace n^2\thinspace(n+2)}\thinspace\xi(n)\thinspace\varepsilon^2\right).
\end{split}
\eeq
As mentioned in Sec.~\ref{framework}, an entropy current formalism has not been developed for the system \eqref{action1} so we do not have a first principle method to compute the total entropy of the fluid-elastic system. However, this is not necessary in order to check if 1st order data can account for black hole charges to 2nd order neither if the action \eqref{action1} with the identification \eqref{idbf} is the correct effective description of higher-dimensional black rings in the blackfold regime. Since the action \eqref{actionbf} can be interpreted as the free energy (see Sec.~\ref{0thorder}) we can obtain the product $\left(TS\right)_{1}$ using $\left(TS\right)_{1}=M-\Omega J-F$. Furthermore, since the charges $M$ and $J$ are supposed to be the ones associated with a stationary black hole in asymptotically flat space time, they must satisfy the Smarr relation
\beq
\left(TS\right)_{2}=\frac{(n+1)}{(n+2)}M-\Omega J~~.
\eeq
A simple exercise tells us that $\left(TS\right)_{1}=\left(TS\right)_{2}+\mathcal{O}\left(\varepsilon^4\right)$ for any value of $n$. Therefore, we conclude that 1st order data in the case of black rings is enough to have full predictability of black hole charges to 2nd order. We will discuss in the next section other possible cases for which 1st order data is enough to predict charges to 2nd order.


\section{Discussion} \label{discussion}
In this section we summarize the main results found in this work and discuss various open problems. We begin by stating the key results. In Sec.~\ref{actions} we have found the most general action quadratic in the extrinsic curvature as well as in the extrinsic twist potential and in second order worldvolume derivatives. For co-dimension-1 surfaces it was required, to the same order, to consider terms proportional to worldvolume derivatives of the extrinsic curvature. It was shown that the equations of motion obtained from this type of actions provide a relativistic generalization of classical elasticity theory of thin membranes when bending effects as well as spin effects are taken into account \eqref{elasticity}. Since the well studied case of fluid membranes is described by an action of the type \eqref{action1}, our work ended up generalizing the Helfrich-Canham bending energy \eqref{HCenergy} to the case in which the fluid living on the membrane is stationary and for non-trivial response coefficients. In such cases, for co-dimension-1 surfaces (not necessarily two-dimensional), there exists 4 extra contributions to second order  than those considered previously in the literature described by the response coefficients $\lambda_{3}(\textbf{k})~,\upsilon_1(\textbf{k}),~\upsilon_3(\textbf{k}),~\vartheta_3(\textbf{k})$. Some of these response coefficients could potentially be measured in a physical experiment involving fluids moving on cellular membranes. 

In general, the results of Sec.~\ref{actions} indicate that the study of hydrodynamics of fluids living on surfaces of arbitrary co-dimension is of increased complexity when compared to the hydrodynamics of space-filling fluids. In fact, for neutral stationary fluids it was found a set of 3 response coefficients \cite{Banerjee:2012iz, Jensen:2012jh} while for non-dissipative fluids a total of 5 transport coefficients \cite{Bhattacharya:2012zx}. Here instead, for surfaces of co-dimension greater than one and according to certain assumptions\footnote{We remind the reader that in Sec.~\ref{intro} we stated that terms constructed with the Levi-Civita symbol would be ignored.}, we have found a total of 7 independent response coefficients, while for co-dimension-1 surfaces we found a total of 8 independent response coefficients. In this counting we have ignored the spin contribution $\varpi_2(\textbf{k})$ as it does not fit into the pole-dipole approximation of Sec.~\ref{poledipole}, which suggests that it violates spin conservation. We also ignored the elastic contributions $\lambda_{4}(\textbf{k}),\lambda_5(\textbf{k})$ and $\vartheta_2(\textbf{k})$ since they can be removed by a change of basis and a field redefinition (see Sec.~\ref{relations}). We have also shown in Sec.~\ref{hydro} how the elastic and hydrodynamic modes couple to each other using the Gauss-Codazzi equation and in which conditions we can regard each of the different contributions as independent.

The work presented in Sec.~\ref{actions} also indicates that all the techniques and methodologies used to study hydrodynamics of space-filling fluids can be applied to the case of fluid branes when the elastic modes \eqref{elastic2} are taken into account. The requirement of stationarity can be relaxed and the method used here can also be applied to non-dissipative fluids as in \cite{Bhattacharya:2012zx}. In this case, the material space introduced in \cite{Bhattacharya:2012zx}, where the fluid variables are defined, must be formulated with respect to the $p$ spatial directions of the worldvolume $\mathcal{W}_{p+1}$.

We have found in Sec.~\ref{poledipole} that all the corrections studied in Sec.~\ref{actions} can be accounted for by the formalism of Vasilic-Vojinovic \cite{Vasilic:2007wp} where a multipole expansion of the stress-energy tensor is carried out to pole-dipole order, except for the terms \eqref{cod12} which require an extension of these ideas to pole-quadrupole order. Indeed, the formalism constructed by these authors provides the most general equations of motion that take into account finite thickness effects of curved branes, regardless of the existence of any underlying effective action. Having established this connection, we note that most of the finite thickness corrections to brane effective actions considered in the literature \cite{Polyakov1986406, Kleinert1986335, Letelier1992, Carter:1994yt, Carter:1997pb,Capovilla:1994bs, Arreaga:2000mr} fit into the formalism of Sec.~\ref{poledipole}. In particular, we established a precise map between extrinsic curvature corrections and the bending moment of the brane as well as extrinsic twist corrections and the spin in transverse directions to the brane. It would be interesting to understand if this map can be useful for the effective description of long strings \cite{Aharony:2013ipa}.

In this work we have generalized extrinsic curvature corrections to brane effective actions to the case of non-extremal branes and connected the formalism of Vasilic-Vojinovic \cite{Vasilic:2007wp} with the formalism of Carter \cite{Carter:1997pb} and the formalism of Capovilla-Guven \cite{Capovilla:1994bs}. The establishment of this connection is of particular usefulness as it allows us to connect effective theories of fluid branes with gravity. In Sec.~\ref{gravity} we have shown that the Young modulus measured from bending neutral black branes falls into the class predicted by the effective action analysis of Sec.~\ref{actions} and used this fact to predict the corrected horizon angular velocity for thin black rings. This gives further motivation for studying the elastic expansion of higher-dimensional black holes via the blackfold approach since now, uncovering the various response coefficients of stationary black branes simultaneously implies the uncovering of the possible response coefficients and the structure of the free energy of real fluid membranes. This fact can be put into a broader context: this is another instance where the study of gravitational physics can shed light into the physics of an apparent unrelated system. Therefore, since thin black branes seem to behave like the membranes of living cells, gravity can be used as a laboratory for uncovering properties of fluid membranes. \\ \\
We now turn to several of the open problems encountered in this work:

\paragraph{The spin/elastic contributions we missed:} As mentioned in Sec.~\ref{intro}, we have ignored terms constructed from the Levi-Civita symbol, in particular, we have ignored dimension-specific corrections of the form
\begin{equation} \label{spin4}
\textbf{k}^{a}\epsilon_{\hat{i}\hat{j}}{\omega_{a}}^{\hat{i}\hat{j}}~~,~~{\epsilon^{ab}}_{ij}{\Omega_{ab}}^{ij}~~,~~{\epsilon^{ab}}_{ij}{R_{ab}}^{ij}~~,~~{\epsilon^{ab}}_{ij}{K_{ac}}^{i}{K_b}^{cj}~~.
\end{equation}
The second term above has been studied in the context of cosmic strings \cite{Carter:1997pb}. Further, the last 3 contributions are coupled to each other due to the Ricci integrability condition \eqref{RC}. It would be interesting to understand if this last set of terms can be accounted for by the formalism of Sec.~\ref{poledipole} to pole-dipole order. The first of these terms is of particular interest as it gives rise to the spin current measured from Myers-Perry branes \cite{Armas:2011uf}. Here, we have introduced the Levi-Civita symbol on a transverse two-plane labelled by the indices $\hat{i},\hat{j}$. To make this statement precise, note that for this case $\mathcal{S}^{a\hat{i}\hat{j}}=\textbf{k}^{a}\epsilon^{\hat{i}\hat{j}}$. If one compares this with the result for Myers-Perry branes found in \cite{Armas:2011uf}, this is exactly the type of correction needed for describing black holes spinning in transverse directions to the worldvolume. It would be interesting to study the wrapping of Myers-Perry branes from this effective action perspective and to construct doubly-spinning black rings to 2nd order in the derivative expansion. This problem will be addressed in a future publication \cite{Armas:2014}.

\paragraph{The entropy current and charged fluid branes:} In this work we have lacked a first principle computation of the entropy for the fluid-elastic system \eqref{action1}. However, using the methods of \cite{Loganayagam:2008is, Banerjee:2012iz} together with some inspiration from the formulation of viscoelasticity of Fukuma-Sakatani \cite{Fukuma:2011pr, Fukuma:2012ws} it is possible to classify the several terms in the entropy current as well as to obtain a first principle computation of the total entropy. A related problem is to generalize the action \eqref{action1} to the case where the fluid carries either a $q=0$, $q=1$ or $q=p$-brane charge as in the cases studied in \cite{Caldarelli:2010xz, Emparan:2011hg}. For $q=0$ and for $p=q$-brane charge the action takes essentially the same form but for $p\ne q$ and $q>0$ new contributions need to be added. This would allow to predict the structure of the piezoeletric moduli measured for charged black branes \cite{Armas:2012ac}. This problem is related to the entropy current formulation because in both cases it requires obtaining from the action a conserved current without additional corrections to the action \eqref{action1} itself. This issue is now under investigation and it will be published elsewhere \cite{Armas:2014}.

\paragraph{Constraints on the response coefficients:} During the analysis of the mode coupling in Sec.~\ref{hydro} we only looked at relations that arise from geometric constraints, such as the Gauss-Codazzi equation \eqref{GC}. However, there may be stability and thermodynamic constraints imposed by elasticity theory or the entropy current analysis in the spirit of \cite{Bhattacharyya:2012nq} that further restrict the set of response coefficients obtained in Sec.~\ref{actions}. Furthermore, it would be interesting to understand the physical meaning of each of the elastic contributions in \eqref{elastic2} by obtaining, for example, the corrected speed of propagation of elastic waves or the corrections to the elasticity tensor of fluid branes introduced in \eqref{elasticity1}. The lack of such knowledge is unsatisfactory and deserves further study.

\paragraph{Elastic corrections to black holes:} In Sec.~\ref{gravity} we applied the 2nd order effective action to the case of black rings in asymptotically flat space and found that the black hole charges were consistent with the free energy interpretation and the Smarr relation. This is compelling evidence that the effective description of higher dimensional black holes in the blackfold regime is given by an action of type \eqref{action1} with the identification \eqref{idbf}. It is also important to refer that the predictions made in \cite{Armas:2011uf, Armas:2012ac} for the corrected horizon angular velocity of (charged) black rings were not accurate enough as they did not take into account the contributions to the monopole stress-energy tensor $T^{ab}$ given in the table of Sec.~\ref{2nd}. We note that we have not obtained a complete prediction of all the corrected black hole thermodynamic quantities. This is because we have not developed a first principle computation of the total entropy of the system. Once this is done, the formalism presented here allows for the full predictability of black hole charges to order $\mathcal{O}\left(\varepsilon^2\right)$ in the derivative expansion. We note, however, that due to the lack of 2nd order data, at the moment it is only possible to accurately obtain the corrected horizon angular velocity for black holes made of bent strings such as the black rings of Sec.~\ref{gravity} and the helical rings and strings found in \cite{Emparan:2009vd}. This is because for embeddings with non-vanishing $\mathcal{R}_{abcd}$~, 2nd order information about the response coefficients of the hydrodynamic modes \eqref{hydro2} is required. In fact, a simple exercise tells us that applying the same prescription as for the case of black rings in Sec.~\ref{gravity} to the case of black odd-spheres \cite{Emparan:2009vd} does not lead to consistent results. The study of the bending of the black branes \eqref{ds0} to 2nd order would be interesting since it would uncover these extra response coefficients. The bending of strings to $\mathcal{O}\left(\varepsilon^2\right)$ is an easier task since the hydrodynamic modes \eqref{hydro2} will vanish but it would still be a worthy endeavour as it would allow to confirm the corrections to the monopole stress-energy tensor found in Sec.~\ref{2nd} as well as the prediction \eqref{omega2}.
\\ \\
We end this work by noting some interesting facts about fluid membranes. When applying the considerations of Sec.~\ref{actions} to fluid membranes one does not have to be concerned with backreaction effects and hence the effective description of the system is that given by \eqref{action1}. If we take the simplest case of a membrane embedded in a flat background, then there is a total of 5 response coefficients, 2 of which are related to stationary flows and have not been considered previously in the literature. While some of these may be subject to local constraints for real cell membranes, as the scalar associated with $\vartheta_1(\textbf{k})$ is, and ignoring viscous effects, according to the analysis of Sec.~\ref{actions}, this is how fluids bend.


\section*{Acknowledgements}
I am greatly indebted to Jyotirmoy Bhattacharya for long lasting discussions from where several of the key ideas in this work originated from and for hospitality at IPMU. I am also grateful to Yuki Sato and Shinji Hirano for hospitality at Nagoya University, to Shigeki Sugimoto for hospitality at IPMU and to Masafumi Fukuma and Yuho Sakatani for exciting discussions on viscoelastic fluids and hospitality at Kyoto University. I also acknowledge NBI for hospitality and CERN during the workshop \textbf{Black Hole Horizons and Quantum Information} where part of this work was done. I am also thankful for discussions with Niels Obers at an earlier stage in this project during the workshop \textbf{Bits, Branes and Black Holes} at KITP. I am also grateful for discussions with Matthias Blau, Joan Camps, Jemal Guven, Troels Harmark, Niels Obers, Mukund Rangamani and Marko Vojinovic. I am especially grateful to Matthias Blau for the post-doc opportunity. Moreover, I am extremely grateful to Joan Camps for many comments and discussions on an earlier draft of this manuscript and also to Jyotirmoy Bhattacharya for more useful comments and discussions. Finally, I want to thank Fabrikken/Christiania for support and interesting discussions. This work was partly supported by the Innovations- und Kooperationsprojekt C-13 of the Schweizerische Universit\"{a}tskonferenz (SUK/CRUS). 

\appendix

\section{Notation, geometry and variations} \label{geometry}
In this section we write down in detail the notation used in this work and their relation to the work of other authors. We consider surfaces of $(p+1)$-dimensional worldvolume $\mathcal{W}_{p+1}$ embedded in a background space-time endowed with metric $g_{\mu\nu}(x^{\alpha})$. The coordinates $x^{\alpha}$, $\alpha=0,...,D-1$ are space-time coordinates. To the worldvolume we assign a set of coordinates $\sigma^{a}$, $a=0,...,p$~, where $p$ is the total number of spatial dimensions of the worldvolume. The worldvolume $\mathcal{W}_{p+1}$ is located at the space-time surface $x^\mu=X^{\mu}(\sigma^{a})$, where $X^{\mu}(\sigma^{a})$ is a set of mapping functions describing the position of the surface in the ambient space-time. Introducing the projector along worldvolume directions ${u_{a}}^{\mu}=\partial_{a}X^{\mu}$ we can construct the induced metric on the worldvolume as $\gamma_{ab}=g_{\mu\nu}{u_{a}}^{\mu}{u_{b}}^{\nu}$, as well as its extrinsic curvature ${K_{ab}}^{\rho}=\nabla_{a}{u_{b}}^{\rho}$ symmetric in the indices $a,b$. The covariant derivative along worldvolume directions $\nabla_{a}={u^{\rho}}_{a}\nabla_{\rho}$, compatible with both the worldvolume metric $\gamma_{ab}$ and the space-time metric $g_{\mu\nu}$, acts on an arbitrary tensor $V^{c\mu}$ as
\beq \label{covd}
\nabla_{a}V^{c\mu}=\partial_{a}V^{c\mu}+{\gamma_{ab}}^{c}V^{b\mu}+\Gamma^{\mu}_{\nu\lambda}{u^{\nu}}_{a}V^{c\lambda}~~,
\eeq
where the Christoffel symbols ${\gamma_{ac}}^{b}$ are computed with respect to the induced metric $\gamma_{ab}$ and the Christoffel symbols $\Gamma^{\mu}_{\nu\lambda}$ with respect to the space-time metric $g_{\mu\nu}$. Any space-time vector $v^{\mu}$ can be projected along the worldvolume directions using ${u_{a}}^{\mu}$ such that $v^{a}={u^{a}}_{\mu}v^{\mu}$. We further introduce a set of projectors ${n^{i}}_{\mu}$, $i=1,...,D-p-1$ onto the transverse space to the worldvolume defined by $g_{\mu\nu}{n_{i}}^{\mu}{n_{j}}^{\nu}=\delta_{ij}$ and ${n^{i}}_{\mu}{u_{a}}^{\mu}=0$. Any space-time vector can be projected along orthogonal directions such that $v^{i}={n^{i}}_{\mu}v^{\mu}$. The extrinsic curvature by definition is transverse in its third index such that ${K_{ab}}^{\rho}={K_{ab}}^{i}{n_{i}}^{\rho}$. Given the normal projectors we can define the extrinsic twist potential via ${\omega_{a}}^{ij}=-{n_{\mu}}^{j}\nabla_{a}{n^{\mu i}}$, which is anti-symmetric in the indices $i,j$. Given the extrinsic twist potential, we can define the outer curvature associated with it through the relation \cite{Capovilla:1994bs}\footnote{Note that here we are using the opposite conventions compared to the ones used in \cite{Capovilla:1994bs}. That is, the extrinsic curvature and extrinsic twist potential defined in \cite{Capovilla:1994bs} should be multiplied by a minus sign to match the conventions used here.}
\beq
{\Omega_{ab}}^{ij}=\nabla_{a}{\omega_{b}}^{ij}-\nabla_{b}{\omega_{a}}^{ij}+{\omega_{a}}^{ik}{\omega_{bk}}^{j}-{\omega_{b}}^{ik}{\omega_{ak}}^{j}~~.
\eeq
It is also useful to work with space-time indices and still keeping track of tangential and perpendicular components to the worldvolume. For that matter, we introduce the first fundamental tensor $\gamma^{\mu\nu}$ such that $\gamma^{\mu\nu}={u^{\mu}}_{a}{u^{\nu}}_{b}\gamma^{ab}$. Using $\gamma^{\mu\nu}$, any space-time tensor can be projected along the worldvolume while still keeping space-time indices. Similarly, we introduce a perpendicular projector such that $\perp_{\mu\nu}=g_{\mu\nu}-\gamma_{\mu\nu}$, which satisfies $\gamma^{\mu\nu}{\perp_{\mu}}^{\rho}=0$. The worldvolume projector ${u_{a}}^{\mu}$ and the transverse projector ${n^{i}}_{\mu}$ are by definition parallel and perpendicular respectively in their space-time index, i.e., ${u_{a}}^{\mu}={u_{a}}^{\nu}{\gamma_{\nu}}^{\mu}$ and ${n^{i}}_{\mu}={n^{i}}_{\nu}{\perp^{\nu}}_{\mu}$. Note that both space-time projectors can be defined in terms of the projectors ${u^{\mu}}_{a}$ and ${n^{\mu}}_{i}$ such that ${\gamma^{\mu}}_{\nu}={u^{\mu}}_{a}{u_{\nu}}^{a}$ and ${\perp^{\mu}}_{\nu}={n^{\mu}}_{i}{n_{\nu}}^{i}$.

\subsubsection*{Variations \emph{\`{a} la Capovilla-Guven}}
In \cite{Capovilla:1994bs}, Capovilla-Guven introduced a covariant derivative $\tilde{\nabla}_{a}$ that preserves covariance under rotations of the normal vectors, such that
\beq \label{torsion}
\tilde\nabla_{a}V^{ci}=\mathcal{D}_{a}V^{ci}+{{\omega_{a}}^{i}}_{j}V^{cj}~~,
\eeq
where $\mathcal{D}_{a}$ is the worldvolume covariant derivative compatible with $\gamma_{ab}$. With the definition \eqref{torsion}, one can easily act with it on the transverse indices $i$ but at the expense of introducing torsion. Note that the worldvolume covariant derivative $\mathcal{D}_{a}$ can be replaced by the covariant derivative $\nabla_{a}$ introduced in \eqref{covd} if one remembers that $\nabla_{a}$ does not act on the transverse indices $i$. In the work presented above we have avoided using this terminology by noting that
\beq \label{trickW}
\tilde\nabla_{a}V^{ai}={n^{i}}_{\mu}\nabla_{a}\left(V^{cj}{n_{j}}^{\mu}\right)=\nabla_{a}V^{ci}-V^{cj}{n_{j}}^{\mu}\nabla_{a}{n^{i}}_{\mu}=\nabla_{a}V^{ci}+{{\omega_{a}}^{i}}_{k}V^{ck}~~,
\eeq
where in the last equality we have used the definition of the extrinsic twist potential. With the definition \eqref{torsion} the variation of the extrinsic curvature \eqref{perpK} can actually be written as \cite{Capovilla:1994bs}
\beq 
\delta_\perp {K_{ab}}^{i}=\tilde\nabla_{a}\tilde\nabla_{b}\Phi^{i}-{R^{i}}_{baj}\Phi^{j}-{K_{ac}}^{i}{K^{c}}_{bj}\Phi^{j}~.
\eeq
Similarly, for the extrinsic twist potential we have that \cite{Capovilla:1994bs}
\beq 
\delta_{\perp} {\omega_{a}}^{ij}=-{K_{ab}}^{i}\tilde \nabla^{b}\Phi^{k}+{K_{ab}}^{j}\tilde\nabla^{b}\Phi^{k}+{R^{ij}}_{ka}\Phi^{k}~.
\eeq
For variations along the worldvolume directions, the different fields transform with the Lie derivative such that covariance under normal rotations is preserved. For example, the tangential variation of the twist is
\beq
\delta_{||} {\omega_{a}}^{ij}=\Phi^{b}\tilde\nabla_{b}{\omega_{a}}^{ij}+{\omega_{b}}^{ij}\nabla_{a}\Phi^{b}~~,
\eeq
which when using \eqref{trickW} leads to \eqref{deltaW}. An interesting application of this covariant derivative is in the rewriting of the spin conservation equation \eqref{scons} which now takes the form
\beq
\tilde\nabla_{a}j^{aij}=0~~.
\eeq
We have also considered the variation of the worldvolume Christoffel symbols ${\gamma_{ac}}^{b}$ as well as of the worldvolume Riemann tensor $\mathcal{R}_{abcd}$. These can be analyzed through the variation of the Christoffel symbols,
\beq \label{Dchris}
\delta {\gamma_{ab}}^{c}=\frac{1}{2}\gamma^{cd}\left(\nabla_{b}\delta\gamma_{ad}+\nabla_{a}\delta\gamma_{bd}-\nabla_{d}\delta\gamma_{ab}\right)~.
\eeq
Using this, one has that
\beq
\delta {\mathcal{R}^{a}}_{bcd}=\nabla_{c}\delta {\gamma_{bd}}^{a}-\nabla_{d}\delta {\gamma_{bc}}^{a}~~,~~\delta \mathcal{R}_{ab}=\nabla_{c}\delta {\gamma_{ab}}^{c}-\nabla_b\delta {\gamma_{ac}}^{c}~~,
\eeq
and finally the variation of the worldvolume Ricci scalar
\beq
\delta {\mathcal{R}}=\nabla_a\left(\gamma^{bc}\delta {\gamma_{bc}}^{a}\right)-\nabla^{b}\left(\delta {\gamma_{ab}}^{a}\right)+\mathcal{R}_{ab}\delta\gamma^{ab}~~.
\eeq
We have also considered variations of the worldvolume derivative of the extrinsic curvature for co-dimension-1 surfaces. This can be obtained using the chain rule,
\beq
\delta \nabla_{a}K_{bc}=\nabla_{a}\delta K_{bc}+\delta {\gamma_{ab}}^{d}{K_{dc}}+\delta {\gamma_{ac}}^{d}K_{bd}~~.
\eeq

\subsubsection*{The third fundamental tensor}
In \cite{Carter:1992vb} Carter defined the third fundamental tensor with space-time indices as
\beq \label{3rd}
{\Xi_{\kappa\lambda\mu}}^{\nu}={\gamma_{\lambda}}^{\rho}{\gamma_{\mu}}^{\sigma}{\perp_{\tau}}^{\nu}\bar\nabla_{\kappa}{K_{\rho\sigma}}^{\tau}~~,
\eeq
where the connection $\bar\nabla_{\kappa}$ denotes the covariant derivative along the worldvolume written with space-time indices. We have avoided this notation in the work above by using instead the definition
\beq
\bar\nabla_{\nu}V^{\mu}={\gamma^{\lambda}}_{\nu}\nabla_{\lambda}V^{\mu}~~.
\eeq
From the third fundamental tensor \eqref{3rd} we can obtain the Codazzi-Mainardi equation simply by taking the appropriate anti-symmetrization \cite{Carter:1992vb}
\beq
2{\Xi_{[\kappa\lambda]\mu}}^{\nu}={\gamma^{\rho}}_{\kappa}{\gamma^{\sigma}}_{\lambda}{\gamma^{\tau}}_{\mu}{\perp^{\nu}}_{\alpha}{{R_{\rho\sigma}}^{\alpha}}_{\tau}~~.
\eeq
Contracting this with ${u^{\lambda}}_{a}{u^{\kappa}}_{b}{u^{\mu}}_{c}{n^{i}}_{\nu}$ yields Eq.~\eqref{CM}.

\subsubsection*{Variations \emph{\`{a} la Carter}}
For the geodynamic-type models constructed by Carter \cite{Carter:1994yt, Carter:1997pb}, a different type of variational principle was used than the one used by Capovilla-Guven \cite{Capovilla:1994bs}. It is a Lagrangian variation in which the background metric $g_{\mu\nu}$ is displaced by an infinitesimal vector such that $\delta_{L}g_{\mu\nu}=2\nabla_{(\mu}\Phi_{\nu)}$. We now show that this type of variations yields the same results as the ones presented in this work. In order to do so we require the Lagrangian variations of the first fundamental tensor $\delta_L \gamma^{\mu\nu}=-2{\gamma_{\sigma}}^{(\mu}\bar \nabla^{\nu)}\Phi^{\sigma}$ and of the second fundamental tensor \cite{Carter:1997pb},
\beq
\begin{split}
\delta_L {K_{\mu\nu}}^{\rho}=&{\perp^{\rho}}_{\lambda}\left(\bar\nabla_{(\mu}\bar\nabla_{\nu)}\Phi^{\lambda}-{\gamma^{\sigma}}_{(\mu}{\gamma^{\tau}}_{\nu)}{R^{\lambda}}_{\sigma\tau\rho}\Phi^{\rho}-{K^{\sigma}}_{(\mu\nu)}\bar\nabla_{\sigma}\Phi^{\lambda}\right) \\
&+\left(2{\perp^{\sigma}}_{(\mu}{K_{\nu)\tau}}^{\rho}-{g^{\rho}}_{\tau}{K_{\mu\nu}}^{\sigma}\right)\left(\nabla_{\sigma}\Phi^{\tau}+\bar\nabla^{\tau}\Phi_{\sigma}\right)~~.
\end{split}
\eeq
Let us consider the simple case of the elastic contribution $\lambda_{1}(\textbf{k})$ given in \eqref{elastic2}. The contribution to the action is of the form
\beq \label{actioncarter}
I\thinspace[X^\mu]=\int_{\mathcal{W}_{p+1}}~\sqrt{-\gamma}\lambda_{1}(\textbf{k})K^{\rho}K_{\rho}~~,
\eeq
where $\gamma$ here should be understood as the determinant of $\gamma_{\mu\nu}$ and $\textbf{k}=|-\gamma_{\mu\nu}\textbf{k}^{\mu}\textbf{k}^{\nu}|^{1/2}$. We define the stress-energy tensor $T^{\mu\nu}$ and the dipole moment ${\mathcal{D}^{\mu\nu}}_{\rho}$ as before:
\beq
T^{\mu\nu}=\frac{2}{\sqrt{-\gamma}}\frac{\delta \mathcal{L}}{\delta\gamma_{\mu\nu}}~~,~~{\mathcal{D}^{\mu\nu}}_{\rho}=\frac{1}{\sqrt{-\gamma}}\frac{\delta \mathcal{L}}{\delta{K_{\mu\nu}}^{\rho}}~~,
\eeq
where $\mathcal{L}=\sqrt{-\gamma}\lambda_{1}(\textbf{k})K^{\rho}K_{\rho}$. The Lagrangian variations of \eqref{actioncarter} are thus of the form
\beq \label{dLcarter}
\delta_L I\thinspace[X^{\mu}]=\int_{\mathcal{W}_{p+1}}\sqrt{-\gamma}\left(\frac{1}{2}T^{\mu\nu}\delta_L \gamma_{\mu\nu}+\lambda_{1}(\textbf{k})K^{\mu}K^{\nu}\delta_L g_{\mu\nu}+{\mathcal{D}^{\mu\nu}}_{\rho}\delta_L{K_{\mu\nu}}^{\rho}\right)~~.
\eeq
The difference here in comparison with \eqref{var1} is the appearance of the middle term above. However, it is always cancelled by the last of the terms appearing from the contraction ${\mathcal{D}^{\mu\nu}}_{\rho}\delta_L{K_{\mu\nu}}^{\rho}$. The end result of the variation for all corrections quadratic in the extrinsic curvature is thus
\beq \label{dcarter}
\begin{split}
\delta_L I\thinspace[X^{\mu}]=\int_{\mathcal{W}_{p+1}}\sqrt{-\gamma}&\left(\bar\nabla_{\nu}\Big[\Phi_\lambda\left(T^{\nu\lambda}-\bar\nabla_\mu\mathcal{D}^{\mu\nu\lambda}\right)+\mathcal{D}^{\mu\nu\lambda}\bar\nabla_{\mu}\Phi_{\lambda}\right] \\
&-\Phi_{\lambda}\left(\bar\nabla_{\nu}\left(T^{\nu\lambda}-\bar\nabla_{\mu}\mathcal{D}^{\mu\nu\lambda}\right)+\mathcal{D}^{\sigma\tau\rho}{R_{\rho\sigma\tau}}^{\lambda}\right)\Big)~.
\end{split}
\eeq
The advantage of this approach is that it yields in the end the equations of motion and boundary conditions in the form \eqref{pdc} and makes it easy to identify the linear momentum $\mathcal{P}^{\nu\lambda}$. Careful inspection of this and comparison with \eqref{var1} leads one to conclude that the variations are the same. For the particular case of \eqref{actioncarter} we have
\beq
T^{\mu\nu}=\lambda_{1}(\textbf{k})K^{\rho}K_{\rho}\gamma^{\mu\nu}-\lambda_{1}'(\textbf{k})\textbf{k}u^{\mu}u^{\nu}K^{\rho}K_{\rho}-4\lambda_{1}(\textbf{k}){K^{\mu\nu\rho}}K_{\rho}~~,~~\mathcal{D}^{\mu\nu\rho}=2\lambda_{1}(\textbf{k})\gamma^{\mu\nu}K^{\rho}~~,
\eeq
which when contracted with ${u^{a}}_{\mu}{u^{b}}_{\nu}$ yields the contributions $\tau^{ab}_{1}$ and $\mathcal{D}^{abi}_{1}$ stated in the table given in Sec.~\ref{2nd}. We have mentioned that the middle term in \eqref{dLcarter} is always cancelled by a contraction involving the last term. We briefly show how this is the case. Take the case of the most general action quadratic in the extrinsic curvature
\beq \label{actioncarter2}
I\thinspace[X^\mu]=\int_{\mathcal{W}_{p+1}}\sqrt{-\gamma}\thinspace\frac{1}{2}\thinspace\mathcal{Y}^{\mu\nu\lambda\rho}{K_{\mu\nu}}^{\sigma}{K_{\lambda\rho\sigma}}~~,
\eeq
where the Young modulus $\mathcal{Y}^{\mu\nu\lambda\rho}$ only depends on $\textbf{k}$, $\textbf{k}^a$ and $\gamma^{\mu\nu}$. Note that $\mathcal{D}^{\mu\nu\rho}=\mathcal{Y}^{\mu\nu\lambda\sigma}{K_{\lambda\sigma}}^{\rho}$. Now introduce a dummie (meaningless in this case) tensor $\mathcal{B}^{\mu\nu}$ defined as
\beq
\mathcal{B}^{\mu\nu}=\frac{1}{\sqrt{-\gamma}}\frac{\delta \mathcal{L}}{\delta\perp_{\mu\nu}}~~,
\eeq
where $\mathcal{L}=(1/2)\sqrt{-\gamma}\thinspace\mathcal{Y}^{\mu\nu\lambda\rho}{K_{\mu\nu}}^{\sigma}{K_{\lambda\rho\sigma}}$. Then for an action of the type \eqref{actioncarter2} this is simply given by $\mathcal{B}^{\mu\nu}=(1/2)\mathcal{Y}^{\sigma\kappa\lambda\rho}{K_{\sigma\kappa}}^{\mu}{K_{\lambda\rho}}^{\nu}$. Further note that the last term in the contraction ${\mathcal{D}^{\mu\nu}}_{\rho}\delta_L{K_{\mu\nu}}^{\rho}$ is always $-\mathcal{D}^{\mu\nu\rho}{K_{\mu\nu}}^{\sigma}\nabla_{\sigma}\Phi_{\rho}$. Therefore, $\mathcal{B}^{\mu\nu}\delta_Lg_{\mu\nu}$ always cancels this term.

\subsubsection*{Linear momentum from the action}
As mentioned in Sec.~\ref{currents} it is possible to obtain the conserved surface currents directly from \eqref{actioncarter} by requiring the action to be invariant under space-time translations along Killing directions. To see this note that for an infinitesimal shift along a Killing direction $\Phi^{\mu}=\beta \thinspace\textbf{k}^{\mu}$ for an arbitrary constant $\beta$. Then \eqref{dcarter} yields
\beq \label{lcarter}
\delta_L I\thinspace[X^{\mu}]=\beta\int_{\mathcal{W}_{p+1}}\sqrt{-\gamma}\left(\bar\nabla_{\nu}\left(\mathcal{P}^{\nu\lambda}\textbf{k}_{\lambda}+\mathcal{D}^{\nu\mu\lambda}\bar\nabla_{\mu}\textbf{k}_{\lambda}\right)-\textbf{k}_{\lambda}\mathcal{E}^{\lambda}(X^{\mu})\right)~~.
\eeq
Here we have used the definition of linear momentum applied to the case $j^{a\mu\nu}=0$ as given in \eqref{pdipole} with the identification \eqref{idelastic}. Furthermore, we have written the term inside the parenthesis in the second line of \eqref{dcarter} as $\mathcal{E}^{\lambda}(X^{\mu})$, which when equated to zero yields the equations of motion \eqref{pdeq1}-\eqref{pdeq2} when $j^{a\mu\nu}=0$. For \eqref{lcarter} to be invariant under these translations when the equations of motion are satisfied $\mathcal{E}^{\lambda}(X^{\mu})=0$ we must have that
\beq \label{ccarter}
\bar\nabla_{\nu}\left(\mathcal{P}^{\nu\lambda}\textbf{k}_{\lambda}+\mathcal{D}^{\nu\mu\lambda}\bar\nabla_{\mu}\textbf{k}_{\lambda}\right)=0~~.
\eeq
Indeed, this is recognized as the conservation of the surface current introduced in \eqref{ansatzP} for the case $j^{a\mu\nu}=0$. This is the reason why $\mathcal{P}^{\nu\lambda}$ is called the linear momentum. Moreover, from \eqref{ccarter}, since $\mathcal{P}^{[\nu\lambda]}=-\bar{\nabla}_{\mu}\mathcal{D}^{\mu[\nu\lambda]}$, we have that
\beq
\textbf{k}_{\lambda}\left(\bar\nabla_{\nu}\mathcal{P}^{\nu\lambda}+\mathcal{D}^{\nu\mu\rho}{R^{\lambda}}_{\nu\mu\rho}\right)=0~~.
\eeq
Therefore, along any Killing direction we obtain the equations of motion as written in \eqref{pdc}. Finally, we note that when $\mathcal{E}^{\lambda}(X^{\mu})=0$, integrating the first term in \eqref{lcarter} into a boundary term yields the definition of conserved surface charge \eqref{Qs}. This type of reasoning had been applied by Guven et al. in \cite{Arreaga:2000mr} for extremal branes in flat space. Here we have generalized it for non-extremal branes in curved space. 

\section{Boundary conditions for hydrodynamic modes} \label{boundary}
In this appendix we analyze the boundary conditions for the hydrodynamic corrections presented in \eqref{hydro2} and \eqref{cod12}. We begin with the hydrodynamic scalars given in \eqref{hydro}. Due to the variation of the Christoffel symbols \eqref{Dchris} the hydrodynamic scalars contribute with additional terms to the boundary equations derived in \eqref{bound1}. We denote the extra contribution by $\mathcal{B}_{\alpha}$~ in terms of $\delta\gamma_{ab}$ given in \eqref{dgamma} and $\delta{\gamma_{ab}}^{c}$ given in \eqref{Dchris}, which should be added to the last line of Eq.~\eqref{bound1}. These contributions are summarized in the following table:
\\
\newcolumntype{C}[1]{>{\centering\let\newline\\\arraybackslash\hspace{0pt}}m{#1}}
\renewcommand{\arraystretch}{1.7}
\begin{center} 
    \begin{tabular}{ | c | C{15cm} |}
    \hline
    \color{red}{Scalar} & \color{red}{$\mathcal{B}_{\alpha}$} \\ \hline
    $\upsilon_1(\textbf{k})\mathcal{V}_1$ & $-\eta_a\upsilon_1(\textbf{k})\left(\frac{1}{2}\nabla^{a}\left(\textbf{k}u^{b}u^{c}\delta\gamma_{bc}\right) +\gamma^{ab}\delta\gamma_{bc}\nabla^{c}\textbf{k}\right)+\frac{\eta_a}{2}\left(\textbf{k}u^{b}u^{c}\delta\gamma_{cd}\nabla^{a}\upsilon_1(\textbf{k})+\upsilon_1(\textbf{k})\gamma^{bc}\delta\gamma_{bc}\nabla^{a}\textbf{k}\right)$ \\   \hline
     $\upsilon_2(\textbf{k})\mathcal{V}_2$ &$-\eta_a\upsilon_2(\textbf{k})\left(\gamma^{ac}\delta{\gamma_{bc}}^{b}-\gamma^{bc}\delta{\gamma_{bc}}^{a}\right)-\eta_a\left(\gamma^{ac}\delta\gamma_{cd}\nabla^{b}\upsilon_2(\textbf{k})-\gamma^{bc}\delta\gamma_{bc}\nabla^{a}\upsilon_2(\textbf{k})\right)$ \\   \hline
      $\upsilon_3(\textbf{k})\mathcal{V}_3$ & $-\eta_a\left(\frac{1}{2}\nabla^{a}\left(\upsilon_3(\textbf{k})\textbf{k}^{b}\textbf{k}^{c}\right)\delta\gamma_{bc}-\nabla^{c}\left(\upsilon_3(\textbf{k})\textbf{k}^{a}\textbf{k}^{b}\right)\delta\gamma_{bc}+\frac{1}{2}\gamma^{bc}\delta\gamma_{bc}\nabla_{b}\left(\upsilon_3(\textbf{k})\textbf{k}^{a}\textbf{k}^{b}\right)\right)+\eta_a\upsilon_3(\textbf{k})\left(\textbf{k}^{b}\textbf{k}^{c}\delta{\gamma_{bc}}^{a}-\textbf{k}^{a}\textbf{k}^{b}\delta{\gamma_{bc}}^{c}\right)$ \\   \hline
     \end{tabular}
      \label{tab:hydro}
\end{center} 
\vskip 0.3cm
The complexity of these contributions is useless. For the hydrodynamic corrections \eqref{hydro} to fit the pole-dipole boundary conditions \eqref{pdb}, one must require all the above terms to vanish at the brane boundary. 

\subsubsection*{Boundary contributions due to the quadrupole moment}
We now turn into the case of the boundary contributions of the terms \eqref{cod12} due to the presence of the quadrupole moment. Besides the boundary terms obtained from \eqref{var1} there is a contribution to the variation of the action \eqref{var1} of the form:
\beq \label{varq}
\begin{split}
\delta I\thinspace\!=\!\!\int_{\mathcal{W}_{p+1}}\!\!\!\!\!\!\!\sqrt{-\gamma}\thinspace\nabla_{a}\Big(&\mathcal{D}^{abc}_{\perp}\delta {K}_{bc}-\nabla_{b}\mathcal{D}^{bac}_{\perp}\nabla_c\Phi^{i}-\Phi^{i}\nabla_b\nabla_c\mathcal{D}^{cba}_{\perp}-2\mathcal{D}^{bac}{K^{d}}_{c}{K_{bd}}^{i}\Phi_{i} \\
&-2\mathcal{D}^{abc}_{\perp}{K^{d}}_{c}{K_{bd}}^{i}\Phi_{i}-2\mathcal{D}^{dbc}_{\perp}{K^{a}}_{c}{K_{bd}}^{i}\Phi_{i}+\mathcal{D}^{bac}_{\perp}{K^{d}}_{c}\nabla_{(b}\Phi_{d)}+2\mathcal{D}^{abc}_{\perp}{K^{d}}_{c}\nabla_{(b}\Phi_{d)} \\
&-2\mathcal{D}^{dbc}_{\perp}{K^{a}}_{c}\nabla_{(d}\Phi_{b)}+\Phi^d\mathcal{D}^{abc}_{\perp}\nabla_{d}K_{bc}+2\mathcal{D}^{bac}_{\perp}\Phi^{d}\nabla_{d}K_{dc}\Big)~~.
\end{split}
\eeq
Note that we used here the index $i$ but one should remember that for co-dimension-1 surfaces there is only one transverse direction. Note also that the first term in \eqref{varq} introduces boundary terms proportional to the background Riemann tensor via \eqref{perpK}. This is a generic effect of quadrupole corrections to the equations of motion in the spirit of Sec.~\ref{poledipole}. Deriving the pole-quadrupole equations of motion is of intrinsic interest and it will be published elsewhere.

\addcontentsline{toc}{section}{References}
\footnotesize
\providecommand{\href}[2]{#2}\begingroup\raggedright\endgroup


\end{document}